\documentclass[12pt]{article}

\usepackage[totalwidth=460truept,totalheight=624truept]{geometry}
\usepackage{latexsym,graphicx,amsfonts,amssymb,amsmath,xcolor,changepage}
\usepackage[hypertexnames=false,hidelinks]{hyperref}

\def\theequation{\arabic{section}.\arabic{equation}}

\renewcommand{\theequation}{\thesection.\arabic{equation}}
\global\arraycolsep=1truept
\renewcommand{\theequation}{\arabic{section}.\arabic{equation}}
\null

\begin{document}

\medskip

\begin{center}
{\Huge \textbf{Quantum Field Theory of}}

\vskip.5truecm

{\Huge \textbf{Physical and Purely Virtual Particles}}

\vskip.5truecm

{\Huge \textbf{in a Finite Interval of\ Time}}

\vskip.7truecm

{\Huge \textbf{on a Compact Space Manifold:}}

\vskip.5truecm

{\Huge \textbf{Diagrams, Amplitudes and Unitarity}}
\end{center}

\vskip.5truecm

\begin{center}
\textsl{Damiano Anselmi}

\vskip.1truecm

{\small \textit{Dipartimento di Fisica \textquotedblleft
E.Fermi\textquotedblright , Universit\`{a} di Pisa, Largo B.Pontecorvo 3,
56127 Pisa, Italy}}

{\small \textit{INFN, Sezione di Pisa, Largo B. Pontecorvo 3, 56127 Pisa,
Italy}}

{\small damiano.anselmi@unipi.it}

\vskip .5truecm

\textbf{Abstract}
\end{center}

We provide a diagrammatic formulation of perturbative quantum field theory
in a finite interval of time $\tau $, on a compact space manifold $\Omega $.
We explain how to compute the evolution operator $U(t_{\text{f}},t_{\text{i}%
})$ between the initial time $t_{\text{i}}$ and the final time $t_{\text{f}%
}=t_{\text{i}}+\tau $, study unitarity and renormalizability, and show how
to include purely virtual particles, by rendering some physical particles
(and all the ghosts, if present) purely virtual. The details about the
restriction to finite $\tau $ and compact $\Omega $ are moved away from the
internal sectors of the diagrams (apart from the discretization of the
three-momenta), and coded into external sources. Unitarity is studied by
means of the spectral optical identities, and the diagrammatic version of
the identity $U^{\dag }(t_{\text{f}},t_{\text{i}})U(t_{\text{f}},t_{\text{i}%
})=1$. The dimensional regularization is extended to finite $\tau $ and
compact $\Omega $, and used to prove, under general assumptions, that
renormalizability holds whenever it holds at $\tau =\infty $, $\Omega =%
\mathbb{R}^{3}$. Purely virtual particles are introduced by removing the
on-shell contributions of some physical particles, and the ghosts, from the
core diagrams, and trivializing their initial and final conditions. The
resulting evolution operator $U_{\text{ph}}(t_{\text{f}},t_{\text{i}})$ is
unitary, but does not satisfy the more general identity $U_{\text{ph}%
}(t_{3},t_{2})U_{\text{ph}}(t_{2},t_{1})=U_{\text{ph}}(t_{3},t_{1})$. As a
consequence, $U_{\text{ph}}(t_{\text{f}},t_{\text{i}})$ cannot be derived
from a Hamiltonian in a standard way, in the presence of purely virtual
particles.

\vfill\eject

\section{Introduction}

\label{intro}\setcounter{equation}{0}

The success of perturbative quantum field theory relies on the theory of
scattering, and the tests of its predictions in colliders. The $S$ matrix
amplitudes describe scattering processes among \textquotedblleft asymptotic
states\textquotedblright , which are free, and far from the interaction
region. Nevertheless, quantum field theory is much more than the $S$ matrix,
and can in principle make predictions about all types of processes. For
example, we can consider the effects of a scattering among particles that
are still interacting. One day, we might want to build colliders to test
those predictions.

While there is no conceptual difficulty in formulating quantum field theory
in a finite interval of time $\tau $ and on a compact space manifold $\Omega 
$, and various approaches can be found in the literature, it is worth to
make an effort to identify the formulation that is closer to the one we are
accustomed to at $\tau =\infty $, $\Omega =\mathbb{R}^{3}$. If so, we can
generalize the known properties and theorems with a minimum effort,
efficiently study key principles like unitarity and renormalizability, and
possibly extend to formulation to purely virtual particles \cite{PVP20}. It
may be challenging to distinguish what is virtual from what is real, what is
on the mass shell and what is not, in a finite interval of time, and on a
compact manifold, so the investigation may hold intriguing surprises.

The first task is to relate as much as possible the diagrams of perturbative
quantum field theory in a finite interval of time $\tau $, and on a compact
space manifold $\Omega $, to the usual diagrams of the $S$ matrix
amplitudes. We achieve this goal by removing (almost all) the details about
the restriction to finite $\tau $ and compact $\Omega $ from the internal
sectors of the diagrams, and dumping them on appropriate external sources
coupled to the vertices. Only the discretization of the momenta\footnote{%
Throughout this paper, we use a nonrelativistic terminology, where
\textquotedblleft momentum\textquotedblright\ means three-momentum (in four
spacetime dimensions), or $(D-1)$-momentum (in $D$ spacetime dimensions).
Only the momenta are discretized, while\ the energies are not.}, due to the
restriction to a compact $\Omega $, enters the loop integrals. This
\textquotedblleft contamination\textquotedblright\ is the maximum allowed to
generalize the study of unitarity along the lines of ref. \cite%
{diagrammarMio}, that is to say, by means of spectral optical identities,
which are purely albegraic and hold threshold by threshold, for arbitrary
frequencies, before integrating on the loop momenta (or summing on their
discretized versions, on a compact $\Omega $). In the end, the diagrams look
like ordinary Feynman diagrams, apart from the discretization of the loop
momenta, and the insertion of an external source for every vertex. The usual
diagrammatic properties and techniques hold unmodified, or can be extended
easily.

These goals are achieved efficiently in the approach based on coherent
states \cite{cohe}. In every other approach they require more effort, but it
is always possible to obtain equivalent results by means of a change of
basis, starting from the coherent-state approach.

We consider theories with Hermitian Lagrangians. The free Hamiltonians may
be bounded from below or not, depending on whether the theory contains only
physical particles, or includes ghosts (particles with kinetic terms
multiplied by the wrong signs). If the theory just contains physical
particles, the evolution operator $U(t_{\text{f}},t_{\text{i}})$ between the
initial time $t_{\text{i}}$ and the final time $t_{\text{f}}=t_{\text{i}%
}+\tau $ is unitary: $U^{\dag }(t_{\text{f}},t_{\text{i}})U(t_{\text{f}},t_{%
\text{i}})=1$. If ghosts are present, an analogous identity holds (called
pseudounitarity equation), but cannot be interpreted as unitarity. A theory
of physical particles and ghosts also satisfies the more general identity%
\begin{equation}
U(t_{3},t_{2})U(t_{2},t_{1})=U(t_{3},t_{1}),  \label{unitar}
\end{equation}%
for arbitrary $t_{1}$, $t_{2}$ and $t_{3}$.

We study these properties diagrammatically. Specifically, we decompose (\ref%
{unitar}) into Cutkosky-Veltman identities \cite{cutkoskyveltman} (see also 
\cite{thooft}). Then, we further decompose those identities into spectral
optical identities, by separating the thresholds from one another, following
ref. \cite{diagrammarMio}. At that point, it is relatively straightforward
to turn a physical particle (or a ghost) into a purely virtual particle,
when needed, by trivializing its initial and final conditions, and removing
the contributions to the spectral optical identities where the particle
would be on shell. This can be done according to the procedure outlined in
ref. \cite{diagrammarMio}, or by replacing the cores of the diagrams with
appropriate non time-ordered versions, as explained in ref. \cite{PVP20}.
Interestingly enough, the physical evolution operator $U_{\text{ph}}(t_{%
\text{f}},t_{\text{i}})$ of a theory that contains both physical and purely
virtual particles turns out to be unitary for arbitrary initial and final
times: $U_{\text{ph}}^{\dag }(t_{\text{f}},t_{\text{i}})U_{\text{ph}}(t_{%
\text{f}},t_{\text{i}})=1$. However, it does not satisfy (\ref{unitar}), and
cannot be derived from a Hamiltonian in a standard way.

Purely virtual particles are particles that cannot exist on the mass shell
at any order of the perturbative expansion. They are not physical particles,
nor ghosts, but sort of \textquotedblleft fake\textquotedblright\ particles.
It is possible to introduce them by removing all the on-shell contributions
due to a physical particle or a ghost in one of the following three
equivalent ways: $i$) a nonanalytic Wick rotation \cite{Piva,LWgrav}, $ii$)
a certain manipulation of the spectral optical identities, to remove the
unwanted on-shell contributions as explained in ref. \cite{diagrammarMio},
and $iii$)\ the use of non-time-ordered diagrams, instead of the standard
diagrams \cite{PVP20}. In all cases, the basic ingredients are two: $a$) a 
\textit{prescription} to modify the interiors of the diagrams, and $b$) a 
\textit{projection} to drop the unwanted particles from the external states.
The final theory is unitary, provided all the ghosts are rendered purely
virtual. It is important to stress that I) both physical particles and
ghosts can be rendered purely virtual, and II) purely virtual particles are
not \cite{LWfakeons} Lee-Wick ghosts \cite{leewick}\footnote{%
For Lee-Wick ghosts in quantum gravity, see \cite{tomboulis}.}, so they do
not need to have nonvanishing widths, and decay. The main application of the
idea is the formulation of a theory of quantum gravity \cite{LWgrav}, which
provides testable predictions \cite{ABP} in inflationary cosmology \cite%
{CMBStage4}. The diagrammatic calculations are not much more difficult than
with physical particles, and it is possible to implement them in softwares
like FeynCalc, FormCalc, LoopTools and Package-X \cite{calc}. At the
phenomenological level, purely virtual particles open interesting
possibilities, because they evade many constraints that are typical of
normal particles (see \cite{PivaMelis} and references therein).

We show that, whenever a theory is renormalized at $\tau =\infty $, $\Omega =%
\mathbb{R}^{3}$, it is also renormalized at finite $\tau $ and on a compact
space manifold $\Omega $. The counterterms are the same at the Lagrangian
level, up to total derivatives (which are not renormalized). These results
are not surprising, considering that the ultraviolet divergences are local,
and concern the behaviors of the correlation functions at infinitesimal
distances and intervals of time: renormalization should know nothing about
global restrictions on $\tau $ and $\Omega $. To prove the statements just
made, we first extend the analytic \cite{anreg} and dimensional \cite{dimreg}
regularization techniques to finite $\tau $ and compact $\Omega $. Then we
use the extended techniques to show that everything works as expected, apart
from minor changes that do not modify the final outcome.

\bigskip

We recall that the coherent states \cite{cohe} are the eigenstates of the
annihilation operator. In the functional-integral (Lagrangian) approach, the
switch to coherent states\ simply amounts to making a change of variables
from coordinates and momenta $q$, $p$ to $z\sim q+ip$, $\bar{z}\sim q-ip$
(and similarly for the fields), and setting the initial conditions on $z$,
the final conditions on $\bar{z}$. For convenience, we keep referring to the
new variables $z$ and $\bar{z}$ by means of the Hamiltonian terminology
\textquotedblleft coherent states\textquotedblright\ \footnote{%
Details on the correspondence between the operatorial coherent-state
approach and the functional integral can be found in the paragraph 9-1-2 of 
\cite{Itzykson-Zuber}.}.

Ultimately, the formalism we develop in this paper gives a diagrammatic
interpretation of the evolution operator $U=\mathrm{e}^{-iHt}$. As such, it
is supposed to work even for the scattering of particles with long-range
interactions, or if the timescale of the experiment is short enough so that
the process is not well-approximated by the $S$ matrix. At the same time, it
retains the perturbative character of the standard approaches to the $S$
matrix amplitudes. One has to check, on a case by case basis, whether the
perturbative expansion is effectively useful, i.e., whether the radiative
corrections are smaller or bigger than the contributions they are supposed
to correct. It may be possible to choose the space manifold $\Omega $ in
order to reduce the effective range of the interactions, and identify new
situations where it makes sense to compare the experimental results with the
predictions obtained by truncating the perturbative expansion to the first
few orders. That said, any time it makes sense to use the evolution operator 
$U=\mathrm{e}^{-iHt}$ perturbatively around the free limit, the results of
this paper provide a diagrammatic way to do it systematically.

The $S$ matrix amplitudes are built by switching to the interaction picture,
and changing the basis to (in and out) asymptotic states, identified as
residues of the propagators of the external legs, which are then amputated.
This part, which is necessary to deal with the asymptotic limit, is not
affected by our discussion.

\bigskip

It is convenient to list here the main properties of the diagrammatic rules
we find, starting from those that apply to the coherent-state framework.

\begin{itemize}
\item {The frequencies are discrete, while the energies are continuous:
Fourier series are used for momenta, while Fourier transforms are used for
energies.}

\item {In theories with physical particles and possibly ghosts:}

\begin{itemize}
\item {The cores of the diagrams are variants of the usual Feynman diagrams,
where the momenta are discretized, and a suitable external source $K$ is
attached to each vertex.}

\item {The sources $K$ and the discretization of the momenta are the sole
information about the restriction to finite $\tau $ and compact $\Omega $.}

\item {The analytic/dimensional regularization technique can be generalized
to finite $\tau $ and compact $\Omega $.}

\item {Once a theory is renormalized at $\tau =\infty $, $\Omega =\mathbb{R}%
^{3}$, it is renormalized at finite $\tau $ and compact $\Omega $, and the
counterterms are the same.}

\item Unitarity, pseudounitarity and the identity (\ref{unitar}) can be
translated diagrammatically into Cutkosky-Veltman identities, \textit{\`{a}
la} \cite{cutkoskyveltman}.

\item The Cutkosky-Veltman identities can be decomposed threshold by
threshold into algebraic, spectral optical identities, \textit{\`{a} la} 
\cite{diagrammarMio}.
\end{itemize}

\item Purely virtual particles can be introduced by rendering some physical
particles (and all the ghosts, if any are present) purely virtual. The
features of the theories of physical and purely virtual particles are:

\begin{itemize}
\item The cores of the diagrams are replaced by appropriate non time ordered
diagrams.

\item Equivalently, the contributions to the spectral optical identities
where the purely virtual particles would be on shell are removed.

\item No external states are associated with purely virtual particles.

\item The initial and final conditions obeyed by purely virtual particles
are trivial. However, their boundary conditions (referring to the boundary
of $\Omega $) need not be trivial.

\item The physical evolution operator $U_{\text{ph}}(t_{\text{f}},t_{\text{i}%
})$ is unitary.

\item The more general identity (\ref{unitar}) does not hold.

\item It is not possible to derive $U_{\text{ph}}(t_{\text{f}},t_{\text{i}})$
from a Hamiltonian in a standard way.
\end{itemize}
\end{itemize}

In a generic approach (not based on coherent states), the propagators have
additional \textquotedblleft on-shell\textquotedblright\ contributions and
infinitely many singularities. A change of basis from the coherent-state
approach to an arbitrary one ensures that all the singularities mutually
cancel out, and any property we prove with coherent states is general.

Although it is possible to perform the Wick rotation to Euclidean space, we
always work in Minkowski spacetime, because unitarity is better studied
there. The connection with finite temperature quantum field theory is not
obvious, and should be worked out separately.

We mostly work in four spacetime dimensions, or in quantum mechanics, but
the results hold in an arbitrary number $D$ of spacetime dimensions. When we
dimensionally regularize, we understand that $D$ is a complex parameter.

Throughout the paper, we work with scalar bosons. The generalization to
fermions is straightforward. In the cases of gauge theories and gravity, we
can apply the techiques developed here with convenient gauge choices, such
as the Feynman gauge. A more general setting (working with arbitrary gauges
and arbitrary gauge-fixing parameters is useful to prove the gauge
independence of physical quantities, and, in practical computations, make
checks of the results) requires to overcome certain technical obstacles,
which are dealt with in a separate paper \cite{MQQG}. 

We work in infinite volume $\Omega =\mathbb{R}^{3}$ till section \ref{volume}%
, where we switch to a compact $\Omega $.

The closest approach to ours that we have found in the literature is the one
of ref. \cite{japan}, where the basic diagrammatics of the coherent-state
approach in a finite interval of time are layed out. Beyond that, we
restrict to an arbitrary compact space manifold $\Omega $, develop the
systematics of regularization and renormalization, study unitarity, the
identity (\ref{unitar})\ and the spectral optical identities
diagrammatically, and extend the formulation to purely virtual particles.

The paper is organized as follows. In sections \ref{setup} and \ref{position}
we consider the approach based on position eigenstates at finite $\tau $ ($%
\Omega =\mathbb{R}^{3}$), and describe its main difficulties. In section \ref%
{coherent} we switch to the approach based on coherent states, still on $%
\Omega =\mathbb{R}^{3}$. In section \ref{volume} we switch to a compact
space manifold $\Omega $. In section \ref{renorma} we generalize the
analytic/dimensional regularization technique and study the renormalization
of the theory. In section \ref{unitari} we study unitarity, while in section %
\ref{unitareq} we work out the unitarity equations in diagrammatic form. In
section \ref{PV} we extend the formulation to purely virtual particles.
Section \ref{conclusions} contains the conclusions. In appendix \ref%
{diagrammar20} we compute some quantities needed in the paper.

\section{Position-eigenstate approach}

\label{setup}\setcounter{equation}{0}

We begin by working with position eigenstates, and their field analogues,
which have an intuitive interpretation. Unfortunately, they lead to
unnecessary complications. For the moment, we restrict time to a finite
interval $\tau $, but keep $\mathbb{R}^{3}$ as the space manifold.

\subsection{Amplitudes}

Let $\phi $ denote scalar bosonic fields. In the operatorial and
functional-integral formulations, the transition amplitude between initial
and final states $\phi _{\text{i}}(\mathbf{x})$ and $\phi _{\text{f}}(%
\mathbf{x})$ at times $t_{\text{i}}$ and $t_{\text{f}}$ (with $t_{\text{i}%
}<t_{\text{f}}$) reads 
\begin{equation}
\langle \phi _{\text{f}},t_{\text{f}}\hspace{0.01in}|\hspace{0.01in}\phi _{%
\text{i}},t_{\text{i}}\rangle =\langle \phi _{\text{f}}\hspace{0.01in}|%
\mathrm{e}^{-iH_{\lambda }\tau }|\hspace{0.01in}\phi _{\text{i}}\rangle
=\!\!\!\!\!\!\!\!\!\underset{_{\substack{ \phi (t_{\text{i}},\mathbf{x}%
)=\phi _{\text{i}}(\mathbf{x})  \\ \phi (t_{\text{f}},\mathbf{x})=\phi _{%
\text{f}}(\mathbf{x})}}}{\int }\!\!\!\!\!\!\!\!\![\mathrm{d}\phi ]\exp
\left( i\int_{t_{\text{i}}}^{t_{\text{f}}}\mathrm{d}t\int \mathrm{d}^{3}%
\mathbf{x\hspace{0.01in}}L_{\lambda }(\phi (t,\mathbf{x}))\right) ,
\label{ampli}
\end{equation}%
where $\tau =t_{\text{f}}-t_{\text{i}}$, $H_{\lambda }$ is the Hamiltonian
and $\mathbf{\hspace{0.01in}}L_{\lambda }$ is the Lagrangian. We assume that 
$L_{\lambda }$ has the form 
\begin{equation}
L_{\lambda }(\phi )=L_{0}(\phi )+L_{I}(\phi ),\qquad L_{0}(\phi )=\frac{1}{2}%
(\partial _{\mu }\phi )(\partial ^{\mu }\phi )-\frac{m^{2}}{2}\phi ^{2},
\label{lagra}
\end{equation}%
where the interaction term $L_{I}(\phi )$ is proportional to some coupling $%
\lambda $, to be treated perturbatively. In various steps, it may be useful
to assume, as usual, that the squared mass has a small negative imaginary
part ($m^{2}\rightarrow m^{2}-i\epsilon $, $\epsilon >0$).

Let $\phi _{0}(t,\mathbf{x})$ denote the solution of the Klein-Gordon
equation $\partial _{\mu }\partial ^{\mu }\phi _{0}+m^{2}\phi _{0}=0$ with
initial and final conditions $\phi _{0}(t_{\text{i}},\mathbf{x})=\phi _{%
\text{i}}(\mathbf{x})$, $\phi _{0}(t_{\text{f}},\mathbf{x})=\phi _{\text{f}}(%
\mathbf{x})$. We write%
\begin{equation}
\phi (t,\mathbf{x})=\phi _{0}(t,\mathbf{x})+\varphi (t,\mathbf{x}),
\label{espans}
\end{equation}%
so the quantum fluctuation $\varphi (t,\mathbf{x})$ has the simpler boundary
conditions $\varphi (t_{\text{i}},\mathbf{x})=\varphi (t_{\text{f}},\mathbf{x%
})=0$. The action reads%
\begin{equation}
S_{\lambda }(\phi )\equiv \int_{t_{\text{i}}}^{t_{\text{f}}}\!\!\mathrm{d}%
t\!\int \!\mathrm{d}^{3}\mathbf{x\hspace{0.01in}\hspace{0.01in}}L_{\lambda
}(\phi (t,\mathbf{x}))=S_{\lambda }(\phi _{0})+S_{\lambda }(\varphi ,\phi
_{0}),\quad S_{\lambda }(\varphi ,\phi _{0})\equiv \int_{t_{\text{i}}}^{t_{%
\text{f}}}\!\!\mathrm{d}t\!\int \!\mathrm{d}^{3}\mathbf{x\hspace{0.01in}}%
L_{\lambda }(\varphi ,\phi _{0}),  \label{espac}
\end{equation}%
where%
\begin{eqnarray}
S_{\lambda }(\phi _{0}) &=&\frac{1}{2}\int \mathrm{d}^{3}\mathbf{x\hspace{%
0.01in}}\left[ \phi _{\text{f}}(\mathbf{x})\dot{\phi}_{0}(t_{\text{f}},%
\mathbf{x})-\phi _{\text{i}}(\mathbf{x})\dot{\phi}_{0}(t_{\text{i}},\mathbf{x%
})\right] +\int_{t_{\text{i}}}^{t_{\text{f}}}\mathrm{d}t\int \mathrm{d}^{3}%
\mathbf{x\hspace{0.01in}}L_{I}(\phi _{0}),  \notag \\
L_{\lambda }(\varphi ,\phi _{0}) &\equiv &L_{0}(\varphi )+L_{I}(\varphi
,\phi _{0}),\qquad L_{I}(\varphi ,\phi _{0})\equiv L_{I}(\phi _{0}+\varphi
)-L_{I}(\phi _{0}).  \label{pfint}
\end{eqnarray}%
Although the $\varphi $ interaction Lagrangian $L_{I}(\varphi ,\phi _{0})$
may contain $\phi _{0}$-dependent terms that are linear or quadratic in $%
\varphi $, we treat them perturbatively, since they are proportional to $%
\lambda $.

For $t_{\text{i}}>t_{\text{f}}$ we define and compute the amplitudes by
means of the identity 
\begin{equation}
\langle \phi _{\text{f}},t_{\text{f}}\hspace{0.01in}|\hspace{0.01in}\phi _{%
\text{i}},t_{\text{i}}\rangle =\left( \langle \phi _{\text{f}},t_{\text{f}}%
\hspace{0.01in}|\hspace{0.01in}\phi _{\text{i}},t_{\text{i}}\rangle ^{\ast
}\right) ^{\ast }=\langle \phi _{\text{i}},t_{\text{i}}\hspace{0.01in}|%
\hspace{0.01in}\phi _{\text{f}},t_{\text{f}}\rangle ^{\ast },
\label{conjugat}
\end{equation}%
where $\langle \phi _{\text{i}},t_{\text{i}}\hspace{0.01in}|\hspace{0.01in}%
\phi _{\text{f}},t_{\text{f}}\rangle $ is the same as in (\ref{ampli}) with
i $\leftrightarrow $ f. Note that the time ordering becomes anti-time
ordering under complex conjugation. For this reason, the complex conjugation
acts on $m^{2}$ as well, when the prescription $i\epsilon $ is attached to
it. If the fields are not real, we have $\phi _{\text{i}}^{\ast }$ and $\phi
_{\text{f}}^{\ast }$ on the right-hand side.

\subsection{Correlation functions and generating functionals}

As usual, it is convenient to introduce an external source $J$ coupled to
the field $\phi $, by making the replacement $L_{\lambda }\rightarrow
L_{\lambda }+J\phi $ in (\ref{ampli}). This allows us to define the
correlation functions as functional derivatives with respect to $J$. We can
write%
\begin{equation}
\langle \phi _{\text{f}},t_{\text{f}}\hspace{0.01in}|\hspace{0.01in}\phi _{%
\text{i}},t_{\text{i}}\rangle _{J}=Z_{\lambda }(J)\exp \left( iS_{\lambda
}(\phi _{0})+i\int_{t_{\text{i}}}^{t_{\text{f}}}\mathrm{d}t\int \mathrm{d}%
^{3}\mathbf{x}\hspace{0.01in}J\phi _{0}\right) ,  \label{amplJ}
\end{equation}%
where%
\begin{equation}
Z_{\lambda }(J)=\mathrm{e}^{iW_{\lambda }(J)}\equiv \!\!\!\!\!\!\!\!\!\!\!\!%
\underset{\varphi (t_{\text{i}},\mathbf{x})=\varphi (t_{\text{f}},\mathbf{x}%
)=0}{\int }\!\!\!\!\!\!\!\!\!\!\!\![\mathrm{d}\varphi ]\exp \left(
iS_{\lambda }(\varphi ,\phi _{0})+\int_{t_{\text{i}}}^{t_{\text{f}}}\mathrm{d%
}t\int \mathrm{d}^{3}\mathbf{x\hspace{0.01in}\hspace{0.01in}}J\varphi
\right) .  \label{zj}
\end{equation}%
We can reduce the effort to working out the correlation functions encoded in 
$Z_{\lambda }(J)$, since the factor in front of it in (\ref{amplJ}) is under
control.

The $Z_{\lambda }$ correlation functions%
\begin{equation}
\langle \varphi (x_{1})\cdots \varphi (x_{n})\rangle _{\lambda }=\left.
Z_{\lambda }^{-1}(0)\frac{\delta ^{n}Z_{\lambda }(J)}{i\delta J(x_{1})\cdots
i\delta J(x_{n})}\right\vert _{J=0}  \label{correl}
\end{equation}%
collect all the diagrams, including the disconnected ones and the reducible
ones. Mimicking standard arguments, we can prove that $W(J)$ is the
generating functional of the connected diagrams. Its Legendre transform%
\begin{equation*}
\Gamma (\Phi )=W(J)-\int_{t_{\text{i}}}^{t_{\text{f}}}\mathrm{d}t\int 
\mathrm{d}^{3}\mathbf{x\hspace{0.01in}}J\Phi ,\qquad \Phi =\frac{\delta W}{%
\delta J},
\end{equation*}%
is the generating functional of the amputated, one-particle irreducible
diagrams.

First, it is useful to show that the functional integral of a functional
total derivative vanishes. That is to say, the identity 
\begin{equation}
\!\!\!\!\!\!\!\!\!\!\!\!\underset{\varphi (t_{\text{i}},\mathbf{x})=\varphi
(t_{\text{f}},\mathbf{x})=0}{\int }\!\!\!\!\!\!\!\!\!\!\!\![\mathrm{d}%
\varphi ]\frac{\delta }{\delta \varphi (y)}\left[ X(\varphi )\exp \left(
iS_{\lambda }(\varphi ,\phi _{0})+\int_{t_{\text{i}}}^{t_{\text{f}}}\mathrm{d%
}t\int \mathrm{d}^{3}\mathbf{x\hspace{0.01in}\hspace{0.01in}}J\varphi
\right) \right] =0  \label{totald}
\end{equation}%
holds, where $t_{\text{i}}<y^{0}<t_{\text{f}}$ and $X(\varphi )$ is a
product of local functionals.

To prove (\ref{totald}), we define%
\begin{equation}
\lbrack X]=\!\!\!\!\!\!\!\!\!\!\!\!\underset{\varphi (t_{\text{i}},\mathbf{x}%
)=\varphi (t_{\text{f}},\mathbf{x})=0}{\int }\!\!\!\!\!\!\!\!\!\!\!\![%
\mathrm{d}\varphi ]\hspace{0.01in}X(\varphi )\exp \left( iS_{\lambda
}(\varphi ,\phi _{0})+\int_{t_{\text{i}}}^{t_{\text{f}}}\mathrm{d}t\int 
\mathrm{d}^{3}\mathbf{x\hspace{0.01in}\hspace{0.01in}}J\varphi \right)
\label{X}
\end{equation}%
and let $[X]_{\alpha }$ denote the same expression upon making the change of
variables $\varphi (t\mathbf{,x})\rightarrow \varphi (t\mathbf{,x})+\alpha (t%
\mathbf{,x})$, where $\alpha (t,\mathbf{x})$ is assumed to vanish
everywhere, but in a neighborhood of $y$. This assumption ensures that we
can integrate the $\alpha $-dependent corrections by parts, without worrying
about boundary contributions. Since $[X]_{\alpha }=[X]$, the left-hand side
of (\ref{totald}) (which is the functional derivative of $[X]_{\alpha }$
with respect to $\alpha $, calculated at $\alpha =0$) must vanish.

Using (\ref{totald}), we can derive functional equations for the generating
functionals. Noting that the $W$ equation is connected and the $\Gamma $
equation is irreducible, we can prove that the solutions $W$ and $\Gamma $
share the same properties. The restriction to finite $\tau $ does not pose
difficulties about this. For details see, for example, \cite{renormalization}%
.

\subsection{Propagator}

The two-point function%
\begin{equation}
G_{\lambda }(x,y)=\langle \varphi (x)\varphi (y)\rangle _{\lambda }=\left.
Z_{\lambda }^{-1}(0)\frac{\delta ^{2}Z_{\lambda }(J)}{i\delta J(x)i\delta
J(y)}\right\vert _{J=0}  \label{twopt}
\end{equation}%
defines the propagator. In the free-field limit, $G_{0}(x,y)$ is uniquely
determined by the problem%
\begin{eqnarray}
(\square _{x}+m^{2})G_{0}(x,y) &=&-i\delta ^{(4)}(x-y),\qquad
G_{0}(y,x)=G_{0}(x,y),  \notag \\
G_{0}(x,y) &=&0\text{ for }x^{0}=t_{\text{i}}\text{, }x^{0}=t_{\text{f}}%
\text{, }y^{0}=t_{\text{i}}\text{, }y^{0}=t_{\text{f}},  \label{Cprobl}
\end{eqnarray}%
where $\square =\partial _{\mu }\partial ^{\mu }$ and the subscript $x$
specifies that the partial derivatives are calculated with respect to $x$.
The Klein-Gordon equation is derived from (\ref{totald}) with $X(\varphi
)=\varphi (x)$, $\lambda =0$ and $J=0$. The second line follows from $%
\varphi (t_{\text{i}},\mathbf{x})=\varphi (t_{\text{f}},\mathbf{x})=0$.

At the practical level, we solve (\ref{Cprobl}) starting from the Feynman
propagator, or any other solution of the Klein-Gordon equation. Then we add
the most general solution of the homogeneous equation, and determine its
arbitrary coefficients from the symmetry property $G_{0}(y,x)=G_{0}(x,y)$
and the conditions that appear in the second line of (\ref{Cprobl}). The
result is reported in formula (\ref{propag}), after Fourier transforming the
space coordinates.

The generating functional of the connected correlation functions in the
free-field limit is%
\begin{equation}
iW_{0}(J)=iW_{0}(0)-\int \mathrm{d}^{4}x\mathrm{\hspace{0.01in}}\hspace{%
0.01in}J_{\bot }(x)G_{0}(x,y)J_{\bot }(y)\hspace{0.01in}\mathrm{d}^{4}x,
\label{proje}
\end{equation}%
where $J_{\bot }(x)=\theta (t_{\text{f}}-x^{0})\theta (x^{0}-t_{\text{i}%
})J(x)$. The constant $W_{0}(0)$ is worked out in appendix \ref{diagrammar20}%
. Formula $Z_{0}(J)=\mathrm{e}^{iW_{0}(J)}$ shows that the Wick theorem
works as usual.

Note that there is no need to project the propagator $G_{0}$ to the interval 
$t_{\text{i}}<t<t_{\text{f}}$, inside the diagrams, since it is always
sandwiched in between vertices or sources $J_{\bot }$, which are already
projected.

\subsection{Interactions}

Expanding $L_{\lambda }(\varphi ,\phi _{0})$ in powers of $\varphi $, we
find $\phi _{0}$-dependent vertices, which can be viewed as local composite
fields coupled to external sources. More explicitly, we can write%
\begin{equation}
\int_{t_{\text{i}}}^{t_{\text{f}}}\mathrm{d}t\int \mathrm{d}^{3}\mathbf{x%
\hspace{0.01in}}L_{I}(\varphi ,\phi _{0})=\int \mathrm{d}^{4}x\sum_{n=1}^{%
\infty }\sum_{\alpha }K_{n\alpha }(x)V_{n\alpha }(\varphi (x)),
\label{vertici}
\end{equation}%
where $V_{n\alpha }(\varphi )$ is a monomial of degree $n$ in $\varphi $ and
its derivatives, $\alpha $ is an extra label to distinguish the various
cases, and $K_{n\alpha }(x)$ are appropriate functions, which we can
interpret as external sources. They collect the projector $\theta (t_{\text{f%
}}-x^{0})\theta (x^{0}-t_{\text{i}})$ onto the interval $\tau $, as well as
the dependence on $\phi _{0}$. The latter is encoded in the shift (\ref%
{expaf}) of the field, which transfers directly into the generating
functional $\Gamma (\Phi )$ as an identical shift of $\Phi $.

\section{Quantum mechanics}

\label{position}\setcounter{equation}{0}

To work out explicit formulas, it is convenient to Fourier transform the
space coordinates, and reduce the problem to a continuum of oscillators in
quantum mechanics. It is then possible to focus on a single oscillator at a
time.

In this section we consider the anharmonic oscillator with Lagrangian%
\begin{equation}
L_{\lambda }(q)=\frac{1}{2}\left( \dot{q}^{2}-\omega ^{2}q^{2}\right)
-V_{\lambda }(q),  \label{ll}
\end{equation}%
where $V_{\lambda }(q)$ is proportional to some coupling $\lambda $. The
amplitude we want to study is%
\begin{equation*}
\langle q_{\text{f}},t_{\text{f}}\hspace{0.01in}|\hspace{0.01in}q_{\text{i}%
},t_{\text{i}}\rangle =\langle q_{\text{f}}\hspace{0.01in}|e^{-iH\tau }|%
\hspace{0.01in}q_{\text{i}}\rangle =\!\!\!\!\!\!\!\!\!\underset{_{q(t_{\text{%
i}})=q_{\text{i}},\hspace{0.02in}q(t_{\text{f}})=q_{\text{f}}}}{\int }%
\!\!\!\!\!\!\!\!\![\mathrm{d}q]\exp \left( i\int_{t_{\text{i}}}^{t_{\text{f}%
}}\mathrm{d}t\mathbf{\hspace{0.01in}}L_{\lambda }(q(t))\right) ,
\end{equation*}%
where $|q\rangle $ denotes the position eigenstate.

As before, we shift $q$ to $q_{0}+q$, where 
\begin{equation}
q_{0}(t)=\frac{q_{\text{f}}\sin ((t-t_{\text{i}})\omega )+q_{\text{i}}\sin
((t_{\text{f}}-t)\omega )}{\sin (\omega \tau )}\qquad  \label{q0}
\end{equation}%
is the solution of the classical equations of motion with boundary
conditions $q(t_{\text{i}})=q_{\text{i}}$, $q(t_{\text{f}})=q_{\text{f}}$.
After the shift, the functional integral is done on fluctuations (still
called $q$) with boundary conditions $q(t_{\text{i}})=q(t_{\text{f}})=0$.

The generating functional (\ref{zj}) is 
\begin{equation}
Z_{\lambda }(J)=\mathrm{e}^{iW_{\lambda }(J)}=\!\!\!\!\!\!\!\underset{_{q(t_{%
\text{f}})=q(t_{\text{i}})=0}}{\int }\!\!\!\!\!\!\![\mathrm{d}q]\exp \left(
i\int_{t_{\text{i}}}^{t_{\text{f}}}\mathrm{d}t\int \mathrm{d}^{3}\mathbf{x%
\hspace{0.01in}}\left( L_{\lambda }(q,q_{0}))+\mathbf{\hspace{0.01in}}%
J(t)q(t)\right) \right) ,  \label{genf}
\end{equation}%
where $L_{\lambda }(q_{0},q)=L_{0}(q)-V_{\lambda }(q_{0}+q)+V_{\lambda
}(q_{0})=L_{0}(q)+\mathcal{O}(\lambda )$.

From (\ref{Cprobl}), we see that the free two-point function $%
G_{0}(t,t^{\prime })=\langle q(t)q(t^{\prime })\rangle _{0}$ is the solution
of the problem%
\begin{equation}
\left( \frac{\mathrm{d}^{2}}{\mathrm{d}t^{2}}+\omega ^{2}\right)
G_{0}(t,t^{\prime })=-i\delta (t-t^{\prime }),\qquad G_{0}(t^{\prime
},t)=G_{0}(t,t^{\prime }),\qquad G_{0}(t_{\text{i}},t^{\prime })=G_{0}(t_{%
\text{f}},t^{\prime })=0.  \label{equa}
\end{equation}%
We start from the Feynman propagator $\mathrm{e}^{-i\omega |t-t^{\prime
}|}/(2\omega )$, and add the (symmetrized) solutions%
\begin{equation*}
a\mathrm{e}^{i\omega (t+t^{\prime })}+b\mathrm{e}^{-i\omega (t+t^{\prime
})}+c(\mathrm{e}^{i\omega (t-t^{\prime })}+\mathrm{e}^{-i\omega (t-t^{\prime
})})
\end{equation*}%
of the homogeneous equation, multiplied by arbitrary coefficients $a$, $b$, $%
c$. Then, we determine these constants from the boundary conditions that
appear to the right of (\ref{equa}). The result is%
\begin{equation}
G_{0}(t,t^{\prime })=i\theta (t-t^{\prime })\frac{\sin (\omega (t_{\text{f}%
}-t))\sin (\omega (t^{\prime }-t_{\text{i}}))}{\omega \sin (\omega \tau )}%
+(t\leftrightarrow t^{\prime }),  \label{propag}
\end{equation}%
where $t_{\text{i}}<t,t^{\prime }<t_{\text{f}}$.

To check the limit $t_{\text{f}}\rightarrow +\infty $, $t_{\text{i}%
}\rightarrow -\infty $, we must assume, as usual, that $\omega $ has a small
negative imaginary part ($\omega \rightarrow \tilde{\omega}=\omega
-i\epsilon $, $\epsilon >0$). As expected, the propagator tends to the
Feynman one,%
\begin{eqnarray*}
\lim_{t_{\text{f}}\rightarrow +\infty }G_{0}(t,t^{\prime }) &=&\frac{\theta
(t-t^{\prime })}{2\omega }\mathrm{e}^{-i\omega (t-t^{\prime })}\left( 1-%
\mathrm{e}^{-2i\omega (t^{\prime }-t_{\text{i}})}\right) +(t\leftrightarrow
t^{\prime }), \\
\lim_{t_{\text{i}}\rightarrow -\infty }\lim_{t_{\text{f}}\rightarrow +\infty
}G_{0}(t,t^{\prime }) &=&\frac{1}{2\omega }\mathrm{e}^{-i\omega |t-t^{\prime
}|}.
\end{eqnarray*}

It is also interesting to derive the Fourier transform of (\ref{propag}),
defined by extending its expression to arbitrary times $t$ and $t^{\prime }$
(instead of restricting it to the interval $t_{\text{i}}<t,t^{\prime }<t_{%
\text{f}}$). We find%
\begin{eqnarray}
&&\tilde{G}_{0}(e,e^{\prime })\equiv \int_{-\infty }^{+\infty }\mathrm{d}%
t\int_{-\infty }^{+\infty }\mathrm{d}t^{\prime }G_{0}(t,t^{\prime })\mathrm{e%
}^{i(et+e^{\prime }t^{\prime })}=(2\pi )\delta (e+e^{\prime })\frac{i}{%
e^{2}-\omega ^{2}}  \notag \\
&&\quad -\frac{i\pi ^{2}\mathrm{e}^{-i\omega \tau }}{\omega \sin (\omega
\tau )}\left[ \delta (e+e^{\prime })2\omega \delta (e^{2}-\omega
^{2})-\delta (e-e^{\prime })(\mathrm{e}^{-2i\omega t_{\text{i}}}\delta
(e+\omega )+\mathrm{e}^{2i\omega t_{\text{f}}}\delta (e-\omega ))\right]
.\,\qquad  \label{propos}
\end{eqnarray}%
In addition to the Feynman propagator, we have two \textquotedblleft on
shell\textquotedblright\ contributions, including one proportional to $%
\delta (e-e^{\prime })$. The reason is that the boundary conditions break
the invariance under time translations, which causes a \textquotedblleft
spontaneous\textquotedblright\ symmetry breaking\ of energy conservation.

When we use the propagator (\ref{propos}) inside the loop diagrams, the
integrals on the loop energies are straightforward, but the integrals on the
loop momenta may be challenging. The infinitely many singularities located
at $\omega \tau =n\pi $, $n\in \mathbb{Z}$, cause further complications.
Yet, the final result is well defined. It is not easy to prove this fact in
the position-eigenstate framework, or a generic framework. Yet, it emerges
quite naturally in the coherent-state approach. Once it is evident there, it
extends directly to the position-eigenstate approach, as well as every other
approach that can be reached from the coherent-state one by means of a
change of basis.

Note that we are using Fourier transforms (\ref{propos}) in time, rather
than Fourier series, because the latter make calculations much harder, and
do not allow us to take advantage of the Wick rotation. It is consistent to
use Fourier transforms, for the following reason. The projection onto the
finite time interval $t_{\text{i}}<t<t_{\text{f}}$ acts on the quadratic
part of the Lagrangian, as well as the interaction part. Inside the loop
diagrams, the propagators are sandwiched in between vertices, which are
projected. Moreover, we can attach projected sources $J_{\bot }$ to the
external legs, as in (\ref{proje}). Provided we do this, we can ignore the
projectors on the propagators. Thus, the simplest option is to extend
formula (\ref{propag}) to arbitrary times $t$ and $t^{\prime }$, after which
we can use the Fourier transform (\ref{propos}).

The denominator $Z_{\lambda }(0)$ of (\ref{correl}) is worked out in
appendix \ref{diagrammar20} at $\lambda =0$.

\section{Coherent-state approach}

\label{coherent}\setcounter{equation}{0}

The main virtue of the coherent-state approach \cite{cohe} is that it moves
all the details of the restriction to finite $\tau $ outside the diagrams.
The cores of the diagrams are then the same as usual, so the final results
are always well defined. The key properties also survive the restriction to
a compact space manifold $\Omega $, where the internal sectors of the
diagrams are affected, but only in a minor way.

In this section we lay out the basic properties of the approach, starting by
recalling how it works in the case of the harmonic oscillator of frequency $%
\omega $ and Lagrangian%
\begin{equation}
L_{0}(q,\dot{q})=\frac{1}{2}\left( \dot{q}^{2}-\omega ^{2}q^{2}\right)
\label{l1}
\end{equation}%
at $\tau =\infty $ ($t_{\text{f}}=+\infty $, $t_{\text{i}}=-\infty $), $%
\Omega =\mathbb{R}^{3}$. Introducing the momentum $p=\partial L_{0}/\partial 
\dot{q}$, we can consider the equivalent Lagrangian%
\begin{equation}
L_{0}^{\prime }(q,\dot{q},p,\dot{p})=\frac{1}{2}(p\dot{q}-q\dot{p}%
-p^{2}-\omega ^{2}q^{2}).  \label{l2}
\end{equation}%
The change of variables\footnote{%
The notation we use differs from the popular ones, to save factors $\sqrt{2}$
in various places and reduce the number of spurious nonlocalities brought in
by the factors $\omega $ (in quantum field theory).}%
\begin{equation}
z=\frac{1}{2}\left( q+i\frac{p}{\omega }\right) ,\qquad \bar{z}=\frac{1}{2}%
\left( q-i\frac{p}{\omega }\right) ,\qquad q=z+\bar{z},\qquad p=-i\omega (z-%
\bar{z}),  \label{chvcoh}
\end{equation}%
turns $L_{0}^{\prime }$ into%
\begin{equation}
\mathcal{L}_{0}(z,\bar{z})\equiv i\omega (\bar{z}\dot{z}-\dot{\bar{z}}%
z)-2\omega ^{2}\bar{z}z=\frac{1}{2}\eta ^{\text{t}}Q\eta ,  \label{L0}
\end{equation}%
where%
\begin{equation}
Q=2\omega \left( 
\begin{tabular}{cc}
$0$ & $-i\frac{\mathrm{d}}{\mathrm{d}t}-\omega $ \\ 
$i\frac{\mathrm{d}}{\mathrm{d}t}-\omega $ & $0$%
\end{tabular}%
\right) ,\qquad \eta =\left( 
\begin{tabular}{l}
$z$ \\ 
$\bar{z}$%
\end{tabular}%
\right) .  \label{Q}
\end{equation}%
We call the functions $z$ and $\bar{z}$ coherent \textquotedblleft
states\textquotedblright\ by analogy with the operatorial approach, even if
they are just functions in the Lagrangian approach.

The free propagators are 
\begin{equation}
\langle z(t)\hspace{0.01in}\bar{z}(t^{\prime })\rangle _{0}=\theta
(t-t^{\prime })\frac{\mathrm{e}^{-i\omega (t-t^{\prime })}}{2\omega },\qquad
\langle z(t)\hspace{0.01in}z(t^{\prime })\rangle _{0}=\langle \bar{z}(t)%
\hspace{0.01in}\bar{z}(t^{\prime })\rangle _{0}=0.  \label{propaz}
\end{equation}

When we include the interactions, the momentum $p$ and the Hamiltonian $%
H_{\lambda }(p,q)$, which are 
\begin{equation*}
p=\frac{\partial L_{\lambda }}{\partial \dot{q}},\qquad H_{\lambda }(p,q)=p%
\dot{q}-L_{\lambda }(q,\dot{q})=H_{0}(p,q)+H_{I}(p,q),
\end{equation*}%
allow us to replace (\ref{ll}) with%
\begin{equation}
L_{\lambda }^{\prime }(q,\dot{q},p,\dot{p})=\frac{1}{2}(p\dot{q}-q\dot{p}%
)-H_{\lambda }(p,q).  \label{h2}
\end{equation}%
As before, we assume that the interaction term $H_{I}(p,q)$ is proportional
to some coupling $\lambda $.

Expressing $p$ and $q$ as in (\ref{chvcoh}), we obtain the interaction
Lagrangian $\mathcal{L}_{I}(z,\bar{z})=-H_{I}(p,q)$, which does not depend
on the\ time derivatives of $z$ and $\bar{z}$. The total Lagrangian is thus%
\begin{equation}
\mathcal{L}_{\lambda }(z,\bar{z})=\mathcal{L}_{0}(z,\bar{z})+\mathcal{L}%
_{I}(z,\bar{z}).  \label{Lcoho}
\end{equation}

For various purposes, it may be convenient to switch back and forth between
the variables $p$, $q$ and $z$, $\bar{z}$. For example, when we upgrade from
quantum mechanics to quantum field theory, (\ref{l2}) is local, while (\ref%
{Lcoho}) may contain spurious nonlocalities due to the dependence of $\omega 
$ on the momentum.

Note that the propagator of $z+\bar{z}$ is the usual Feynman one,%
\begin{equation}
\langle \lbrack z(t)+\bar{z}(t)]\hspace{0.01in}[z(t^{\prime })+\bar{z}%
(t^{\prime })]\rangle _{0}=\frac{\mathrm{e}^{-i\omega |t-t^{\prime }|}}{%
2\omega }=\int \frac{\mathrm{d}e}{2\pi }\int \frac{\mathrm{d}e^{\prime }}{%
2\pi }\frac{i(2\pi )\delta (e+e^{\prime })\mathrm{e}^{-i(et+e^{\prime
}t^{\prime })}}{e^{2}-\omega ^{2}+i\epsilon }.  \label{propz}
\end{equation}

It is convenient to couple $z$ and $\bar{z}$ to independent sources $\bar{%
\zeta}$ and $\zeta $ and write the functional integral as%
\begin{equation*}
Z(\zeta ,\bar{\zeta})=\int [\mathrm{d}z\mathrm{d}\bar{z}]\mathrm{\exp }%
\left( \int_{t_{\text{i}}}^{t_{\text{f}}}\mathrm{d}t\hspace{0.01in}\mathcal{L%
}_{\lambda }(z,\bar{z})+i\int (\bar{\zeta}z+\bar{z}\zeta )\mathrm{d}t\right)
.
\end{equation*}%
The reason is that the change of variables (\ref{chvcoh}) amounts to
lowering the number of time derivatives of the kinetic terms from two to
one, and doubling the number of propagating independent fields: from $q$
only to $z$ and $\bar{z}$ (or $q$ and $p$). This way, the particle and
antiparticle poles $e=\pm \omega $ in (\ref{propz}) are assigned to
different fields. Doubling the sources as well, we can distinguish the poles 
$e=\pm \omega $ on the external legs of the diagrams.

\subsection{Finite time interval}

When we restrict to a finite time interval $\tau =t_{\text{f}}-t_{\text{i}}$%
, the action becomes%
\begin{equation}
S_{\lambda }(z,\bar{z})=-i\omega \left( \bar{z}_{\text{f}}z(t_{\text{f}})+%
\bar{z}(t_{\text{i}})z_{\text{i}}\right) +\int_{t_{\text{i}}}^{t_{\text{f}}}%
\mathrm{d}t\hspace{0.01in}\mathcal{L}_{\lambda }(z,\bar{z}),  \label{accoho}
\end{equation}%
with initial and final conditions $z(t_{\text{i}})=z_{\text{i}}$, $\bar{z}%
(t_{\text{f}})=\bar{z}_{\text{f}}$, where $\mathcal{L}_{\lambda }(z,\bar{z})$
is given by (\ref{Lcoho}). The corrections to the integral of $\mathcal{L}%
_{\lambda }$ that appear on the right-hand side must be included in order to
have the right classical variational problem. Indeed, the variation of those
corrections compensates the contributions due to the total derivative
contained in the expression 
\begin{equation*}
\delta \mathcal{L}_{0}(z,\bar{z})=i\omega \frac{\mathrm{d}}{\mathrm{d}t}(%
\bar{z}\delta z-z\delta \bar{z})+2i\omega \delta \bar{z}(\dot{z}+i\omega
z)-2i\omega (\dot{\bar{z}}-i\omega \bar{z})\delta z,
\end{equation*}%
where $\delta z$ and $\delta \bar{z}$ denote the variations of $z$ and $\bar{%
z}$. The cancellation just mentioned is crucial: without it, the variational
problem gives the extra conditions\ $\bar{z}_{\text{f}}=z_{\text{i}}=0$,
which trivialize the set of solutions of the classical equations of motion.
Note that the interaction Lagrangian $\mathcal{L}_{I}(z,\bar{z})$ does not
generate total derivatives, since it does not depend on $\dot{z}$ and $\dot{%
\bar{z}}$.

Introducing the sources $\zeta $ and $\bar{\zeta}$, the transition amplitude
is 
\begin{equation}
\langle \bar{z}_{\text{f}},t_{\text{f}};z_{\text{i}},t_{\text{i}}\rangle
_{\zeta ,\bar{\zeta}}=\!\!\!\!\!\!\!\!\!\!\!\!\!\underset{z(t_{\text{i}})=z_{%
\text{i}},\,\bar{z}(t_{\text{f}})=\bar{z}_{\text{f}}}{\int }%
\!\!\!\!\!\!\!\!\!\!\!\![\mathrm{d}z\mathrm{d}\bar{z}]\mathrm{\exp }\left(
iS_{\lambda }(z,\bar{z})+\int_{t_{\text{i}}}^{t_{\text{f}}}\mathrm{d}t%
\hspace{0.01in}(\bar{\zeta}z+\bar{z}\zeta )\right) .  \label{amplitu}
\end{equation}

By means of the change of variables%
\begin{equation}
z(t)=z_{0}(t)+w(t),\qquad \bar{z}(t)=\bar{z}_{0}(t)+\bar{w}(t),
\label{split}
\end{equation}%
we shift the trajectories $z(t)$, $\bar{z}(t)$ by the solutions 
\begin{equation}
z_{0}(t)=z_{\text{i}}\mathrm{e}^{-i\omega (t-t_{\text{i}})},\qquad \bar{z}%
_{0}(t)=z_{\text{f}}\mathrm{e}^{-i\omega (t_{\text{f}}-t)},  \label{z0}
\end{equation}%
of the classical problem at $\lambda =0$, which is%
\begin{equation*}
\left( i\frac{\mathrm{d}}{\mathrm{d}t}-\omega \right) z_{0}(t)=0,\qquad
z_{0}(t_{\text{i}})=z_{\text{i}},\qquad \left( -i\frac{\mathrm{d}}{\mathrm{d}%
t}-\omega \right) \bar{z}_{0}(t)=0,\qquad \bar{z}_{0}(t_{\text{f}})=\bar{z}_{%
\text{f}}.
\end{equation*}%
Then the fluctuations $w(t)$, $\bar{w}(t)$ are integrated with the simpler
conditions $w(t_{\text{i}})=0$, $\bar{w}(t_{\text{f}})=0$.

The functional integral (\ref{amplitu}) turns into%
\begin{equation}
\langle \bar{z}_{\text{f}},t_{\text{f}};z_{\text{i}},t_{\text{i}}\rangle
_{\zeta ,\bar{\zeta}}=\exp \left( iS_{\lambda }(z_{0},\bar{z}_{0})+\int_{t_{%
\text{i}}}^{t_{\text{f}}}\mathrm{d}t\hspace{0.01in}\left( \bar{\zeta}z_{0}+%
\bar{z}_{0}\zeta \right) \right) Z_{\lambda }(\zeta ,\bar{\zeta}),
\label{amplicohe}
\end{equation}%
where%
\begin{eqnarray}
S_{\lambda }(z_{0},\bar{z}_{0}) &=&-2i\omega \bar{z}_{\text{f}}\mathrm{e}%
^{-i\omega \tau }z_{\text{i}}+\int_{t_{\text{i}}}^{t_{\text{f}}}\mathrm{d}t%
\hspace{0.01in}\mathcal{L}_{I}(z_{0},\bar{z}_{0}),  \notag \\
Z_{\lambda }(\zeta ,\bar{\zeta})=\mathrm{e}^{iW_{\lambda }(\zeta ,\bar{\zeta}%
)}=\!\!\!\!\!\!\!\!\!\!\!\! &&\underset{w(t_{\text{i}})=\bar{w}(t_{\text{f}%
})=0}{\int }\!\!\!\!\!\!\!\!\!\![\mathrm{d}w\mathrm{d}\bar{w}]\mathrm{\exp }%
\left( iS_{\lambda }(w,\bar{w},z_{0},\bar{z}_{0})+\int_{t_{\text{i}}}^{t_{%
\text{f}}}\mathrm{d}t\hspace{0.01in}\left( \bar{\zeta}w+\bar{w}\zeta \right)
\right) ,\quad  \label{Lw}
\end{eqnarray}%
while the action $S_{\lambda }$ of the fluctuations $w$ and $\bar{w}$, and
its Lagrangian are 
\begin{eqnarray}
S_{\lambda }(w,\bar{w},z_{0},\bar{z}_{0}) &=&\int_{t_{\text{i}}}^{t_{\text{f}%
}}\mathrm{d}t\hspace{0.01in}\mathcal{L}_{\lambda }(w,\bar{w},z_{0},\bar{z}%
_{0}),  \notag \\
\mathcal{L}_{\lambda }(w,\bar{w},z_{0},\bar{z}_{0}) &=&\mathcal{L}_{0}(w,%
\bar{w})+\mathcal{L}_{I}(z_{0}+w,\bar{z}_{0}+\bar{w})-\mathcal{L}_{I}(z_{0},%
\bar{z}_{0}).  \label{Lw2}
\end{eqnarray}

As in (\ref{totald}), the functional integral of a functional total
derivative vanishes:%
\begin{equation}
\!\!\!\!\!\!\!\!\!\underset{w(t_{\text{i}})=\bar{w}(t_{\text{f}})=0}{\int }%
\!\!\!\!\!\!\!\!\![\mathrm{d}w\mathrm{d}\bar{w}]\frac{\delta }{\delta \tilde{%
w}(t^{\prime })}\left[ X(w,\bar{w})\exp \left( iS_{\lambda }(w,\bar{w},z_{0},%
\bar{z}_{0})+\int_{t_{\text{i}}}^{t_{\text{f}}}\mathrm{d}t\mathbf{\hspace{%
0.01in}}\left( \bar{\zeta}w+\bar{w}\zeta \right) \right) \right] =0,
\label{totaldcoh}
\end{equation}%
where $t_{\text{i}}<t^{\prime }<t_{\text{f}}$, $\tilde{w}$ can stand for $w$
or $\bar{w}$, and $X(w,\bar{w})$ is a product of local functionals. Standard
arguments show that $W(\zeta ,\bar{\zeta})$ is the generating functional of
the connected diagrams, and its Legendre transform $\Gamma $ is the
generating functional of the amputated, one-particle irreducible diagrams.

To study $Z_{\lambda }(\zeta ,\bar{\zeta})$ and $W_{\lambda }(\zeta ,\bar{%
\zeta})$, it is sufficient to consider the diagrams of $w$ and $\bar{w}$. We
start from the free theory%
\begin{equation*}
Z_{0}(\zeta ,\bar{\zeta})=\exp \left( iW_{0}(\zeta ,\bar{\zeta})\right)
=\!\!\!\!\!\!\!\!\!\!\!\!\underset{w(t_{\text{i}})=\bar{w}(t_{\text{f}})=0}{%
\int }\!\!\!\!\!\!\!\!\!\![\mathrm{d}w\mathrm{d}\bar{w}]\mathrm{\exp }\left(
i\int_{t_{\text{i}}}^{t_{\text{f}}}\mathrm{d}t\hspace{0.01in}\left( \mathcal{%
L}_{0}(w,\bar{w})+\bar{\zeta}w+\bar{w}\zeta \right) \right) .
\end{equation*}%
The key property of the coherent-state approach is that the free propagators
of the quantum fluctuations $w$, $\bar{w}$ coincide with those of $z$, $\bar{%
z}$ at $\tau =\infty $, given in formula (\ref{propaz}):%
\begin{equation}
\langle w(t)\hspace{0.01in}\bar{w}(t^{\prime })\rangle _{0}=\theta
(t-t^{\prime })\frac{\mathrm{e}^{-i\omega (t-t^{\prime })}}{2\omega },\qquad
\langle w(t)\hspace{0.01in}w(t^{\prime })\rangle _{0}=\langle \bar{w}(t)%
\hspace{0.01in}\bar{w}(t^{\prime })\rangle _{0}=0.  \label{propaw}
\end{equation}%
Indeed, they can be worked out by solving the same equations (which follow
from (\ref{totaldcoh}) with $X(w,\bar{w})=w$ or $\bar{w}$), with the initial
and final conditions $w(t_{\text{i}})=\bar{w}(t_{\text{f}})=0$. The theta
functions in (\ref{propaw}) ensure that the solutions do not depend on $t_{%
\text{i}}$ and $t_{\text{f}}$, so the propagators at $\tau =\infty $ and $%
\tau <\infty $ are exactly the same.

In other words, in the coherent-state approach the propagators know nothing
about $t_{\text{f}}$ and $t_{\text{i}}$, and all the features due the
restriction to finite $\tau $ can be removed from the internal sectors of
the diagrams, and dumped on the external sectors. In the approach based on
position eigenstates, instead, the propagators include on-shell corrections
that depend on $t_{\text{f}}$ and $t_{\text{i}}$ in a complicated way, as
shown by formulas (\ref{propos}).

We have%
\begin{equation}
iW_{0}(\zeta ,\bar{\zeta})=-\frac{i\omega \tau }{2}-\int_{t_{\text{i}}}^{t_{%
\text{f}}}\mathrm{d}t\int_{t_{\text{i}}}^{t_{\text{f}}}\mathrm{d}t^{\prime }%
\hspace{0.01in}\bar{\zeta}(t)\theta (t-t^{\prime })\frac{\mathrm{e}%
^{-i\omega (t-t^{\prime })}}{2\omega }\zeta (t^{\prime }),  \label{Wcohe}
\end{equation}%
where $W_{0}(0,0)=-\omega \tau /2$ is calculated in appendix \ref%
{diagrammar20}.

Now we explain how to treat the vertices. From (\ref{Lw2}) we see that,
expanding $\mathcal{L}_{I}(z_{0}+w,\bar{z}_{0}+\bar{w})-\mathcal{L}%
_{I}(z_{0},\bar{z}_{0})$ in powers of $w$ and $\bar{w}$, the vertices have
the form%
\begin{equation*}
V_{n,\bar{n},n^{\prime },\bar{n}^{\prime }}=\int_{t_{\text{i}}}^{t_{\text{f}%
}}\mathrm{d}t\hspace{0.01in}w^{n}(t)\bar{w}^{\bar{n}}(t)\dot{w}^{n^{\prime
}}(t)\dot{\bar{w}}^{\bar{n}^{\prime }}(t)f_{V}(t)=\int_{-\infty }^{+\infty }%
\mathrm{d}t\hspace{0.01in}w^{n}(t)\bar{w}^{\bar{n}}(t)\dot{w}^{n^{\prime
}}(t)\dot{\bar{w}}^{\bar{n}^{\prime }}(t)F_{V}(t),
\end{equation*}%
where $f_{V}(t)$ is a certain function built with the solutions $z_{0}$ and $%
\bar{z}_{0}$, while $F_{V}(t)=f_{V}(t)\theta (t_{\text{f}}-t)\theta (t-t_{%
\text{i}})$. Performing the Fourier transform, we obtain%
\begin{equation*}
V_{n,\bar{n},n^{\prime },\bar{n}^{\prime }}=\int_{-\infty }^{+\infty }\left(
\prod_{i=1}^{N}\frac{\mathrm{d}e_{i}}{2\pi }\tilde{w}_{i}(e_{i})\right) 
\hspace{0.01in}K(-E),\qquad N=n+\bar{n}+n^{\prime }+\bar{n}^{\prime },\qquad
E=\sum_{i=1}^{N}e_{i},
\end{equation*}%
where $\tilde{w}_{i}\hspace{0.01in}$ denotes the Fourier transform of $w$, $%
\bar{w}$, $\dot{w}$ or $\dot{\bar{w}}$, depending on the case, and $K(e)$ is
the Fourier transform of $F_{V}(t)$. We obtain an ordinary vertex coupled to
an external source $K$. To emphasize this fact, we write the interaction
Lagrangian as%
\begin{equation*}
\int_{t_{\text{i}}}^{t_{\text{f}}}\mathrm{d}t\left( \mathcal{L}_{I}(z_{0}+w,%
\bar{z}_{0}+\bar{w})-\mathcal{L}_{I}(z_{0},\bar{z}_{0})\right) \equiv
\int_{-\infty }^{+\infty }\mathrm{d}t\mathcal{L}_{IK}^{w}(w,\bar{w},K),
\end{equation*}%
where $\mathcal{L}_{IK}^{w}$ collects the vertices coupled to the sources $K$%
. From now on, we understand that the integration limits of the integrals
are $\pm \infty $, when they are not specified.

We can write%
\begin{equation}
\exp \left( iW_{\lambda }(\zeta ,\bar{\zeta})\right) =\mathrm{\exp }\left(
i\int \mathrm{d}t\mathcal{L}_{IK}^{w}\left( \frac{\delta }{i\delta \bar{\zeta%
}}+\frac{\delta }{i\delta \zeta },K\right) \right) \exp \left( iW_{0}(\zeta ,%
\bar{\zeta})\right) .  \label{inta}
\end{equation}%
Since the vertices of $\mathcal{L}_{IK}^{w}$ are projected to the interval $%
t_{\text{i}}\leqslant t\leqslant t_{\text{f}}$, it may be convenient to
extend the propagators to arbitrary times as explained before. To do so, we
replace $W_{0}(\zeta ,\bar{\zeta})$ with%
\begin{equation*}
\tilde{W}_{0}(\zeta ,\bar{\zeta})=-\frac{i\omega \tau }{2}-\int \mathrm{d}%
t\int \mathrm{d}t^{\prime }\hspace{0.01in}\bar{\zeta}(t)\theta (t-t^{\prime
})\frac{\mathrm{e}^{-i\omega (t-t^{\prime })}}{2\omega }\zeta (t^{\prime })
\end{equation*}%
and define%
\begin{equation}
\exp \left( i\tilde{W}_{\lambda }(\zeta ,\bar{\zeta})\right) =\mathrm{\exp }%
\left( i\int \mathrm{d}t\mathcal{L}_{IK}^{w}\left( \frac{\delta }{i\delta 
\bar{\zeta}},\frac{\delta }{i\delta \zeta },K\right) \right) \exp \left( i%
\tilde{W}_{0}(\zeta ,\bar{\zeta})\right) .  \label{Wt}
\end{equation}%
Then,%
\begin{equation}
W_{\lambda }(\zeta ,\bar{\zeta})=\tilde{W}_{\lambda }(\zeta _{\bot },\bar{%
\zeta}_{\bot }),  \label{Wperp}
\end{equation}%
where $\zeta _{\bot }(t)=$ $\zeta (t)\theta (t_{\text{f}}-t)\theta (t-t_{%
\text{i}})$, $\bar{\zeta}_{\bot }(t)=$ $\bar{\zeta}(t)\theta (t_{\text{f}%
}-t)\theta (t-t_{\text{i}})$.

Formula (\ref{Wperp}) shows that, in order to work out $W_{\lambda }(\zeta ,%
\bar{\zeta})$, it is sufficient to calculate $\tilde{W}_{\lambda }(\zeta ,%
\bar{\zeta})$ and restrict its correlation functions to the time interval $%
t_{\text{i}}<t<t_{\text{f}}$. In addition, formula (\ref{Wt}) shows that we
can compute the correlation functions of $\tilde{W}_{\lambda }$ by means of
the usual diagrammatic rules, with standard propagators%
\begin{equation}
\langle w(e)\hspace{0.01in}\bar{w}(-e)\rangle _{0}=\frac{i}{2\omega
(e-\omega +i\epsilon )}  \label{propawfree}
\end{equation}%
(after Fourier transform\footnote{%
With an abuse of notation, we use the same symbols for the fields and their
Fourier transforms, when it is possible to understand which is which from
their arguments. So, the functions $w(e_{1})$ and $\bar{w}(e_{2})$ denote
the Fourier transforms of $w(t_{1})$ and $\bar{w}(t_{2})$, respectively. In $%
\langle w(e_{1})\hspace{0.01in}\bar{w}(e_{2})\rangle _{0}$, we omit a factor 
$2\pi $ and the delta function for the overall energy conservation. This
gives (\ref{propawfree}).}), using the vertices encoded in $\mathcal{L}%
_{IK}^{w}$.

\begin{figure}[t]
\begin{center}
\includegraphics[width=14truecm]{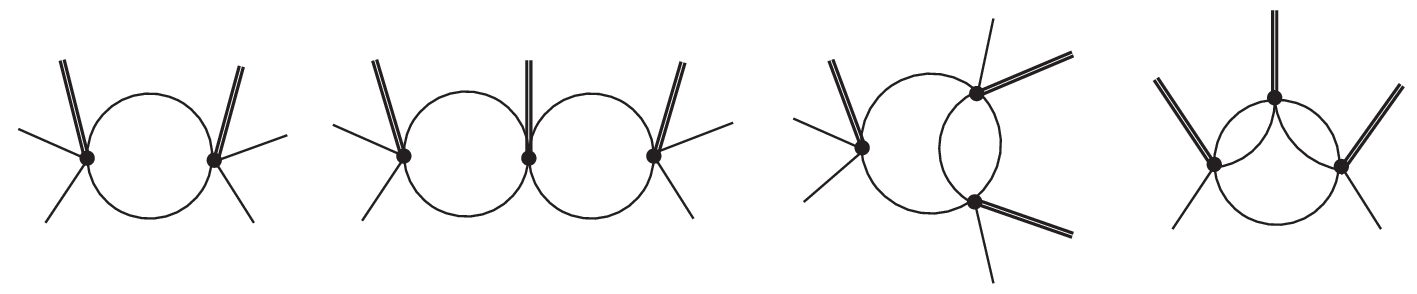}
\end{center}
\par
\vskip -.7truecm
\caption{Diagrams at finite $\protect\tau $: every vertex (denoted by a dot)
has a source $K$ attached to it (denoted by a double line), besides internal
and possibly external legs (denoted by single lines)}
\label{BasicDiagra}
\end{figure}

The correlation functions are the functional derivatives of $\langle \bar{z}%
_{\text{f}},t_{\text{f}};z_{\text{i}},t_{\text{i}}\rangle _{\zeta ,\bar{\zeta%
}}$ with respect to $i\bar{\zeta}(t)$ and $i\zeta (t)$, calculated at $\bar{%
\zeta}(t)=\zeta (t)=0$. They give diagrams that look like the ones at $\tau
=\infty $, internally, but carry an important difference externally: every
vertex is attached to a source $K$, which takes care of the restriction to
finite $\tau $. In some sense, there exist no truly internal vertices.
Examples of diagrams are shown in fig. \ref{BasicDiagra}.

For example, the bubble diagram (first diagram of fig. \ref{BasicDiagra})
may give (once we amputate the external $w$, $\bar{w}$ legs)%
\begin{equation*}
\Pi (t)G^{2}(t,t^{\prime })\Pi (t^{\prime }),
\end{equation*}%
where $G(t,t^{\prime })=\langle w(t)\hspace{0.01in}\bar{w}(t^{\prime
})\rangle _{0}$ is a propagator (\ref{propaw}) and $\Pi (t)=\theta (t_{\text{%
f}}-t)\theta (t-t_{\text{i}})$ is the projector onto the interval $\tau $.
Switching to Fourier transforms, we find%
\begin{equation*}
\int \frac{\mathrm{d}e^{\prime }}{2\pi }K(e_{1}-e^{\prime })B(e^{\prime
})K(e_{2}+e^{\prime }),\qquad B(e^{\prime })=\int \frac{\mathrm{d}e}{2\pi }%
\tilde{G}(e)\tilde{G}(e^{\prime }-e),
\end{equation*}%
where $\tilde{G}(e)=\langle w(e)\hspace{0.01in}\bar{w}(-e)\rangle _{0}$ as
in (\ref{propawfree}), $K(e)$ is the Fourier transform of $\Pi (t)$, and $%
e_{1,2}$ are the external energies. We see that the core diagram $%
B(e^{\prime })$ is the same as usual, while the external sources $K$ take
care of the restriction to finite $\tau $.

At $\tau =\infty $ we are accustomed to express the transition amplitudes by
means of correlation functions on the vacuum state, which describe
scattering processes between arbitrary incoming and outgoing particles,
through the LSZ\ reduction formulas \cite{LSZ}. At finite $\tau $, instead,
the initial and final configurations of the amplitudes $\langle \bar{z}_{%
\text{f}},t_{\text{f}};z_{\text{i}},t_{\text{i}}\rangle _{0,0}$, calculated
at vanishing sources $\zeta $ and $\bar{\zeta}$, are enough to cover all the
physical situations. This means that, strictly speaking, we could limit
ourselves to consider the diagrams that have no external legs, which know
about the initial and final configurations through the external sources $K$.
Yet, those diagrams are better studied by introducing the sources $\zeta $
and $\bar{\zeta}$, hence the correlation functions, since propagators and
subdiagrams are particular cases of diagrams that do contain external legs. 

What we have done so far in quantum mechanics extends straightforwardly to
quantum field theory (at finite $\tau $, on $\Omega =\mathbb{R}^{3}$). The
formulas written for a specific frequency $\omega $ can be generalized by
assuming that each field depends on the position $\mathbf{x}$ ($%
z(t)\rightarrow z(t,\mathbf{x})$, $w(t)\rightarrow w(t,\mathbf{x})$, etc.),
and that every interaction term is a product among fields, sources $K$ and
their derivatives, located at the same point $\mathbf{x}$, integrated in $%
\mathrm{d}^{3}\mathbf{x}$ on $\mathbb{R}^{3}$.

As far as the quadratic Lagrangian (\ref{L0}) is concerned, we must
interpret it as 
\begin{equation}
\mathcal{L}_{0}(w,\bar{w})\rightarrow \int \mathrm{d}^{3}\mathbf{x}\left[ i%
\bar{w}\sqrt{-\triangle +m^{2}}\dot{w}-i\dot{\bar{w}}\sqrt{-\triangle +m^{2}}%
w-2\bar{w}(-\triangle +m^{2})w\right] ,  \label{lwo}
\end{equation}%
where $\triangle $ denotes the Laplacian. The first two terms in the square
brackets are nonlocal, and so may be the interaction terms $\mathcal{L}%
_{IK}^{w}$, due to $p$ dependence in (\ref{h2}). However, these
nonlocalities are spurious, because they disappear if we switch to the
variables%
\begin{equation}
Q\equiv w+\bar{w}=q-z_{0}-\bar{z}_{0},\qquad P\equiv -i\omega (w-\bar{w}%
)=p+i\omega (z_{0}-\bar{z}_{0}),  \label{QP}
\end{equation}%
and view the dependencies on $z_{0}$ and $\bar{z}_{0}$ as external sources.

It may be convenient to switch back and forth between the variables $Q$, $P$
and $w$, $\bar{w}$. The former are more convenient for renormalization,
because they have a standard power counting. The latter are more clearly
related to the initial and final conditions.

Ultimately, the difference between the diagrammatics of quantum field theory
at $\tau =\infty $ and the one at $\tau <\infty $ is limited to the external
sources $K$. The result is that the diagrams are the same as usual,
internally, and obey all the known theorems. They even allow us to
generalize the prescription/projection to purely virtual particles, which we
discuss in section \ref{PV}. In the next section we show that the key
properties also survive the restriction to a compact space manifold $\Omega $%
.

\section{Compact space manifold}

\label{volume}\setcounter{equation}{0}

Now we study quantum field theory in a finite interval of time, and on a
compact, smooth space manifold $\Omega $. We can study manifolds with a
nontrivial boundary $\partial \Omega $, or closed manifolds $\Omega $, such
as the sphere or the torus (equivalent to the box with periodic boundary
conditions). When $\partial \Omega $ is nontrivial, we assume that the
fields $\phi $ satisfy Dirichlet boundary conditions%
\begin{equation}
\phi (t,\mathbf{x}_{\partial \Omega })=f(t,\mathbf{x}_{\partial \Omega })
\label{bou}
\end{equation}%
on $\partial \Omega $, where $f$ are regular functions and $\mathbf{x}%
_{\partial \Omega }$ are the space variables restricted to $\partial \Omega $%
. Problems may appear when $\Omega $ is not smooth (as in the case $\Omega $
= cone), or the boundary conditions (\ref{bou}) are singular. These
situations must be treated case by case.

We assume that the Lagrangian density depends only on the field $\phi $ and
its first derivatives $\partial _{\mu }\phi $, and that each Lagrangian term
contains at most two derivatives. Although we write formulas for scalar
fields, our formulation is general, and applies to bosons as well as
fermions, with obvious modifications. In the case of bosons of higher spins,
it is sufficient to view $\phi $ as a multiplet. In the case of gravity,
where the curvature tensors $R_{\mu \nu \rho \sigma }$, $R_{\mu \nu }$ and $%
R $ involve two derivatives $\partial _{\rho }\partial _{\sigma }\phi _{\mu
\nu }$ of the fluctuation $\phi _{\mu \nu }$ of the metric tensor $g_{\mu
\nu }$ around flat space, we must eliminate them by adding total derivatives
to the Lagrangian. This is always possible, since we are assuming that the
latter does not depend on higher derivatives of $\phi _{\mu \nu }$, and does
not contain terms with more than two derivatives. Later we show how to
obtain the correct classical variational problem, once the Lagrangian is
rearranged as explained.

For example, in quantum gravity with purely virtual particles, we should not
use the higher-derivative formulation \textquotedblleft $R+R^{2}+C^{2}$%
\textquotedblright\ of \cite{LWgrav}\ (where $C_{\mu \nu \rho \sigma }$ is
the Weyl tensor, and $C^{2}$ stands for $C_{\mu \nu \rho \sigma }C^{\mu \nu
\rho \sigma }$), which contains Lagrangian terms with four derivatives or
less, but the two-derivative formulation of \cite{Absograv}, obtained
through the introduction of extra fields. Moreover, we should include the
total derivatives mentioned above, to make sure that terms like $\phi
_{1}\cdots \phi _{n-1}\partial \partial \phi _{n}$ are eliminated in favor
of terms like $\phi _{1}\cdots \phi _{n-2}\partial \phi _{n-1}\partial \phi
_{n}$. The two-derivative formulation of quantum gravity with purely virtual
particles is still renormalizable (at $\tau =\infty $, $\Omega =\mathbb{R}%
^{3}$; for its renormalizability at $\tau <\infty $, $\Omega $ = compact
manifold, see section \ref{renorma}), although not manifestly.

The boundary conditions (\ref{bou}) are not straightforward to deal with,
since we do not know how to Fourier expand the field. It is better to first
shift $\phi $ by any background\ field\ $\phi _{0}$ that satisfies the same
conditions:%
\begin{equation}
\phi (t,\mathbf{x})=\phi _{0}(t,\mathbf{x})+\varphi (t,\mathbf{x}),\qquad
\phi _{0}(t,\mathbf{x}_{\partial \Omega })=f(t,\mathbf{x}_{\partial \Omega
}),  \label{expaf}
\end{equation}%
so that the difference $\varphi (t,\mathbf{x})$ satisfies the simplified
Dirichlet boundary conditions $\varphi (t,\mathbf{x}_{\partial \Omega })=0$.
Note that we are not requiring $\phi _{0}$ to be the solution of a
particular differential equation.

Denote the Lagrangian density by 
\begin{equation}
L_{\lambda }(\phi ,\dot{\phi},\nabla \phi )=L_{0}(\phi ,\dot{\phi},\nabla
\phi )+L_{I}(\phi ,\dot{\phi},\nabla \phi ),\qquad L_{0}(\phi ,\dot{\phi}%
,\nabla \phi )=\frac{1}{2}\dot{\phi}^{2}-\frac{1}{2}(\nabla \phi )^{2}-\frac{%
m^{2}}{2}\phi ^{2},  \label{Llambda}
\end{equation}%
where the interaction term $L_{I}$ is proportional to a coupling $\lambda $,
which is treated perturbatively. After the shift (\ref{expaf}), we write $%
L_{\lambda }(\phi ,\dot{\phi},\nabla \phi )$ as $\tilde{L}_{\lambda
}(\varphi ,\dot{\varphi},\nabla \varphi ,\phi _{0})$ and obtain 
\begin{equation}
\tilde{L}_{\lambda }(\varphi ,\dot{\varphi},\nabla \varphi ,\phi
_{0})=L_{\lambda }(\phi _{0},\dot{\phi}_{0},\nabla \phi _{0})+\varphi A(\phi
_{0})+\dot{\varphi}B(\phi _{0})+\nabla (\varphi C(\phi _{0}))+\hat{L}%
_{\lambda }(\varphi ,\dot{\varphi},\nabla \varphi ,\phi _{0}),
\label{lexpaf}
\end{equation}%
for some functions $A(\phi _{0})$, $B(\phi _{0})$ and $C(\phi _{0})$ of the
background field $\phi _{0}$, where the last term $\hat{L}_{\lambda
}(\varphi ,\dot{\varphi},\nabla \varphi ,\phi _{0})=L_{0}(\varphi ,\dot{%
\varphi},\nabla \varphi )+\mathcal{O}(\lambda )$ is at least quadratic in $%
\varphi $. The first three terms on the right-hand side go unmodified to the
generating functional $\Gamma $, while the fourth one disappears when it is
integrated on $\Omega $.

\subsection{Fourier expansion}

Now we expand the shifted field $\varphi (t,\mathbf{x})$ in a basis of
eigenfunctions of the Laplacian on $\Omega $.

Let $e_{\mathbf{n}}(\mathbf{x})$, where $\mathbf{n}$ is some label ranging
in some set $\mathcal{U}$, denote a complete set of orthonormal eigenstates
of the operator $-\triangle +m^{2}$ on $\Omega $, defined by the Dirichlet
boundary conditions $e_{\mathbf{n}}(\mathbf{x}_{\partial \Omega })=0$ on $%
\partial \Omega $ (if $\partial \Omega \neq \oslash $). Let $\omega _{%
\mathbf{n}}^{2}$ denote their eigenvalues, which are real and positive. The $%
\varphi $ expansion and the orthonormality relations read%
\begin{equation}
\varphi (t,\mathbf{x})=\sum_{\mathbf{n}\in \mathcal{U}}\varphi _{\mathbf{n}%
}(t)e_{\mathbf{n}}(\mathbf{x})\text{,\qquad }\int_{\Omega }\mathrm{d}^{3}%
\mathbf{x\hspace{0.01in}}e_{\mathbf{n}^{\prime }}^{\ast }(\mathbf{x})e_{%
\mathbf{n}}(\mathbf{x})=\delta _{\mathbf{nn}^{\prime }}.  \label{orto}
\end{equation}

Since we are working with real fields $\varphi $, we can choose a basis of
real eigenfunctions. However, in various cases, complex eigenfunctions may
be more convenient, because they can highlight the momentum conservation at
the vertices. In that case, the complex conjugate $e_{\mathbf{n}}^{\ast }(%
\mathbf{x})$ of $e_{\mathbf{n}}(\mathbf{x})$ is an eigenfunction with the
same eigenvalue $\omega _{\mathbf{n}}^{2}$. Thus, there exists an $\mathbf{n}%
^{\ast }\in \mathcal{U}$ such that $e_{\mathbf{n}}^{\ast }(\mathbf{x})=$ $e_{%
\mathbf{n}^{\ast }}(\mathbf{x})$.

In typical cases, $\mathbf{n}^{\ast }=-\mathbf{n}$, but here we want to stay
as general as possible. Note that a real $\varphi $ has $\varphi _{\mathbf{n}%
}^{\ast }(t)=\varphi _{\mathbf{n}^{\ast }}(t)$. The formulas we write look
the same with real or complex eigenfunctions: we just have to interpret the
range of $\mathbf{n}$ appropriately.

For example, if $\Omega $ is a three torus $T^{3}$, that is to say, a box
with periodic boundary conditions, we have%
\begin{equation}
e_{\mathbf{n}}(\mathbf{x})=\frac{\mathrm{e}^{i\mathbf{\bar{n}\cdot x}}}{%
\sqrt{|\Omega |}},\qquad \omega _{\mathbf{n}}=\sqrt{\mathbf{\bar{n}}%
^{2}+m^{2}},  \label{tor}
\end{equation}%
where $|\Omega |=L_{1}L_{2}L_{3}$ is the volume of $\Omega $, $\mathbf{n}%
=(n_{1},n_{2},n_{3})\in \mathbb{Z}^{3}$, $\mathbf{\bar{n}}=2\pi
(n_{1}/L_{1},n_{2}/L_{2},n_{3}/L_{3}\mathbf{)}$ and $L_{1}$, $L_{2}$ and $%
L_{3}$ are the lengths of the sides of $T^{3}$.

If $\Omega $ is a generic box (with boundary), let $L_{1}/2$, $L_{2}/2$ and $%
L_{3}/2$ denote the lengths of its sides. The Dirichlet boundary conditions $%
e_{\mathbf{n}}(\mathbf{x}_{\partial \Omega })=0$ on $\partial \Omega $ give%
\begin{equation}
e_{\mathbf{n}}(\mathbf{x})=\sqrt{\frac{8}{|\Omega |}}\prod_{a=1}^{3}\sin
\left( \bar{n}_{a}x_{a}\right) ,\qquad \omega _{\mathbf{n}}=\sqrt{\mathbf{%
\bar{n}}^{2}+m^{2}},  \label{Dbox}
\end{equation}%
where $\mathbf{n}\in \mathbb{N}_{+}^{3}$.

If $\Omega $ is the sphere $S^{3}$ of radius $R$, we expand $\varphi $ into
spherical harmonics $Y^{klm}$ with frequencies%
\begin{equation*}
\omega _{klm}=\frac{1}{R}\sqrt{k(k+2)+m^{2}R^{2}},
\end{equation*}%
where $(k,l,m)\in \mathbb{Z}^{3}$, $k\geqslant l\geqslant 0$, $-l\leqslant
m\leqslant l$ \cite{spherical}.

If $\Omega $ is a ball of radius $R$, $e_{\mathbf{n}}(\mathbf{x})$ are
proportional to the usual spherical harmonics $Y^{lm}$, times Bessel
functions of the first kind:%
\begin{equation*}
r^{1/2}J_{(2l+1)/2}(kr)Y^{lm}(\theta ,\phi ),\qquad l\geqslant 0\text{, }%
-l\leqslant m\leqslant l,
\end{equation*}%
where $k\equiv \sqrt{\omega ^{2}-m^{2}}$ is fixed by the boundary conditions
at $r=R$.

Let us briefly outline the plan from now, before entering into further
details. After the Fourier expansion (\ref{tor}), we switch to coherent
states, to deal with the restriction to finite $\tau $. We obtain a
propagator, for the quantum fluctuations, that is unaffected by the initial
and final conditions at $t_{\text{i}}$ and $t_{\text{f}}$, and is affected
by $\Omega $ only in a minor way. So doing, we manage to develop a formalism
that does not alter the spectral optical identities of \cite{diagrammarMio}
in a significant way (this aspect will become clear only in section \ref{PV}%
). Specifically, we move all the details about the restriction to finite $%
\tau $ and compact $\Omega $ to the external sectors of the diagrams (apart
from the discretizations of the loop momenta, due to the Fourier expansion).
The formulation we obtain allows us to study unitarity via the spectral
optical identities, and extend the formulation to purely virtual particles.

Before dealing with the complete theory, we treat the quadratic part, and
show that the propagator has the form we want.

\subsection{Free field theory}

For the moment, we concentrate on the free Lagrangian $L_{0}(\varphi ,\dot{%
\varphi},\nabla \varphi )$ for the fluctuation $\varphi $, which coincides
with the Lagrangian $\hat{L}_{\lambda }(\varphi ,\dot{\varphi},\nabla
\varphi ,\phi _{0})$ of (\ref{lexpaf}) at $\lambda =0$. The integral of $%
L_{0}(\varphi ,\dot{\varphi},\nabla \varphi )$ must be equipped with
\textquotedblleft endpoint corrections\textquotedblright\ $\hat{S}_{\text{e}%
0}$, so that the total gives the correct classical variational problem.

Expanding $\varphi $ as in (\ref{orto}), integrating by parts using the $%
\varphi $ boundary condition $\varphi (t,\mathbf{x}_{\partial \Omega })=0$,
and including unspecified endpoint corrections $\hat{S}_{\text{e}}$, we
consider the free action%
\begin{equation}
\hat{S}_{\text{free}}(\varphi )=\hat{S}_{\text{e}}+\int_{t_{\text{i}}}^{t_{%
\text{f}}}\mathrm{d}t\int_{\Omega }\mathrm{d}^{3}\mathbf{x}\hspace{0.01in}%
L_{0}(\varphi ,\dot{\varphi},\nabla \varphi )=\hat{S}_{\text{e}}+\frac{1}{2}%
\sum_{\mathbf{n}\in \mathcal{U}}\int_{t_{\text{i}}}^{t_{\text{f}}}\mathrm{d}t%
\hspace{0.01in}\left( \dot{\varphi}_{\mathbf{n}^{\ast }}\dot{\varphi}_{%
\mathbf{n}}-\varphi _{\mathbf{n}^{\ast }}\omega _{\mathbf{n}}^{2}\varphi _{%
\mathbf{n}}\right) .  \label{sh0}
\end{equation}%
Then, we switch to coherent states\footnote{%
It may be convenient to switch to real eigenfunctions by splitting the set $%
\mathcal{U}$ as the union $\mathcal{U}_{r}\cup \mathcal{U}_{c}\cup \mathcal{U%
}_{c}^{\ast }$ of $\mathcal{U}_{r}$, which collects the $\mathbf{n}$ such
that $\varphi _{\mathbf{n}^{\ast }}=\varphi _{\mathbf{n}}$, and $\mathcal{U}%
_{c}\cup \mathcal{U}_{c}^{\ast }$, which collects the $\mathbf{n}$ such that 
$\varphi _{\mathbf{n}^{\ast }}\neq \varphi _{\mathbf{n}}$. Putting one
element of the pair $(\mathbf{n},\mathbf{n}^{\ast })$ in $\mathcal{U}_{c}$
and the other in $\mathcal{U}_{c}^{\ast }$, we separate the sum on $\mathbf{n%
}\in \mathcal{U}_{r}$ from the sum on $\mathbf{n}\in \mathcal{U}_{c}\cup 
\mathcal{U}_{c}^{\ast }$. Defining%
\begin{equation*}
\varphi _{\mathbf{n}}=\frac{\psi _{\mathbf{n}}+i\chi _{\mathbf{n}}}{\sqrt{2}}%
\qquad \text{for }\mathbf{n}\in \mathcal{U}_{c},
\end{equation*}%
where $\psi _{\mathbf{n}}$ and $\chi _{\mathbf{n}}$ are real, we find%
\begin{equation*}
\hat{S}_{\text{free}}(\varphi )=\hat{S}_{\text{e}}+\frac{1}{2}\int_{t_{\text{%
i}}}^{t_{\text{f}}}\mathrm{d}t\hspace{0.01in}\left[ \sum_{\mathbf{n}\in 
\mathcal{U}_{r}}\left( \dot{\varphi}_{\mathbf{n}}^{2}-\omega _{\mathbf{n}%
}^{2}\varphi _{\mathbf{n}}^{2}\right) +\sum_{\mathbf{n}\in \mathcal{U}%
_{c}}\left( \dot{\psi}_{\mathbf{n}}^{2}-\omega _{\mathbf{n}}^{2}\psi _{%
\mathbf{n}}^{2}+\dot{\chi}_{\mathbf{n}}^{2}-\omega _{\mathbf{n}}^{2}\chi _{%
\mathbf{n}}^{2}\right) \right] .
\end{equation*}%
At this point, we switch to coherent states by applying the procedure of
section \ref{coherent} to $\varphi _{\mathbf{n}}$, $\psi _{\mathbf{n}}$ and $%
\chi _{\mathbf{n}}$ separately. Switching back to $\varphi _{\mathbf{n}}$, $%
\mathbf{n}\in \mathcal{U}$, at the end, we find that the formulas can be
written in compact notation, summing over $\mathbf{n}\in \mathcal{U}$, as
reported in this section.} 
\begin{equation}
z_{\mathbf{n}}=\frac{1}{2}\left( \varphi _{\mathbf{n}}+i\frac{\pi _{\mathbf{n%
}}}{\omega _{\mathbf{n}}}\right) ,\qquad \bar{z}_{\mathbf{n}}=\frac{1}{2}%
\left( \varphi _{\mathbf{n}}-i\frac{\pi _{\mathbf{n}}}{\omega _{\mathbf{n}}}%
\right) ,  \label{zetan}
\end{equation}%
by introducing the momenta $\pi _{\mathbf{n}}(t)=\dot{\varphi}_{\mathbf{n}%
}(t)$ ($\pi _{\mathbf{n}}^{\ast }(t)=\pi _{\mathbf{n}^{\ast }}(t)$), and
applying the procedure described in section \ref{coherent} to each $\mathbf{n%
}$. We repeat the derivation in detail in the next subsection, when we
include the interactions.

We denote the initial and final conditions by $z_{\mathbf{n}}(t_{\text{i}%
})=z_{\mathbf{n}\text{i}}$, $\bar{z}_{\mathbf{n}}(t_{\text{f}})=\bar{z}_{%
\mathbf{n}\text{f}}$, and follow the arguments that lead to (\ref{accoho}).
Putting a prime on $\hat{S}_{\text{free}}$, to emphasize that we are working
with new variables, the free action is%
\begin{equation}
\hat{S}_{\text{free}}^{\prime }=-i\sum_{\mathbf{n}\in \mathcal{U}}\!\left( 
\bar{z}_{\mathbf{n}^{\ast }\text{f}}\omega _{\mathbf{n}}z_{\mathbf{n}}(t_{%
\text{f}})+\bar{z}_{\mathbf{n}^{\ast }}(t_{\text{i}})\omega _{\mathbf{n}}z_{%
\mathbf{n}\text{i}}\right) +\int_{t_{\text{i}}}^{t_{\text{f}}}\!\!\!\mathrm{d%
}t\!\sum_{\mathbf{n}\in \mathcal{U}}\!\left[ i(\bar{z}_{\mathbf{n}^{\ast
}}\omega _{\mathbf{n}}\dot{z}_{\mathbf{n}}-\dot{\bar{z}}_{\mathbf{n}^{\ast
}}\omega _{\mathbf{n}}z_{\mathbf{n}})-2\bar{z}_{\mathbf{n}^{\ast }}\omega _{%
\mathbf{n}}^{2}z_{\mathbf{n}}\right] \!,  \label{sof}
\end{equation}%
where the first sum on the right-hand side is $\hat{S}_{\text{e}}$.

At this point, we expand the coherent states 
\begin{equation}
\bar{z}_{\mathbf{n}}(t)=\bar{z}_{0\mathbf{n}}(t)+\bar{w}_{\mathbf{n}%
}(t),\qquad z_{\mathbf{n}}(t)=z_{0\mathbf{n}}(t)+w_{\mathbf{n}}(t),
\label{shift2}
\end{equation}%
around particular solutions%
\begin{equation}
z_{0\mathbf{n}}(t)=z_{\mathbf{n}\text{i}}\mathrm{e}^{-i\omega _{\mathbf{n}%
}(t-t_{\text{i}})},\qquad \bar{z}_{0\mathbf{n}}(t)=\bar{z}_{\mathbf{n}\text{f%
}}\mathrm{e}^{-i\omega _{\mathbf{n}}(t_{\text{f}}-t)},  \label{zon}
\end{equation}%
of the free equations with the same initial and final conditions, 
\begin{equation}
i\dot{z}_{0\mathbf{n}}-\omega _{\mathbf{n}}z_{0\mathbf{n}}=0,\qquad -i\dot{%
\bar{z}}_{0\mathbf{n}}-\omega _{\mathbf{n}}\bar{z}_{0\mathbf{n}}=0,\qquad
z_{0\mathbf{n}}(t_{\text{i}})=z_{\mathbf{n}\text{i}},\qquad \bar{z}_{0%
\mathbf{n}}(t_{\text{f}})=\bar{z}_{\mathbf{n}\text{f}},  \label{inico}
\end{equation}%
so that the quantum fluctuations $\bar{w}_{\mathbf{n}}$, $w_{\mathbf{n}}$
satisfy simpler initial and final conditions: 
\begin{equation}
w_{\mathbf{n}}(t_{\text{i}})=\bar{w}_{\mathbf{n}}(t_{\text{f}})=0.
\label{wconda}
\end{equation}

The free action is finally 
\begin{equation}
S_{\text{free}}(w,\bar{w})=-2i\sum_{\mathbf{n}\in \mathcal{U}}\bar{z}_{%
\mathbf{n}^{\ast }\text{f}}\omega _{\mathbf{n}}\mathrm{e}^{-i\omega _{%
\mathbf{n}}\tau }z_{\mathbf{n}\text{i}}+\int_{t_{\text{i}}}^{t_{\text{f}}}%
\mathrm{d}t\sum_{\mathbf{n}\in \mathcal{U}}\hspace{0.01in}\left[ i(\bar{w}_{%
\mathbf{n}^{\ast }}\omega _{\mathbf{n}}\dot{w}_{\mathbf{n}}-\dot{\bar{w}}_{%
\mathbf{n}^{\ast }}\omega _{\mathbf{n}}w_{\mathbf{n}})-2\bar{w}_{\mathbf{n}%
^{\ast }}\omega _{\mathbf{n}}^{2}w_{\mathbf{n}}\right] ,  \label{scoh}
\end{equation}%
and the $w$ propagators read%
\begin{equation}
\langle w_{\mathbf{n}}(e)\bar{w}_{\mathbf{n}^{\prime }}(-e)\rangle _{c}^{%
\text{free}}=\frac{i\delta _{\mathbf{n}^{\ast }\mathbf{n}^{\prime }}}{%
2\omega (e-\omega _{\mathbf{n}}+i\epsilon )},\qquad \langle w_{\mathbf{n}%
}(e)w_{\mathbf{n}^{\prime }}(-e)\rangle _{c}^{\text{free}}=\langle \bar{w}_{%
\mathbf{n}}(e)\bar{w}_{\mathbf{n}^{\prime }}(-e)\rangle _{c}^{\text{free}}=0,
\label{prow}
\end{equation}%
after Fourier transform, where the subscript $c$ means \textquotedblleft
connected\textquotedblright . We have inserted it to use the formulas (\ref%
{prow}) below. Note that, due to finite volume effects (the linear terms of (%
\ref{lexpaf}), which are proportional to $A$ and $B$), the full $w$-$\bar{w}$
free propagator does not coincide with the connected part of the $z$-$\bar{z}
$ one.

As expected, the propagators do not know of the initial and final
conditions. Moreover, they know of $\Omega $ only through the discretization
of the frequencies and the momenta.

\subsection{Interacting theory}

Starting over from (\ref{lexpaf}), the total action can be written as 
\begin{eqnarray}
\mathcal{S}_{\lambda }(\varphi ,\phi _{0}) &=&\mathcal{S}_{\text{e}%
}+\int_{t_{\text{i}}}^{t_{\text{f}}}\!\!\mathrm{d}t\!\int_{\Omega }\!\mathrm{%
d}^{3}\mathbf{x\hspace{0.01in}}L_{\lambda }(\phi ,\dot{\phi},\nabla \phi
)=\int_{t_{\text{i}}}^{t_{\text{f}}}\!\!\mathrm{d}t\!\int_{\Omega }\!\mathrm{%
d}^{3}\mathbf{x\hspace{0.01in}}L_{\lambda }(\phi _{0},\dot{\phi}_{0},\nabla
\phi _{0})+\mathcal{\hat{S}}_{\lambda }(\varphi ,\phi _{0}),  \notag \\
\mathcal{\hat{S}}_{\lambda }(\varphi ,\phi _{0}) &=&\mathcal{S}_{\text{e}%
}+\int_{t_{\text{i}}}^{t_{\text{f}}}\!\!\mathrm{d}t\!\int_{\Omega }\!\mathrm{%
d}^{3}\mathbf{x\hspace{0.01in}}\left[ \hat{L}_{\lambda }(\varphi ,\dot{%
\varphi},\nabla \varphi ,\phi _{0})+\varphi A(\phi _{0})+\dot{\varphi}B(\phi
_{0})\right] ,  \label{redu}
\end{eqnarray}%
where $\mathcal{S}_{\text{e}}$ collects the endpoint and boundary
corrections that must be included to have the correct classical variational
problem.

After the Fourier expansion (\ref{orto}), we have%
\begin{eqnarray*}
\mathcal{\hat{S}}_{\lambda }(\varphi ,\phi _{0}) &=&\mathcal{S}_{\text{e}%
}+\int_{t_{\text{i}}}^{t_{\text{f}}}\!\mathrm{d}t\!\hspace{0.02in}\check{L}%
_{\lambda }(\varphi _{\mathbf{n}},\dot{\varphi}_{\mathbf{n}},\phi _{0}), \\
\check{L}_{\lambda }(\varphi _{\mathbf{n}},\dot{\varphi}_{\mathbf{n}},\phi
_{0}) &\equiv &\int_{\Omega }\!\mathrm{d}^{3}\mathbf{x\hspace{0.01in}}\left[ 
\hat{L}_{\lambda }(\varphi ,\dot{\varphi},\nabla \varphi ,\phi _{0})+\varphi
A(\phi _{0})+\dot{\varphi}B(\phi _{0})\right] .
\end{eqnarray*}%
Defining%
\begin{equation*}
A_{\mathbf{n}^{\ast }}\equiv \int_{\Omega }\!\mathrm{d}^{3}\mathbf{x}\hspace{%
0.01in}A(\phi _{0})e_{\mathbf{n}}(\mathbf{x}),\qquad B_{\mathbf{n}^{\ast
}}\equiv \int_{\Omega }\!\mathrm{d}^{3}\mathbf{x}\hspace{0.01in}B(\phi
_{0})e_{\mathbf{n}}(\mathbf{x}),
\end{equation*}%
and separating the interactions $\check{L}_{I}(\varphi _{\mathbf{n}},\dot{%
\varphi}_{\mathbf{n}},\phi _{0})$ from the rest, we write%
\begin{equation*}
\check{L}_{\lambda }(\varphi _{\mathbf{n}},\dot{\varphi}_{\mathbf{n}},\phi
_{0})\equiv \frac{1}{2}\sum_{\mathbf{n}\in \mathcal{U}}\left( \dot{\varphi}_{%
\mathbf{n}^{\ast }}\dot{\varphi}_{\mathbf{n}}-\varphi _{\mathbf{n}^{\ast
}}\omega _{\mathbf{n}}^{2}\varphi _{\mathbf{n}}+2A_{\mathbf{n}^{\ast
}}\varphi _{\mathbf{n}}+2B_{\mathbf{n}^{\ast }}\dot{\varphi}_{\mathbf{n}%
}\right) +\check{L}_{I}(\varphi _{\mathbf{n}},\dot{\varphi}_{\mathbf{n}%
},\phi _{0}).
\end{equation*}%
Note that $\phi _{0}$, $A(\phi _{0})$ and $B(\phi _{0})$ may not admit an
acceptable expansion in the basis $e_{\mathbf{n}}(\mathbf{x})$, within the
same space of functions as $\varphi $ does. Nevertheless, we do not need to
interpret $A_{\mathbf{n}^{\ast }}$ and $B_{\mathbf{n}^{\ast }}$ as
coefficients of a Fourier expansion. It is enough to view them as the
integrals shown. So doing, we can include the effects of $\phi _{0}$ into
the external sources $K_{0}$ (see below).

Then, we introduce the Hamiltonian%
\begin{equation*}
H_{\lambda }(\pi _{\mathbf{n}},\varphi _{\mathbf{n}},\phi _{0})=\sum_{%
\mathbf{n}\in \mathcal{U}}\pi _{\mathbf{n}^{\ast }}\dot{\varphi}_{\mathbf{n}%
}-\check{L}_{\lambda }(\varphi _{\mathbf{n}},\dot{\varphi}_{\mathbf{n}},\phi
_{0}),
\end{equation*}%
where the momenta are%
\begin{equation*}
\pi _{\mathbf{n}}=\dot{\varphi}_{\mathbf{n}}+B_{\mathbf{n}}+\Delta _{\mathbf{%
n}},\qquad \Delta _{\mathbf{n}}\equiv \frac{\partial \check{L}_{I}(\varphi _{%
\mathbf{n}},\dot{\varphi}_{\mathbf{n}},\phi _{0})}{\partial \dot{\varphi}_{%
\mathbf{n}^{\ast }}},
\end{equation*}%
and switch to the action 
\begin{equation}
\mathcal{\hat{S}}_{\lambda }(\varphi ,\phi _{0})=\mathcal{S}_{\text{e}%
}^{\prime }+\int_{t_{\text{i}}}^{t_{\text{f}}}\!\mathrm{d}t\!\hspace{0.02in}%
L_{\lambda }^{\prime }(\varphi _{\mathbf{n}},\dot{\varphi}_{\mathbf{n}},\pi
_{\mathbf{n}},\dot{\pi}_{\mathbf{n}},\phi _{0}),  \label{switcha}
\end{equation}%
by means of the equivalent Lagrangian 
\begin{eqnarray}
L_{\lambda }^{\prime }(\varphi _{\mathbf{n}},\dot{\varphi}_{\mathbf{n}},\pi
_{\mathbf{n}},\dot{\pi}_{\mathbf{n}},\phi _{0}) &=&\frac{1}{2}\sum_{\mathbf{n%
}\in \mathcal{U}}(\pi _{\mathbf{n}^{\ast }}\dot{\varphi}_{\mathbf{n}}-\dot{%
\pi}_{\mathbf{n}^{\ast }}\varphi _{\mathbf{n}})-H_{\lambda }(\pi _{\mathbf{n}%
},\varphi _{\mathbf{n}},\phi _{0}),  \notag \\
&=&\frac{1}{2}\sum_{\mathbf{n}\in \mathcal{U}}(\pi _{\mathbf{n}^{\ast }}\dot{%
\varphi}_{\mathbf{n}}-\dot{\pi}_{\mathbf{n}^{\ast }}\varphi _{\mathbf{n}%
}-\pi _{\mathbf{n}^{\ast }}\pi _{\mathbf{n}}-\varphi _{\mathbf{n}^{\ast
}}\omega _{\mathbf{n}}^{2}\varphi _{\mathbf{n}})  \label{lpp} \\
&&+\frac{1}{2}\sum_{\mathbf{n}\in \mathcal{U}}(2B_{\mathbf{n}^{\ast }}\pi _{%
\mathbf{n}}+2A_{\mathbf{n}^{\ast }}\varphi _{\mathbf{n}}-B_{\mathbf{n}^{\ast
}}B_{\mathbf{n}})+L_{I}(\pi _{\mathbf{n}},\varphi _{\mathbf{n}},\phi _{0}), 
\notag
\end{eqnarray}%
and possibly different endpoint corrections $\mathcal{S}_{\text{e}}^{\prime
} $. The interaction part reads 
\begin{equation*}
L_{I}(\pi _{\mathbf{n}},\varphi _{\mathbf{n}},\phi _{0})=\check{L}%
_{I}(\varphi _{\mathbf{n}},\dot{\varphi}_{\mathbf{n}},\phi _{0})+\frac{1}{2}%
\sum_{\mathbf{n}\in \mathcal{U}}\Delta _{\mathbf{n}^{\ast }}\Delta _{\mathbf{%
n}}.
\end{equation*}

Finally, we switch to coherent states $z_{\mathbf{n}}$, $\bar{z}_{\mathbf{n}%
} $ by means of (\ref{zetan}), and to the quantum fluctuations $w_{\mathbf{n}%
}$ and $\bar{w}_{\mathbf{n}}$ by means of the shift (\ref{shift2}).

Now we are ready to determine the corrections $\mathcal{S}_{\text{e}%
}^{\prime }$, so as to have the correct classical variational problem. They
may depend on the initial, final and possibly boundary conditions (\ref{bou}%
). Defining%
\begin{equation}
z(t,\mathbf{x})=\sum_{\mathbf{n}\in \mathcal{U}}z_{\mathbf{n}}(t)e_{\mathbf{n%
}}(\mathbf{x}),\qquad \bar{z}(t,\mathbf{x})=\sum_{\mathbf{n}\in \mathcal{U}}%
\bar{z}_{\mathbf{n}}(t)e_{\mathbf{n}}(\mathbf{x}),  \label{zexp}
\end{equation}%
the initial and final conditions are%
\begin{equation}
z(t_{\text{i}},\mathbf{x})=z_{\text{i}}(\mathbf{x})\equiv \sum_{\mathbf{n}%
\in \mathcal{U}}z_{\mathbf{n}\text{i}}e_{\mathbf{n}}(\mathbf{x}),\qquad \bar{%
z}(t_{\text{f}},\mathbf{x})=\bar{z}_{\text{f}}(\mathbf{x})\equiv \sum_{%
\mathbf{n}\in \mathcal{U}}\bar{z}_{\mathbf{n}\text{f}}e_{\mathbf{n}}(\mathbf{%
x}),  \label{endp}
\end{equation}%
where $z_{\text{i}}(\mathbf{x})$ and $\bar{z}_{\text{f}}(\mathbf{x})$ are
given functions on $\Omega $. Note that they vanish on $\partial \Omega $,
which makes them compatible with the boundary conditions (\ref{bou}).

It is easy to check that the correct endpoint action is 
\begin{equation}
\mathcal{S}_{\text{e}}^{\prime }=\hat{S}_{\text{e}}=-i\sum_{\mathbf{n}\in 
\mathcal{U}}\!\left( \bar{z}_{\mathbf{n}^{\ast }\text{f}}\omega _{\mathbf{n}%
}z_{\mathbf{n}}(t_{\text{f}})+\bar{z}_{\mathbf{n}^{\ast }}(t_{\text{i}%
})\omega _{\mathbf{n}}z_{\mathbf{n}\text{i}}\right) .  \label{she}
\end{equation}%
The first thing to notice is that the Lagrangian $L_{\lambda }^{\prime }$ of
(\ref{lpp}) depends on the time derivatives $\dot{\varphi}_{\mathbf{n}}$ in
a very simple way. At the same time, the gradients of the fields have
disappeared after the Fourier expansion ($\pi _{\mathbf{n}}$, $\varphi _{%
\mathbf{n}}$, $z_{\mathbf{n}}$ and $\bar{z}_{\mathbf{n}}$ depend only on
time). In particular, the interaction Lagrangian $L_{I}(\pi _{\mathbf{n}%
},\varphi _{\mathbf{n}},\phi _{0})$ does not contain time derivatives $\dot{z%
}_{\mathbf{n}}$ and $\dot{\bar{z}}_{\mathbf{n}}$, after the switch to
coherent states. Thus, when we study the variations $\delta z_{\mathbf{n}}$, 
$\delta \bar{z}_{\mathbf{n}}$ of $z_{\mathbf{n}}$ and $\bar{z}_{\mathbf{n}}$%
, the endpoint contributions are compensated by the same endpoint
corrections we had in the free-field limit, as in (\ref{sof}).

\subsection{Complete action}

The final action (\ref{redu}) is, from (\ref{switcha}), (\ref{lpp}) and (\ref%
{she}) 
\begin{eqnarray}
&&S_{\lambda }(w,\bar{w})\equiv \mathcal{S}_{\lambda }(\varphi ,\phi
_{0})=\int_{t_{\text{i}}}^{t_{\text{f}}}\mathrm{d}t\int_{\Omega }\mathrm{d}%
^{3}\mathbf{x\hspace{0.01in}}L_{\lambda }(\phi _{0},\dot{\phi}_{0},\nabla
\phi _{0})+\hat{S}_{\text{free}}^{\prime }  \notag \\
&&\qquad +\frac{1}{2}\int_{t_{\text{i}}}^{t_{\text{f}}}\mathrm{d}t\mathbf{%
\hspace{0.01in}}\sum_{\mathbf{n}\in \mathcal{U}}(2B_{\mathbf{n}^{\ast }}\pi
_{\mathbf{n}}+2A_{\mathbf{n}^{\ast }}\varphi _{\mathbf{n}}-B_{\mathbf{n}%
^{\ast }}B_{\mathbf{n}})+\int_{t_{\text{i}}}^{t_{\text{f}}}\mathrm{d}t%
\mathbf{\hspace{0.01in}}\mathcal{L}_{I}(z_{\mathbf{n}},\bar{z}_{\mathbf{n}%
}),\,\qquad  \label{complac}
\end{eqnarray}%
where $\mathcal{L}_{I}(z_{\mathbf{n}},\bar{z}_{\mathbf{n}})=L_{I}(\pi _{%
\mathbf{n}},\varphi _{\mathbf{n}},\phi _{0})$, with the substitutions $\pi _{%
\mathbf{n}}=-i\omega _{\mathbf{n}}(z_{\mathbf{n}}-\bar{z}_{\mathbf{n}})$, $%
\varphi _{\mathbf{n}}=z_{\mathbf{n}}+\bar{z}_{\mathbf{n}}$, and $\hat{S}_{%
\text{free}}^{\prime }$ is the expression of formula (\ref{sof}). It is
understood that the relations between $z_{\mathbf{n}}$, $\bar{z}_{\mathbf{n}%
} $ and $w_{\mathbf{n}}$, $\bar{w}_{\mathbf{n}}$ are still given by the
shift (\ref{shift2}), defined by the free-field solution (\ref{zon}) with
initial/final conditions (\ref{inico}). This way\footnote{%
Another possibility is to define $w_{\mathbf{n}}$, $\bar{w}_{\mathbf{n}}$ by
shifting $z_{\mathbf{n}}$, $\bar{z}_{\mathbf{n}}$ by the solution of the
interacting equations of motion. At the practical level, it does not make
much of a difference, but some formulas would have to be adapted to that
choice.}, $\hat{S}_{\text{free}}^{\prime }$ coincides with the free $w$, $%
\bar{w}$ action $S_{\text{free}}(w,\bar{w})$ of formula (\ref{scoh}).

We see that $S_{\lambda }(w,\bar{w})$ is made of three types of
contributions: $i$) those that go unmodified into the generating functional $%
\Gamma $, which are the first term after the equal sign in the first line,
and the sum in the second line; $ii$) the free part, which is (\ref{scoh})
and gives the propagators (\ref{prow}); $iii$) the interaction part, encoded
in $\mathcal{L}_{I}(z_{\mathbf{n}},\bar{z}_{\mathbf{n}})$. For the
calculations, we can concentrate on the last two terms.

\subsection{Amplitudes, correlation functions, and diagrams}

The amplitudes we want to calculate are%
\begin{eqnarray}
\langle \bar{z}_{\text{f}},t_{\text{f}};z_{\text{i}},t_{\text{i}}\rangle
_{\zeta ,\bar{\zeta}} &=&\!\!\!\!\!\!\!\!\underset{z(t_{\text{i}})=z_{\text{i%
}},\hspace{0.01in}\bar{z}(t_{\text{f}})=\bar{z}_{\text{f}}}{\int }%
\!\!\!\!\!\!\!\![\mathrm{d}z\mathrm{d}\bar{z}]\,\mathrm{\exp }\left(
iS_{\lambda }(w,\bar{w})+i\int_{t_{\text{i}}}^{t_{\text{f}}}\mathrm{d}t%
\hspace{0.01in}\int_{\Omega }\!\mathrm{d}^{3}\mathbf{x}(\bar{\zeta}z+\bar{z}%
\zeta )\right)  \notag \\
&=&\!\!\!\!\!\!\!\!\underset{w(t_{\text{i}})=\bar{w}(t_{\text{f}})=0}{\int }%
\!\!\!\!\!\!\!\![\mathrm{d}w\mathrm{d}\bar{w}]\,\mathrm{\exp }\left(
iS_{\lambda }(w,\bar{w})+i\int_{t_{\text{i}}}^{t_{\text{f}}}\mathrm{d}t%
\hspace{0.01in}\int_{\Omega }\!\mathrm{d}^{3}\mathbf{x}(\bar{\zeta}z_{0}+%
\bar{z}_{0}\zeta +\bar{\zeta}w+\bar{w}\zeta )\right) \!\!,\qquad \quad
\label{amplito}
\end{eqnarray}%
where $w(t,\mathbf{x})$, $\bar{w}(t,\mathbf{x})$, $z_{0}(t,\mathbf{x})$ and $%
\bar{z}_{0}(t,\mathbf{x})$ are defined from their Fourier coefficients in
analogy with (\ref{zexp}). As usual, we have introduced sources $\zeta $, $%
\bar{\zeta}$, to prepare for the diagrammatic approach. The initial and
final \textquotedblleft states\textquotedblright\ are described by the
configurations (\ref{endp}), which are compatible with the boundary
conditions (\ref{bou}).

At $\mathcal{L}_{I}(z_{\mathbf{n}},\bar{z}_{\mathbf{n}})=0$ (which we call
\textquotedblleft free limit\textquotedblright , although some $\lambda $
dependence remains in $A$, $B$ and $L_{\lambda }(\phi _{0},\dot{\phi}%
_{0},\nabla \phi _{0})$), we find, using (\ref{complac}),%
\begin{equation}
\langle \bar{z}_{\text{f}},t_{\text{f}};z_{\text{i}},t_{\text{i}}\rangle
_{\zeta ,\bar{\zeta}}^{\text{free}}=\exp \left( i\tilde{W}_{0}+i\hat{W}%
_{0}(\zeta ^{\prime },\bar{\zeta}^{\prime })+i\int_{t_{\text{i}}}^{t_{\text{f%
}}}\!\!\mathrm{d}t\!\int_{\Omega }\!\mathrm{d}^{3}\mathbf{x\hspace{0.01in}(}%
\bar{\zeta}^{\prime }z_{0}+\bar{z}_{0}\zeta ^{\prime })\right) \equiv 
\mathrm{e}^{iW^{\text{free}}},  \label{freeampl}
\end{equation}%
where%
\begin{equation}
\zeta ^{\prime }=\zeta +A+i\omega B,\qquad \bar{\zeta}^{\prime }=\bar{\zeta}%
+A-iB\omega ,  \label{zetap}
\end{equation}%
and, from (\ref{amplicohe})\ and (\ref{Wcohe}),%
\begin{eqnarray}
\tilde{W}_{0} &=&\int_{t_{\text{i}}}^{t_{\text{f}}}\!\!\mathrm{d}%
t\!\int_{\Omega }\!\mathrm{d}^{3}\mathbf{x}\left[ \mathbf{\hspace{0.01in}}%
L_{\lambda }(\phi _{0},\dot{\phi}_{0},\nabla \phi _{0})-\frac{1}{2}%
B^{2}(\phi _{0})\right] ,  \notag \\
\hat{W}_{0}(\zeta ,\bar{\zeta}) &=&-2i\sum_{\mathbf{n}\in \mathcal{U}}\bar{z}%
_{\mathbf{n}^{\ast }\text{f}}\omega _{\mathbf{n}}\mathrm{e}^{-i\omega _{%
\mathbf{n}}\tau }z_{\mathbf{n}\text{i}}+i\int_{t_{\text{i}}}^{t_{\text{f}%
}}\!\!\mathrm{d}t\!\!\int_{t_{\text{i}}}^{t_{\text{f}}}\!\!\mathrm{d}%
t^{\prime }\!\sum_{\mathbf{n}\in \mathcal{U}}\bar{\zeta}_{\mathbf{n}^{\ast
}}(t)\theta (t-t^{\prime })\frac{\mathrm{e}^{-i\omega _{\mathbf{n}%
}(t-t^{\prime })}}{2\omega _{\mathbf{n}}}\zeta _{\mathbf{n}}(t^{\prime
}).\qquad  \label{WW0}
\end{eqnarray}%
In (\ref{zetap}) $\omega $ stands for the operator $\sqrt{-\triangle +m^{2}}$%
, where the derivatives act away from $B$. In (\ref{WW0}) $\zeta _{\mathbf{n}%
}$ and $\bar{\zeta}_{\mathbf{n}}$ are the coefficients of the Fourier
expansion of $\zeta $ and $\bar{\zeta}$ (which we can assume to make sense,
since the sources couple to $\bar{z}$ and $z$). We have moved the infinite
contribution $-(\tau /2)\sum_{\mathbf{n}\in \mathcal{U}}\omega _{\mathbf{n}}$
into the normalization of the functional integral.

Switching $\mathcal{L}_{I}(z_{\mathbf{n}},\bar{z}_{\mathbf{n}})$ back on,
the amplitudes of the interacting theory are%
\begin{equation}
\langle \bar{z}_{\text{f}},t_{\text{f}};z_{\text{i}},t_{\text{i}}\rangle
_{\zeta ,\bar{\zeta}}=\exp \left( i\int_{t_{\text{i}}}^{t_{\text{f}}}\!\!%
\mathrm{d}t\mathbf{\hspace{0.02in}}\mathcal{L}_{I}\left( \frac{\delta }{%
i\delta \bar{\zeta}_{\mathbf{n}^{\ast }}},\frac{\delta }{i\delta \zeta _{%
\mathbf{n}^{\ast }}}\right) \right) \mathrm{e}^{iW^{\text{free}}}.
\label{amplint}
\end{equation}%
So far, we have tacitly assumed $t_{\text{f}}>t_{\text{i}}$. For $t_{\text{i}%
}>t_{\text{f}}$, we have 
\begin{equation}
\langle \bar{z}_{\text{f}},t_{\text{f}};z_{\text{i}},t_{\text{i}}\rangle
_{\zeta ,\bar{\zeta}}=\langle \bar{z}_{\text{i}},t_{\text{i}};z_{\text{f}%
},t_{\text{f}}\rangle _{\zeta ,\bar{\zeta}}^{\ast },  \label{conj}
\end{equation}%
from (\ref{conjugat}).

The correlation functions are 
\begin{equation}
\langle \bar{z}_{\text{f}},t_{\text{f}}|\hspace{0.01in}\tilde{z}_{\mathbf{n}%
_{1}}(t_{1})\cdots \tilde{z}_{\mathbf{n}_{k}}(t_{k})|\hspace{0.01in}z_{\text{%
i}},t_{\text{i}}\rangle \equiv \left. \frac{\delta ^{k}\langle \bar{z}_{%
\text{f}},t_{\text{f}};z_{\text{i}},t_{\text{i}}\rangle _{\zeta ,\bar{\zeta}}%
}{i\delta \tilde{\zeta}_{\mathbf{n}_{1}^{\ast }}(t_{1})\cdots i\delta \tilde{%
\zeta}_{\mathbf{n}_{k}^{\ast }}(t_{n})}\right\vert _{\zeta =\bar{\zeta}%
=0}\!\!\!\!\!,  \label{corre}
\end{equation}%
where $\tilde{z}_{\mathbf{n}_{j}}(t_{j})$ and $\tilde{\zeta}_{\mathbf{n}%
_{j}^{\ast }}(t_{j})$ may stand for $z_{\mathbf{n}_{j}}(t_{j})$ and $\bar{%
\zeta}_{\mathbf{n}_{j}^{\ast }}(t_{j})$, or $\bar{z}_{\mathbf{n}_{j}}(t_{j})$
and $\zeta _{\mathbf{n}_{j}^{\ast }}(t_{j})$. Although the amplitudes $%
\langle \bar{z}_{\text{f}},t_{\text{f}};z_{\text{i}},t_{\text{i}}\rangle $
have no external legs, since they are evaluated at $\zeta =\bar{\zeta}=0$,
the correlation functions are useful for the diagrammatic calculations,
since propagators and subdiagrams are particular cases of diagrams with
external legs. It is understood that the correlation functions (\ref{corre}%
)\ vanish when an insertion $\tilde{z}_{\mathbf{n}_{j}}(t_{j})$ lies outside
the time interval $(t_{\text{i}},t_{\text{f}})$.

Note that the correlation functions (\ref{corre}) receive contributions from
all the diagrams, including those that factorize subdiagrams with no
external legs. By definition, the connected correlation functions do not
include those types of diagrams.

We see that, ultimately, the $\tilde{\zeta}$ derivatives of (\ref{corre})
and (\ref{amplint})\ act on the free amplitude $\mathrm{e}^{iW^{\text{free}%
}} $, where $W^{\text{free}}$ is encoded in formulas (\ref{freeampl}) and (%
\ref{WW0}). Since $W^{\text{free}}$ is the exponential of a quadratic form
in the sources $\tilde{\zeta}$, the derivatives bring down propagators or
endpoints. The endpoints are the normalized one-point function%
\begin{equation}
\left. \mathrm{e}^{-iW^{\text{free}}}\frac{\delta \mathrm{e}^{iW^{\text{free}%
}}}{i\delta \bar{\zeta}_{\mathbf{n}^{\ast }}(t)}\right\vert _{\zeta =\bar{%
\zeta}=0}=z_{0\mathbf{n}}(t)+i\int_{t_{\text{i}}}^{t_{\text{f}}}\!\!\mathrm{d%
}t^{\prime }\hspace{0.02in}\theta (t-t^{\prime })\frac{\mathrm{e}^{-i\omega
_{\mathbf{n}}(t-t^{\prime })}}{2\omega _{\mathbf{n}}}(A_{\mathbf{n}%
}(t^{\prime })+i\omega _{\mathbf{n}}B_{\mathbf{n}}(t^{\prime })),
\label{endpoint}
\end{equation}%
and a similar expression for $-i\left. \mathrm{e}^{-iW^{\text{free}}}\delta 
\mathrm{e}^{iW^{\text{free}}}/\delta \zeta _{\mathbf{n}}(t)\right\vert
_{\zeta =\bar{\zeta}=0}$. We see that both the restriction to finite $\tau $
and the one to compact $\Omega $ contribute to the endpoints.

\begin{figure}[t]
\begin{center}
\includegraphics[width=16truecm]{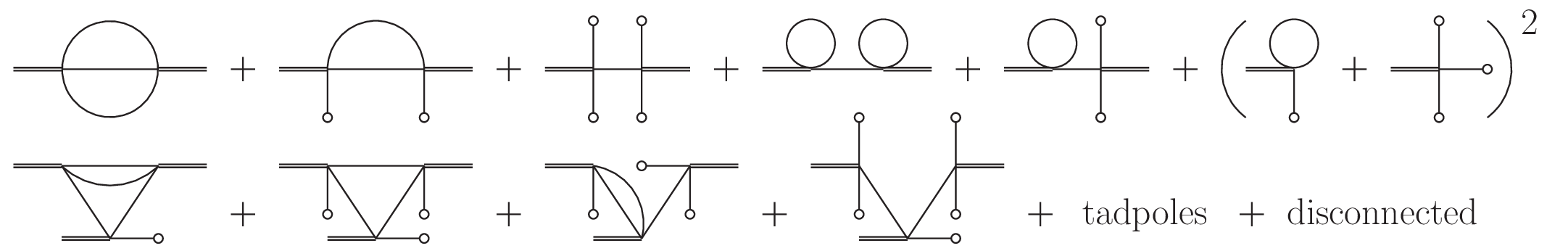}
\end{center}
\par
\vskip -.7truecm
\caption{Diagrams with two and three cubic vertices}
\label{endpointdiagra}
\end{figure}

By repeatedly differentiating $\mathrm{e}^{iW^{\text{free}}}$, we can build
the diagrams, and, ultimately, calculate any amplitude $\langle \bar{z}_{%
\text{f}},t_{\text{f}};z_{\text{i}},t_{\text{i}}\rangle $ we want,
perturbatively and diagrammatically. In fig. \ref{endpointdiagra} we
illustrate the diagrams with two and three cubic vertices, and no external
legs. The double lines stand for the sources $K$, while the little circles
stand for the endpoints (\ref{end}). The internal lines are the propagators (%
\ref{prow}), while the vertices are studied below.

From what we have said, it follows that, in the end, the diagrams look like
the ones we are accustomed to at $\tau =\infty $, $\Omega =\mathbb{R}^{3}$,
at least internally, apart from the discretization of the loop momenta.
Externally, the sources $K$ attached to the vertices take care of the
restrictions to finite $\tau $ and compact $\Omega $. These properties are
going to be extremely useful to study unitarity and extend the formulation
to purely virtual particles.

\subsection{Vertices}

Going through the derivation just outlined, we see that the vertices have
the form%
\begin{equation}
\int_{t_{\text{i}}}^{t_{\text{f}}}\mathrm{d}t\int_{\Omega }\mathrm{d}^{3}%
\mathbf{x\hspace{0.01in}}K(t,\mathbf{x})Q^{i}P^{j}\overset{k}{\overbrace{%
\nabla Q\cdots \nabla Q}}\mathbf{\hspace{0.04in}}\overset{l}{\overbrace{%
\nabla P\cdots \nabla P}},  \label{vv}
\end{equation}%
where, as in (\ref{QP}),%
\begin{equation}
Q(t,\mathbf{x})=\sum_{\mathbf{n}\in \mathcal{U}}\left( w_{\mathbf{n}}(t)+%
\bar{w}_{\mathbf{n}}(t)\right) e_{\mathbf{n}}(\mathbf{x}),\qquad P(t,\mathbf{%
x})=-i\sum_{\mathbf{n}\in \mathcal{U}}\omega _{\mathbf{n}}\left( w_{\mathbf{n%
}}(t)-\bar{w}_{\mathbf{n}}(t)\right) e_{\mathbf{n}}(\mathbf{x}),  \label{PQ}
\end{equation}%
while the source $K(t,\mathbf{x})$ is built with $z_{0}$, $\bar{z}_{0}$ and $%
\phi _{0}$. Defining the constants\footnote{%
Note that we do not need to require that $K(t,\mathbf{x})$ admits a Fourier
expansion in the basis $e_{\mathbf{n}}(\mathbf{x})$, or that it admits one
in the same space of functions as $\varphi $ does.}%
\begin{equation}
C_{\mathbf{n}_{1}\cdots \mathbf{n}_{k}}^{K\mathbf{m}_{1}\cdots \mathbf{m}%
_{l}}(t)=i^{-l}|\Omega |^{(k+l-2)/2}\int_{\Omega }\mathrm{d}^{3}\mathbf{x%
\hspace{0.01in}}K(t,\mathbf{x})e_{\mathbf{n}_{1}}(\mathbf{x})\cdots e_{%
\mathbf{n}_{k}}(\mathbf{x})\nabla e_{\mathbf{m}_{1}}(\mathbf{x})\cdots
\nabla e_{\mathbf{m}_{l}}(\mathbf{x}),  \label{C}
\end{equation}%
the vertices can be arranged as%
\begin{equation}
i^{k+l}|\Omega |^{v}\sum_{\mathbf{n},\mathbf{m}}\Pi \omega \int_{t_{\text{i}%
}}^{t_{\text{f}}}\!\!\mathrm{d}t\hspace{0.01in}C_{\mathbf{n}_{1}\cdots 
\mathbf{n}_{i+j}}^{K\mathbf{m}_{1}\cdots \mathbf{m}_{k+l}}w_{\mathbf{n}%
_{1}}\cdots w_{\mathbf{n}_{i}}\bar{w}_{\mathbf{m}_{1}}\cdots \bar{w}_{%
\mathbf{m}_{k}}w_{\mathbf{n}_{i+1}}\cdots w_{\mathbf{n}_{i+j}}\bar{w}_{%
\mathbf{m}_{k+1}}\cdots \bar{w}_{\mathbf{m}_{k+l}},\quad  \label{vertices}
\end{equation}%
where $v=(2-i-j-k-l)/2$ and $\Pi \omega $ stands for a product of
frequencies $\omega _{\mathbf{n}}$, $\omega _{\mathbf{m}}$.

Let us examine some typical situations, focusing on $\Pi \omega =1$ and $%
K\equiv 1$, and dropping the superscript $K$ in $C$.

If $\Omega $ is a three torus $T^{3}$, we find 
\begin{equation}
C_{\mathbf{n}_{1}\cdots \mathbf{n}_{i}}^{\mathbf{n}_{i+1}\cdots \mathbf{n}%
_{i+j}}=\mathbf{\bar{n}}_{i+1}\cdots \mathbf{\bar{n}}_{i+j}\mathbf{\delta }(%
\mathbf{n}_{1}+\cdots +\mathbf{n}_{i+j}),\qquad \mathbf{\delta }(\mathbf{n}%
)=\left\{ 
\begin{tabular}{l}
$1$ if $\mathbf{n}=\mathbf{0},$ \\ 
$0$ otherwise,%
\end{tabular}%
\right.  \label{evid}
\end{equation}%
so the discretized momentum is conserved at the vertices.

If $\Omega $ is a bounded box, the discretized momentum is not conserved at
the vertices. We can gain a form of momentum conservation by introducing
external sources $K$ to take care of the restriction to finite volume,
similar to the sources $K_{n\alpha }$ introduced in (\ref{vertici}) for the
restriction to finite $\tau $. For simplicity, we focus on vertices that do
not involve gradients, since it is straightforward to generalize the results
to include them.

We double the sides of the box by writing a typical vertex as%
\begin{equation*}
\int_{t_{\text{i}}}^{t_{\text{f}}}\mathrm{d}t\int_{\Omega }\mathrm{d}^{3}%
\mathbf{x\hspace{0.01in}}w^{i}\bar{w}^{j}=\int_{t_{\text{i}}}^{t_{\text{f}}}%
\mathrm{d}t\left( \prod_{a=1}^{3}\int_{0}^{L_{a}}\mathrm{d}x_{a}\right) K(%
\mathbf{x})w^{i}\bar{w}^{j},
\end{equation*}%
where $K(\mathbf{x})=\prod_{a=1}^{3}\theta (L_{a}-2x_{a})$. Moreover, we use 
\begin{equation*}
e_{\mathbf{n}}(\mathbf{x})=\frac{i}{\sqrt{L_{1}L_{2}L_{3}}}\prod_{a=1}^{3}(%
\mathrm{e}^{i\bar{n}_{a}x_{a}}-\mathrm{e}^{-i\bar{n}_{a}x_{a}})\equiv
\sum_{\pi (\mathbf{n})}c_{\pi (\mathbf{n})}f_{\pi (\mathbf{n})}(\mathbf{x}%
),\qquad f_{\mathbf{n}}(\mathbf{x})\equiv \frac{\mathrm{e}^{i\mathbf{\bar{n}%
\cdot x}}}{\sqrt{L_{1}L_{2}L_{3}}},
\end{equation*}%
to switch to the basis $f_{\mathbf{n}}(\mathbf{x})$ with periodic boundary
conditions, where $c_{\pi (\mathbf{n})}=\pm i$ are certain coefficients, and
the label $\pi (\mathbf{n})$ collects all the ways $(\pm n_{1},\pm n_{2},\pm
n_{3})$ of flipping the signs in front of the integer numbers. We obtain 
\begin{equation*}
\int_{\Omega }\mathrm{d}^{3}\mathbf{x\hspace{0.01in}}w^{i}\bar{w}^{j}\propto
\sum_{\mathbf{n}_{0},\mathbf{n}_{1}\cdots \mathbf{n}_{i+j}}\left[ \sum_{\pi (%
\mathbf{n})}\mathbf{\delta}\!\left( \mathbf{n}_{0}+\sum_{k=1}^{i+j}\pi (%
\mathbf{n}_{k})\right) \prod_{k=1}^{i+j}c_{\pi (\mathbf{n}_{k})}\right] K_{%
\mathbf{n}_{0}}w_{\mathbf{n}_{1}}\cdots w_{\mathbf{n}_{i}}\bar{w}_{\mathbf{n}%
_{i+1}}\cdots \bar{w}_{\mathbf{n}_{i+j}},
\end{equation*}%
where 
\begin{equation*}
K_{\mathbf{n}}=\left( \prod_{a=1}^{3}\int_{0}^{L_{a}}\mathrm{d}x_{a}\right) 
\mathbf{\hspace{0.01in}}f_{\mathbf{n}}^{\ast }(\mathbf{x})K(\mathbf{x}).
\end{equation*}%
We see that now we have momentum conservation at the vertices, provided we
take into account the momentum carried by $K_{\mathbf{n}}$.

Each vertex can be imagined as a sum (the one in square brackets) of usual
vertices coupled to external sources. A diagram with $I$ internal legs
splits into a sum of copies that have identical propagators, but different
vertices, which correspond to the choices of signs in front of the
components of the internal momenta.

If $\Omega $ is the sphere $S^{3}$, the boundary is absent. Invariance under
translations on the sphere means that there is no reflection, and the
(angular) momentum is conserved. The coefficients $C_{\mathbf{n}_{1}\cdots 
\mathbf{n}_{i}}^{K\mathbf{m}_{1}\cdots \mathbf{m}_{j}}$ can be worked out
from the decomposition of the product of two (or more) spherical harmonics
in the same basis of spherical harmonics.

If $\Omega $ is a ball $B$, the angular momentum is conserved, but the
boundary originates a reflection. It is easier to first consider the
analogue of this problem in two space dimensions, where $B$ is replaced by a
disc $D_{2}$. Viewing $D_{2}$ as a hemisphere, we double it into the sphere $%
S^{2}$, and introduce a source $K$ to make the vertex vanish in the extra
hemisphere. The boundary of $D_{2}$ is the equator of $S^{2}$, and reflects
the radial component of the momentum.

Going back to the ball $B$ in three space dimensions, we double the ball by
adding the exterior space and the point at infinity, thereby obtaining $%
S^{3} $, and make the vertex vanish in the complement of $B$ by means of an
external source $K$. Then, the boundary of $B$ reflects the radial component
of the momentum.

Further sources $K$ must be introduced to deal with the restriction to
finite $\tau $, as explained in formula (\ref{vertici}).

At the end, we achieve our goal: we move almost every detail about the
restriction to finite $\tau $ and compact $\Omega $ away from the interior
sectors of the diagrams. \textquotedblleft Almost every\textquotedblright\
means every, but for the discretization of the loop momenta. The
discretization does enter the diagrams, since it affects the propagators.
What is important is that it does not affect the spectral optical identities
of ref. \cite{diagrammarMio} in an invasive way, because those identities
hold threshold by threshold, for arbitrary frequencies, without integrating
on the loop momenta, or summing over $\mathbf{n}\in \mathcal{U}$.

Now we are equipped with what we need to proceed. We first regularize and
renormalize the theory, then investigate unitarity, and finally extend the
formulation to theories that include purely virtual particles.

\section{Regularization and renormalization}

\label{renorma}\setcounter{equation}{0}

In this section we discuss the renormalization of quantum field theory in a
finite interval of time $\tau $, on a compact space manifold $\Omega $, and
show that it coincides with the one of the theory at $\tau =\infty $, $%
\Omega =\mathbb{R}^{3}$. The ultraviolet behavior of a correlation function
just depends on its small-distance behavior in coordinate space, which
should know nothing about the restriction to a compact $\Omega $ (as long as 
$\Omega $ is smooth), as well as the restriction to a finite $\tau $.
Specifically, for large values of $\mathbf{n}$, the sums on $\mathbf{n}$
reduce to the usual integrals. The behavior of a diagram at large $\mathbf{n}
$ matches the ultraviolet behavior at $\tau =$ $\infty $, $\Omega =\mathbb{R}%
^{3}$.

The common power counting rules apply. If the theory is equipped with the
counterterms that renormalize its divergences at $\tau =\infty $, $\Omega =%
\mathbb{R}^{3}$, it is also renormalized at finite $\tau $ on a compact $%
\Omega $. Problems could appear if $\Omega $ has singularities, such as the
tip of a cone. These situations must be dealt with on a case by case basis.

\subsection{Regularization}

The simplest regularization procedure amounts to truncating the infinite
sums by means of a cutoff $\mathbf{N}$ on the sum over $\mathbf{n}$. A more
elegant option is to generalize the dimensional regularization technique,
which has the advantage of being manifestly gauge invariant. Before
describing how the generalization is done, it is convenient to briefly
review two variants of the usual technique at $\tau =\infty $, $\Omega =%
\mathbb{R}^{3}$.

We dimensionally regularize the integrals on the loop momenta, by continuing
them to dimension $D-1$, where $D$ is complex. However, we do not
dimensionally regularize the integrals on the loop energies. As far as those
are concerned, we have two options. The first option is to integrate on the
loop energies after integrating on the loop momenta. So doing, the integrals
on the energies are automatically regularized by means of an analytic
regularization\footnote{%
The analytic regularization \cite{anreg} is obtained by raising the free
propagators to a complex power $\delta $, which is treated analytically and
sent to one after removing the divergent parts (which are poles in $\delta
-1 $). Gauge invariance is recovered by means of finite local counterterms,
up to anomalies. The dimensional regularization \cite{dimreg} is a
particular analytic regularization, which uses the number of dimensions as
the regularizing parameter, and has the advantage of being manifestly gauge
invariant (up to anomalies).} (see below for an illustrative example). The
second option is to integrate on the loop energies first. In this respect,
it is important to stress that in the coherent-state approach the integrals
on the loop energies are all convergent (if done first), apart from the
tadpoles. The reason is that, by (\ref{complac}), no time derivatives appear
in the vertices. The simplest way to calculate the energy integrals is by
means of the residue theorem (and a symmetric integration for the tadpoles,
which is justified by the first option of integration).

The first option is more convenient to study the divergent parts of the
diagrams, and their renormalization. The second option is the one we prefer
here, because it is more convenient to study unitarity via the spectral
optical identities of ref. \cite{diagrammarMio}.

We can generalize the regularization techniques just mentioned to finite $%
\tau $ and compact $\Omega $ as follows. First, we observe that the
restriction to finite $\tau $ poses no problem, because it does not enter
the diagrams in the approach we have formulated (based on coherent states
and Fourier transforms for energies). We just need to pay attention to the
effects of the restriction to a compact $\Omega $ inside the diagrams, due
to the discretizations of the loop momenta and the frequencies $\omega _{%
\mathbf{n}}$.

In several cases it may be straightforward to continue the manifold $\Omega $
to $D-1$ dimensions. This occurs, for example, in the cases of the torus,
the box with boundary,\ the sphere and the ball. If we need to separate a
radial coordinate $r$ from angular coordinates $\theta _{i}$, as in the case
of the ball, we dimensionally continue only the angular part, integrate on
that first, and then integrate on $r$. So doing, by an argument similar to
the one used above for the integrals on the loop energies, the $r$ integral
ends up being regularized by means of the analytic regularization.

A more general, and conceptually elegant, possibility is available. We
extend $\Omega $ to $\Omega \times \Omega _{\varepsilon }$ by attaching an
evanescent manifold $\Omega _{\varepsilon }$ to $\Omega $, where $%
\varepsilon =4-D$ if we are interested in four spacetime dimensions, $%
\varepsilon =d-D$ if we want to regularize a theory in $d$ spacetime
dimensions. Since we do not need to restrict the attachment $\Omega
_{\varepsilon }$ to be compact, we just choose the simplest option for it,
which is $\Omega _{\varepsilon }=\mathbb{R}^{-\varepsilon }$. Then we use
Fourier series for the coordinates of $\Omega $, but Fourier transforms for
those of $\mathbb{R}^{-\varepsilon }$. And, of course, Fourier transforms
for times and energies. So doing, the diagrams involve integrals on the loop
energies, integrals on the momenta $\mathbf{p}_{\varepsilon }$ of $\mathbb{R}%
^{-\varepsilon }$, and sums on the labels $\mathbf{n}$ of the $\Omega $
frequencies $\omega _{\mathbf{n}}$.

From the calculational point of view, the first option is to start by
integrating on the momenta $\mathbf{p}_{\varepsilon }$ of $\mathbb{R}%
^{-\varepsilon }$, then sum on the labels $\mathbf{n}$ of $\Omega $ and
finally integrate on the loop energies. The last two operations can be
freely interchanged, since both end up being regularized by the analytic
regularization. For example, let us consider the integral of a power of a
propagator (which might depend on Feynman parameters, if it is originated by
the product of more propagators). After integrating on $\mathbf{p}%
_{\varepsilon }$, we obtain 
\begin{equation}
\int \frac{\mathrm{d}^{-\varepsilon }\mathbf{p}_{\varepsilon }}{(2\pi
)^{\varepsilon }}\frac{1}{(e^{2}-\mathbf{\bar{n}}^{2}-\mathbf{p}%
_{\varepsilon }^{2}-m^{2}+i\epsilon )^{\alpha }}=\frac{\Gamma \left( \alpha +%
\frac{\varepsilon }{2}\right) (-1)^{\alpha }}{(4\pi )^{-\varepsilon
/2}\Gamma (\alpha )(m^{2}+\mathbf{\bar{n}}^{2}-e^{2}-i\epsilon )^{(2\alpha
+\varepsilon )/2}},  \label{dimreg}
\end{equation}%
where $\mathbf{\bar{n}}$ are some functions of the labels $\mathbf{n}$. At
this point, the integral on the energy $e$ and the sum over $\mathbf{n}$ are
analytically regularized by the $\varepsilon $-dependent exponent.

The second option, preferred to study the spectral optical identities, is to
first integrate on the loop energies by means of the residue theorem, with
the help of a symmetric integration (if needed), and then integrate on the
momenta of $\mathbb{R}^{-\varepsilon }$. At the end, we may sum on $\mathbf{n%
}$, if needed. That sum is not necessary for the spectral optical
identities, while it is of course necessary for the calculations of the
amplitudes.

\subsection{The infinite time, infinite volume limit}

Before discussing the renormalization, it is convenient to show that when $%
\tau $ tends to infinity and $\Omega $ tends to $\mathbb{R}^{3}$, we obtain
the results of ordinary quantum field theory, that is to say, the usual
vacuum-to-vacuum amplitudes, and the usual diagrams\footnote{%
Here and below, words such as \textquotedblleft ordinary\textquotedblright\
and \textquotedblleft usual\textquotedblright\ refer to quantum field theory
with $\tau =\infty $ and $\Omega =\mathbb{R}^{3}$.}.

We first give the rules to work out the limit on a generic manifold $\Omega $%
, then consider some explicit cases. We recall that $\mathbf{n}$ is the
label of the eigenvalues of the Laplacian with Dirichlet boundary
conditions. The differences $\Delta \mathbf{n}$ between the labels of two
close eigenvalues are of order unity, and the eigenvalues become a continuum
in the limit $\Omega \rightarrow \mathbb{R}^{3}$.

Recall that the eigenfunctions $e_{\mathbf{n}}(x)$ on $\Omega $ satisfy%
\begin{equation}
-\triangle e_{\mathbf{n}}(x)+m^{2}e_{\mathbf{n}}(x)=\omega _{\mathbf{n}%
}^{2}e_{\mathbf{n}}(x)\text{ in }\Omega \text{,\qquad }e_{\mathbf{n}}(x)=0%
\text{ on }\partial \Omega .  \label{eigen}
\end{equation}%
We make an overall rescaling of $\Omega $ by a factor $\eta $, and denote
the resulting manifold by $\Omega _{\eta }$. Replacing $x$ by $x/\eta $ in
(\noindent \ref{eigen}), we see that the functions $f_{\mathbf{n}}(x)\equiv
e_{\mathbf{n}}(x/\eta )$ are eigenfunctions on $\Omega _{\eta }$, since they
satisfy%
\begin{equation}
-\triangle f_{\mathbf{n}}(x)+m^{2}f_{\mathbf{n}}(x)=\hat{\omega}_{\mathbf{n}%
}^{2}f_{\mathbf{n}}(x)\text{ in }\Omega _{\eta }\text{,\qquad and }f_{%
\mathbf{n}}(x)=0\text{ on }\partial \Omega _{\eta },  \label{igenf}
\end{equation}%
with $\hat{\omega}_{\mathbf{n}}^{2}=m^{2}+(\omega _{\mathbf{n}%
}^{2}-m^{2})/\eta ^{2}$. This means that there exists a $\mathbf{p}(\mathbf{n%
},\eta )$, ranging in some domain $\mathcal{U}_{p}$, such that 
\begin{equation}
\hat{e}_{\mathbf{p}(\mathbf{n},\eta )}(x)=\eta ^{(1-D)/2}e_{\mathbf{n}%
}(x/\eta ),\qquad x\in \Omega _{\eta },  \label{en}
\end{equation}%
is an orthonormal basis of eigenfunction on $\Omega _{\eta }$, where the
power of $\eta $ in front of $e_{\mathbf{n}}$ is fixed to have the right
normalization, and the hat on $e$ emphasizes that $\hat{e}_{\mathbf{p}(%
\mathbf{n},\eta )}$ possibly involves a different notation for the subscript
($\mathbf{n}$ and $\mathbf{p}$ being generic labels, so far), better suited
to study the limit of infinite volume.

At this point, we take the limit $\eta \rightarrow \infty $, with $\mathbf{p}
$ fixed. This means, in particular, that $\mathbf{n}$ is understood as a
function of $\eta $. Let us start from the summation. We can write%
\begin{equation}
\sum_{\mathbf{n}\in \mathcal{U}}\equiv \int_{\mathcal{U}}\mathrm{d}^{D-1}%
\mathbf{n}=\sum_{\mathbf{p}(\mathbf{n},\eta )\in \mathcal{U}_{p}}\equiv
\int_{\mathcal{U}_{p}}\frac{\mathrm{d}^{D-1}\mathbf{p}}{(2\pi )^{D-1}}%
J,\qquad \qquad J\equiv (2\pi )^{D-1}\det \left( \frac{\partial \mathbf{n}}{%
\partial \mathbf{p}}\right) ,  \label{sum}
\end{equation}%
where $J$ is the Jacobian, apart from a normalization. The \textquotedblleft
integrals\textquotedblright\ on $\mathcal{U}$ and $\mathcal{U}_{p}$ are just
other ways to write the sums on $\mathcal{U}$ and $\mathcal{U}_{p}$.

Define constants $c_{J}$ and $d_{J}$ so that $J\simeq c_{J}\eta ^{d_{J}}$
when $\eta $ tends to infinity. On general grounds, we can view the sum on $%
\mathbf{p}$ as the sum on the states obtained after rescaling $\Omega $.
When $\eta $ is large, it is also the sum on the phase space cells. This
means that we can choose variables such that $J\sim |\Omega _{\eta
}|=|\Omega |\eta ^{D-1}$. Then, (\ref{sum}) gives 
\begin{equation}
\lim_{\eta \rightarrow \infty }\frac{\eta ^{1-D}}{|\Omega |}\sum_{\mathbf{n}%
\in \mathcal{U}}=\lim_{\eta \rightarrow \infty }\frac{\eta ^{1-D}}{|\Omega |}%
\sum_{\mathbf{p}(\mathbf{n},\eta )\in \mathcal{U}_{p}}=\int_{\mathbb{R}%
^{D-1}}\frac{\mathrm{d}^{D-1}\mathbf{p}}{(2\pi )^{D-1}}.  \label{inte}
\end{equation}%
Using this formula, we find%
\begin{equation}
\hat{e}_{\mathbf{p}}(\mathbf{x})=\sum_{\mathbf{p}^{\prime }\in \mathcal{U}%
_{p}}\delta _{\mathbf{p},\mathbf{p}^{\prime }}\hat{e}_{\mathbf{p}^{\prime }}(%
\mathbf{x})\simeq |\Omega |\eta ^{D-1}\int_{\mathbb{R}^{D-1}}\frac{\mathrm{d}%
^{D-1}\mathbf{p}}{(2\pi )^{D-1}}\delta _{\mathbf{p},\mathbf{p}^{\prime }}%
\hat{e}_{\mathbf{p}^{\prime }}(\mathbf{x}).  \label{e1}
\end{equation}%
Let $e_{\mathbf{p}}^{\infty }(\mathbf{x})$ denote the basis of the Fourier
transform in $\mathbb{R}^{D-1}$. We clearly have%
\begin{equation}
e_{\mathbf{p}}^{\infty }(\mathbf{x})=\int_{\mathbb{R}^{D-1}}\frac{\mathrm{d}%
^{D-1}\mathbf{p}}{(2\pi )^{D-1}}(2\pi )^{D-1}\delta ^{(D-1)}(\mathbf{p}-%
\mathbf{p}^{\prime })e_{\mathbf{p}^{\prime }}^{\infty }(\mathbf{x}).
\label{e2}
\end{equation}%
Since $e_{\mathbf{p}}^{\infty }/\hat{e}_{\mathbf{p}}\simeq e_{\mathbf{p}%
^{\prime }}^{\infty }/\hat{e}_{\mathbf{p}^{\prime }}$, the comparison
between (\ref{e1}) and (\ref{e2}) gives 
\begin{equation}
\lim_{\eta \rightarrow \infty }|\Omega |\eta ^{D-1}\delta _{\mathbf{n},%
\mathbf{n}^{\prime }}=\lim_{\eta \rightarrow \infty }|\Omega |\eta
^{D-1}\delta _{\mathbf{p},\mathbf{p}^{\prime }}=(2\pi )^{D-1}\delta ^{(D-1)}(%
\mathbf{p}-\mathbf{p}^{\prime }).  \label{delta}
\end{equation}%
Then, we also have 
\begin{eqnarray*}
|\Omega |\eta ^{D-1}\int_{{\Omega }_{\eta }}\mathrm{d}^{D-1}\mathbf{x\hspace{%
0.01in}}\hat{e}_{\mathbf{p}}^{\ast }(\mathbf{x})\hat{e}_{\mathbf{p}^{\prime
}}(\mathbf{x}) &=&|\Omega |\eta ^{D-1}\delta _{\mathbf{p},\mathbf{p}^{\prime
}} \\
&\rightarrow &(2\pi )^{D-1}\delta ^{(D-1)}(\mathbf{p}-\mathbf{p}^{\prime
})=\int_{\mathbb{R}^{D-1}}\mathrm{d}^{D-1}\mathbf{x\hspace{0.01in}}e_{%
\mathbf{p}}^{\infty \ast }(\mathbf{x})e_{\mathbf{p}^{\prime }}^{\infty }(%
\mathbf{x}),
\end{eqnarray*}%
which implies%
\begin{equation}
\lim_{\eta \rightarrow \infty }|\Omega |^{1/2}\eta ^{(D-1)/2}\hat{e}_{%
\mathbf{p}(\mathbf{n},\eta )}(\mathbf{x})=e_{\mathbf{p}}^{\infty }(\mathbf{x}%
).  \label{ep}
\end{equation}%
Next, we use this formula to compare the Fourier expansions of a field $\chi
(t,\mathbf{x})$ before and after the limit,%
\begin{equation*}
\chi (t,\mathbf{x})=\int_{\mathbb{R}^{D-1}}\frac{\mathrm{d}^{D-1}\mathbf{p}}{%
(2\pi )^{D-1}}\chi _{\mathbf{p}}^{\infty }(t)e_{\mathbf{p}}^{\infty }(%
\mathbf{x})=\sum_{\mathbf{p}(\mathbf{n},\eta )\in \mathcal{U}_{p}}\hat{\chi}%
_{\mathbf{p}(\mathbf{n},\eta )}(t)\hat{e}_{\mathbf{p}(\mathbf{n},\eta )}(%
\mathbf{x}).
\end{equation*}%
We find%
\begin{equation}
\lim_{\eta \rightarrow \infty }|\Omega |^{1/2}\eta ^{(D-1)/2}\hat{\chi}_{%
\mathbf{p}(\mathbf{n},\eta )}(t)=\chi _{\mathbf{p}}^{\infty }(t).
\label{chin}
\end{equation}

As far as the vertices are concerned, making the change of variables $%
\mathbf{x}\rightarrow \mathbf{x}/\eta $ in (\ref{C}) and using (\ref{en})
and (\ref{ep}), we obtain%
\begin{equation}
\lim_{\eta \rightarrow \infty }\eta ^{D-1}|\Omega |C_{\mathbf{n}_{1}\cdots 
\mathbf{n}_{k}}^{\mathbf{m}_{1}\cdots \mathbf{m}_{l}}=C_{\mathbf{p}%
_{1}\cdots \mathbf{p}_{k}}^{\infty \mathbf{q}_{1}\cdots \mathbf{q}%
_{l}}\equiv i^{-l}\int_{\mathbb{R}^{D-1}}\mathrm{d}^{D-1}\mathbf{x\hspace{%
0.01in}}e_{\mathbf{p}_{1}}^{\infty }(\mathbf{x})\cdots e_{\mathbf{p}%
_{k}}^{\infty }(\mathbf{x})\nabla e_{\mathbf{q}_{1}}^{\infty }(\mathbf{x}%
)\cdots \nabla e_{\mathbf{q}_{l}}^{\infty }(\mathbf{x}).  \label{Clim}
\end{equation}%
The same steps show that the coefficients $C_{\mathbf{n}_{1}\cdots \mathbf{n}%
_{i}}^{\mathbf{m}_{1}\cdots \mathbf{m}_{j}}$ of $\Omega $ coincide with the
coefficients $\hat{C}_{\mathbf{p}_{1}\cdots \mathbf{p}_{i}}^{\mathbf{q}%
_{1}\cdots \mathbf{q}_{j}}$ of $\Omega _{\eta }$.

\bigskip

The first example we consider is the torus. As explained above, we can
dimensionally regularize it by extending it to $T^{D-1}$ or $T^{3}\times 
\mathbb{R}^{-\varepsilon }$. We adopt the first option, which is more
symmetric. The diagrams on a torus have expressions that are similar to the
usual ones (with external sources attached to the vertices), apart from the
discretizations of the loop momenta and the frequencies.

We rescale each side $L_{i}$ by a factor $\eta $ and denote the rescaled
torus by $T_{\eta }^{D-1}$. Given the labels $\mathbf{n}$, $\mathbf{m}$,
etc., define momenta $\mathbf{p}$, $\mathbf{q}$, etc., through%
\begin{equation}
\mathbf{n}=(n_{i})=\eta \left( \frac{p_{i}L_{i}}{2\pi }\right) \mathbf{%
,\qquad m}=(m_{i})=\eta \left( \frac{q_{i}L_{i}}{2\pi }\right) \mathbf{,}
\label{enne}
\end{equation}%
etc. Clearly, $J=|T_{\eta }^{D-1}|$.

When we sum on $\mathbf{n}$, we sum on values that are separated by a $%
\Delta \mathbf{n}$ of order unity. If we make a change of variables from $%
\mathbf{n}$ to $\mathbf{p}$, we end up by summing on values separated by $%
\mathrm{d}\mathbf{p}=(2\pi /\eta )(\Delta n_{i}/L_{i})$, which becomes
arbitrarily small when the sides of the box tend to infinity. There, by
definition, the sum becomes an integral. This means that we have the relation%
\begin{equation}
\lim_{\eta \rightarrow \infty }\frac{1}{\eta ^{D-1}|T^{D-1}|}\sum_{\mathbf{n}%
\in \mathbb{Z}^{D-1}}=\int \frac{\mathrm{d}^{D-1}\mathbf{p}}{(2\pi )^{D-1}}.
\notag
\end{equation}%
The other relations can be checked similarly: (\ref{en}) follows from (\ref%
{tor}), while (\ref{ep}) gives $e_{\mathbf{p}}^{\infty }(\mathbf{x})=\mathrm{%
e}^{i\mathbf{p\cdot x}}$. In particular, formula (\ref{evid}) shows that $C_{%
\mathbf{n}_{1}\cdots \mathbf{n}_{i}}^{\mathbf{m}_{1}\cdots \mathbf{m}_{j}}=%
\hat{C}_{\mathbf{p}_{1}\cdots \mathbf{p}_{i}}^{\mathbf{q}_{1}\cdots \mathbf{q%
}_{j}}$, and (\ref{Clim}) holds with%
\begin{equation*}
C_{\mathbf{p}_{1}\cdots \mathbf{p}_{i}}^{\infty \mathbf{p}_{i+1}\cdots 
\mathbf{p}_{i+j}}=\mathbf{p}_{i+1}\cdots \mathbf{p}_{i+j}(2\pi )^{D-1}\delta
^{(D-1)}(\mathbf{p}_{1}+\cdots +\mathbf{p}_{i+j}).
\end{equation*}

Another example is the box with Dirichlet boundary conditions. We stick to a
segment for more clarity, since the extension to arbitrary space dimensions
is straightforward. If $L/2$ is the length of the segment, the
eigenfunctions $e_{n}(x)$ can be read from (\ref{Dbox}). Centering around
the origin by means of the shift $x=y+(L/4)$, rescaling $L$ by a factor $%
\eta $, and defining $p=4\pi k/(\eta L)$, $p^{\prime }=2\pi (2k+1)/(\eta L)$%
, the functions $e_{2k}(x)$ and $e_{2k+1}(x)$, $k\in \mathbb{N}_{+}$ give%
\begin{equation*}
\hat{e}_{p}(y)=\frac{2\sin (py)}{\sqrt{\eta L}},\qquad \hat{e}_{p^{\prime
}}(y)=\frac{2\cos (p^{\prime }y)}{\sqrt{\eta L}}.
\end{equation*}%
Then (\ref{ep}) gives%
\begin{equation*}
\lim_{\eta \rightarrow \infty }|\Omega |^{1/2}\eta ^{(D-1)/2}\left\{ 
\begin{tabular}{l}
$\hat{e}_{p}(y)$ \\ 
$\hat{e}_{p^{\prime }}(y)$%
\end{tabular}%
\right. =\left\{ 
\begin{tabular}{l}
$\sqrt{2}\sin (py)$ \\ 
$\sqrt{2}\cos (p^{\prime }y)$%
\end{tabular}%
\right. ,
\end{equation*}%
which is just an unusual basis for the Fourier transform in $\mathbb{R}$.

Finally, we consider the sphere in two dimensions. The kinetic Lagrangian of
a massive scalar field $\chi $ can be written in the form%
\begin{equation}
\int_{-\infty }^{+\infty }\mathrm{d}u\int_{-\pi R}^{+\pi R}\mathrm{d}v\left[ 
\frac{\partial ^{2}\chi }{\partial u^{2}}+\frac{\partial ^{2}\chi }{\partial
v^{2}}-m^{2}\chi ^{2}\left( 1-\tanh ^{2}\left( \frac{u}{R}\right) \right) %
\right] ,  \label{s2}
\end{equation}%
where%
\begin{equation*}
u=R\hspace{0.01in}\text{arctanh}\left( \cos \theta \right) ,\qquad v=R(\phi
-\pi ).
\end{equation*}%
and $R$, $\theta $ and $\phi $ are the usual spherical coordinates. Due to
the function that multiplies $m^{2}$, the eigenfunctions of the kinetic
operator blow up exponentially at infinity, unless the eigenvalues are
restricted to the correct, discrete set. When $R$ is rescaled by $\eta $,
and $\eta $ tends to infinity, the eigenvalues tend to a continuum, and (\ref%
{s2}) tends to the Lagrangian in $\mathbb{R}^{2}$.

Similar arguments hold for the sphere in three dimensions, the ball, the
cylinder and the disc.

\bigskip

As far as the external sources $K$ attached to the vertices are concerned,
we can distinguish the sources $K_{\tau }$ that restrict the time integrals
to the interval $\tau $, and just tend to one in the limits $t_{\text{i}%
}\rightarrow -\infty $, $t_{\text{f}}\rightarrow \infty $, from the sources $%
K_{0}$ due to the solutions $\phi _{0}$ and $z_{0}$, $\bar{z}_{0}$ of
formulas (\ref{expaf}) and (\ref{zon}), which may know about the boundary
function $f$ of (\ref{bou}). The sources $K_{0}$ must tend to whatever we
need to describe transition amplitudes between arbitrary states at $\tau
=\infty $, $\Omega =\mathbb{R}^{3}$.

Normally, we are interested in vacuum-to-vacuum amplitudes at $\tau =\infty $%
, $\Omega =\mathbb{R}^{3}$. Formula (\ref{zon}) shows that $z_{0\mathbf{n}%
}(t)$ and $\bar{z}_{0\mathbf{n}}(t)$ tend to zero, if we assume that $z_{%
\mathbf{n}\text{i}}$ and $\bar{z}_{\mathbf{n}\text{f}}$ are kept constant,
and the prescription $-i\epsilon $ is attached to $\omega _{\mathbf{n}}$.
If, in addition, we make $f$ tend to zero when $\eta \rightarrow \infty $,
we obtain the desired vacuum-to-vacuum amplitudes.

Choosing different behaviors for $z_{\mathbf{n}\text{i}}$ and $\bar{z}_{%
\mathbf{n}\text{f}}$, and keeping a nonvanishing $f$, we can describe
amplitudes between nontrivial states with arbitrary behaviors at infinity.
The convergence of those limits must be studied case by case.

\subsection{Renormalization}

We distinguish the interior parts of the diagrams from the exterior parts.
We know that the restriction to finite $\tau $ does not enter the diagrams,
but only affects the exterior parts, which we discuss later. The restriction
to finite volume affects the interior parts of the diagrams by means of the
discretization of the loop momenta, and the sums on $\mathbf{n}$, which
replace the usual integrals.

The ultraviolet behaviors of the diagrams coincide with those of the usual
diagrams, and the ultraviolet divergences are renormalized by the same
Lagrangian counterterms. The basic reason is as follows. Ultraviolet
divergences may appear when the sums on $\mathbf{n}$ do not converge.
Whenever we vary $\mathbf{n}$ by an amount $\Delta \mathbf{n}$, which is of
order unity, and take $|\mathbf{n|}$ large, the ratio $\Delta \mathbf{n}/%
\mathbf{n}$ becomes infinitesimal, so the sums become integrals. This means
that the large $\mathbf{n}$ behaviors can be studied by means of the
formulas of the previous subsection. All the details about the restriction
to finite volume disappear from the interior parts of the diagrams, and
their divergent parts are the same as usual.

Let us check this statement in a simple example, the bubble diagram on a
torus, regularized as $T^{3}\times \mathbb{R}^{-\varepsilon }$. The diagram
gives an expression proportional to%
\begin{equation*}
\int_{-\infty }^{+\infty }\frac{\mathrm{d}e}{2\pi }\sum_{\mathbf{n}}\int 
\frac{\mathrm{d}^{-\varepsilon }\mathbf{p}_{\varepsilon }}{(2\pi
)^{\varepsilon }}\frac{1}{e^{2}-\mathbf{\bar{n}}^{2}-\mathbf{p}_{\varepsilon
}^{2}-m^{2}+i\epsilon }\frac{1}{(e-e_{\text{ext}})^{2}-(\mathbf{\bar{n}}-%
\mathbf{\bar{n}}_{\text{ext}})^{2}-\mathbf{p}_{\varepsilon
}^{2}-m^{2}+i\epsilon },
\end{equation*}%
where $e_{\text{ext}}$ and $\mathbf{\bar{n}}_{\text{ext}}$ are the energy
and momentum that flow inside the diagram. For the purposes of
renormalization, we introduce a Feynman parameter and integrate on $\mathbf{p%
}_{\varepsilon }$ by means of formula (\ref{dimreg}). Integrating on the
energy as well, we find 
\begin{equation*}
B\equiv \frac{i\Gamma \left( \frac{3+\varepsilon }{2}\right) }{2\sqrt{\pi }%
(4\pi )^{-\varepsilon /2}}\int_{0}^{1}\mathrm{d}x\sum_{\mathbf{n}}\frac{1}{%
(m^{2}+\mathbf{\bar{n}}_{x}^{2}-x(1-x)e_{\text{ext}}^{2}-i\epsilon
)^{(3+\varepsilon )/2}},
\end{equation*}%
where%
\begin{equation*}
\mathbf{\bar{n}}_{x}^{2}=(\mathbf{\bar{n}}-x\mathbf{\bar{n}}_{\text{ext}}%
\mathbf{)}^{2}+x(1-x)\mathbf{\bar{n}}_{\text{ext}}^{2}.
\end{equation*}%
We first work below the threshold ($\left\vert e_{\text{ext}}^{2}-\mathbf{%
\bar{n}}_{\text{ext}}^{2}\right\vert <4m^{2}$). The divergent part can be
isolated from the rest by means of a Schwinger parameter. We approximate $%
\mathbf{\bar{n}}_{x}^{2}-x(1-x)e_{\text{ext}}^{2}$ to $\mathbf{\bar{n}}^{2}$%
, since we are interested in $\mathbf{\bar{n}}$ large, and keep the mass $m$
nonzero, to avoid spurious infrared divergences.

Summing on $\mathbf{n}$ with the help of the theta function $\theta
_{3}(q)=\sum_{n=-\infty }^{+\infty }q^{n^{2}}$, and using $\theta _{3}(%
\mathrm{e}^{-x})\sim \sqrt{\pi /x}$ for $x\rightarrow 0^{+}$, we obtain 
\begin{eqnarray*}
B_{\text{div}} &=&\frac{i(4\pi )^{\varepsilon /2}}{2\sqrt{\pi }}\left.
\!\int_{0}^{\infty }\!\!\!\beta ^{(1+\varepsilon )/2}\mathrm{e}^{-\beta
m^{2}}\mathrm{d}\beta \sum_{\mathbf{n}}\mathrm{e}^{-\beta \mathbf{\bar{n}}%
^{2}}\right\vert _{\text{div}} \\
&=&\frac{i(4\pi )^{\varepsilon /2}}{2\sqrt{\pi }}\left. \int_{0}^{\infty
}\beta ^{(1+\varepsilon )/2}\mathrm{e}^{-\beta m^{2}}\mathrm{d}\beta
\prod_{i=1}^{3}\theta _{3}\left( \mathrm{e}^{-\beta (2\pi
)^{2}/L_{i}^{2}}\right) \right\vert _{\text{div}}=\frac{i|\Omega |}{8\pi
^{2}\varepsilon },
\end{eqnarray*}%
having used (\ref{inte}) to convert the sum into an integral. The divergent
part we have obtained coincides with the usual one. Above the threshold the
finite part changes, but the divergent part remains the same.

If we introduce a cutoff $N$ for the sum, the divergence is clearly
logarithmic in $N$. We find%
\begin{equation*}
B_{\text{div}}=\frac{i|\Omega |}{8\pi ^{2}}\ln N.
\end{equation*}%
The identifications%
\begin{equation*}
N=\frac{\Lambda }{\mu }\frac{|\Omega |^{1/3}}{2\pi },\qquad \frac{1}{%
\varepsilon }=\ln \frac{\Lambda }{\mu },
\end{equation*}%
where $\mu $ is the dynamical scale, show that the counterterm matches the
usual one, apart from a change of scheme, which can be adjusted without
changing the physical quantities.

Sticking to the example of the torus, whenever a momentum $\mathbf{p}$
appears in the usual integral, $\mathbf{\bar{n}}$ appears in the sum. While
the integrals are replaced by sums in the limit $\Omega \rightarrow \mathbb{R%
}^{D-1}$, the integrands are the same as usual with $\mathbf{p}\rightarrow 
\mathbf{\bar{n}}$. Thus, the divergences are the same as usual, with the
same replacement. In particular, they are local and insensitive to total
derivatives (because so they are at $\tau =\infty $, $\Omega =\mathbb{R}%
^{D-1}$).

In this respect, note that at finite $\tau $, on a compact $\Omega $, we are
not allowed to alter the total derivatives of the Lagrangian (unless their
contributions to the action are topological, in which case their variations
vanish), because they are determined by the requirement of having the
correct classical variational problem.

We know that every vertex has an external source $K$ attached to it. This
means that we can view it as a local composite field $\mathcal{O}_{t,\mathbf{%
x}}(w,\bar{w})$, that is to say, a product of fields $w(t,\mathbf{x})$, $%
\bar{w}(t,\mathbf{x})$ at the same spacetime point. The correlation
functions we are considering are thus%
\begin{equation}
\langle w(t_{1},\mathbf{x}_{1})\cdots w(t_{n},\mathbf{x}_{n})\bar{w}%
(t_{1}^{\prime },\mathbf{x}_{1}^{\prime })\cdots \bar{w}(t_{r}^{\prime },%
\mathbf{x}_{r}^{\prime })\mathcal{O}_{t_{1}^{\prime \prime },\mathbf{x}%
_{1}^{\prime \prime }}^{(1)}(w,\bar{w})\cdots \mathcal{O}_{t_{s}^{\prime
\prime },\mathbf{x}_{s}^{\prime \prime }}^{(s)}(w,\bar{w})\rangle .
\label{correco}
\end{equation}%
What it important is that, once we switch to the energy-momentum framework,
the diagrams contributing to these correlation functions are the same as
usual, internally, apart from the discretization of the momenta. Moreover,
their divergent parts are same as usual, because the discretization does not
affect the ultraviolet behavior. So, once a correlation function is equipped
with the right counterterms at $\tau =\infty $, $\Omega =\mathbb{R}^{D-1}$,
it is also well defined at $\tau <\infty $, $\Omega $ = compact manifold.

Externally, the correlation functions (\ref{correco}) are equipped with
sources $K_{\tau }$ and sources $K_{0}$. The former restrict the time
integrals to $\tau $, which presents no difficulty. The latter are due to
the solutions $\phi _{0}$ and $z_{0}$, $\bar{z}_{0}$, introduced by the
shifts (\ref{expaf}) and (\ref{zon}). The particular solutions $z_{0}$, $%
\bar{z}_{0}$ are regular functions of time, to be integrated in the finite
interval $\tau $. Their space dependencies are also regular, since they
describe the initial and final states of the transition amplitude we are
calculating. As far as $\phi _{0}$ is concerned, it must be assumed to the
regular as well, because it encodes the boundary conditions on $\Omega $. It
is not necessary to assume that it admits a Fourier expansion in the same
domain as $w$, $\bar{w}$ do. These remarks prove that the diagrams and the
correlation functions (\ref{correco}) lead to well-defined radiative
corrections.

Since the part where the vertex turns into a composite field may be
confusing, we describe some aspects of the statements made so far in more
detail. The shifts (\ref{expaf}) and (\ref{zon}) generate replicas of the
diagrams, which are automatically renormalized by the same counterterms. For
example, a shift $\phi =\phi _{0}+\varphi $ of a vertex $\phi ^{4}$ gives 
\begin{equation}
\phi ^{4}=\phi _{0}^{4}+4\phi _{0}\varphi ^{3}+6\phi _{0}^{2}\varphi
^{2}+4\phi _{0}^{3}\varphi +\varphi ^{4}.  \label{phi4}
\end{equation}%
There is no substantial difference between using $\varphi ^{4}$ in a
diagram, where two $\varphi $ legs are internal and the other two are
external, and using $\phi _{0}^{2}\varphi ^{2}$ with two $\varphi $ internal
legs and the external factor $\phi _{0}^{2}$. Note that the further factor 6
rearranges the combinatorics as needed. Internally, the diagrams are the
same, so they need the same wave-function renormalization constants
externally.

At the practical level, we start from the usual renormalized Lagrangian and
perform all the operations we have described so far on it, that is to say,
on the renormalized fields. Then the renormalization constants (and,
possibly, the field redefinitions: see below) are distributed correctly.

For example, the renormalized Lagrangian of the $\phi ^{4}$ theory at $\tau
=\infty $, $\Omega =\mathbb{R}^{3}$ is%
\begin{equation*}
L_{\lambda }(\phi )=\frac{Z_{\phi }}{2}\left[ (\partial _{\mu }\phi
)(\partial ^{\mu }\phi )-Z_{m}m^{2}\phi ^{2}\right] -\frac{\lambda
Z_{\lambda }Z_{\phi }^{2}}{4!}\phi ^{4}.
\end{equation*}%
The shift (\ref{expaf}) generates a renormalized Lagrangian where $%
Z_{\lambda }$ and $Z_{m}$ remain the renormalization constants of the
coupling $\lambda $ and the mass $m$, respectively, and $Z_{\phi }^{1/2}$
becomes the wave-function renormalization constant of both $\varphi $ and $%
\phi _{0}$. The correlation functions are then externally equipped with the
right renormalization constants.

Consider, for definiteness, the term $L_{\lambda }(\phi _{0},\dot{\phi}%
_{0},\nabla \phi _{0})$ of (\ref{complac}). Although it does not contain $%
\varphi $ external legs, it contains renormalization constants: they are
those that provide the right counterterms for the diagrams with no $\varphi $
external legs, built with vertices such as those of (\ref{phi4}). In turn,
those diagrams are replicas of the diagrams that do contain $\varphi $
external legs.

Similarly, the contribution%
\begin{equation*}
-2i\sum_{\mathbf{n}\in \mathcal{U}}\bar{z}_{\mathbf{n}^{\ast }\text{f}%
}\omega _{\mathbf{n}}\mathrm{e}^{-i\omega _{\mathbf{n}}\tau }z_{\mathbf{n}%
\text{i}}
\end{equation*}%
to (\ref{scoh}) ends up being equipped with the right renormalization
constants, which subtract divergent parts of the same form.

We see that, not surprisingly, the initial, final and boundary conditions
must be applied to the renormalized fields, rather than the bare ones. For
example, in formula (\ref{amplitu}), the initial and final conditions $z(t_{%
\text{i}})=z_{\text{i}}$, $\,\bar{z}(t_{\text{f}})=\bar{z}_{\text{f}}$
concern the renormalized coherent states $z(t)$ and $\bar{z}(t)$, not the
bare ones.

In conclusion, to ensure that the theory at finite $\tau $ and compact $%
\Omega $ is equipped with the right counterterms, we start from the
classical action, multiply the couplings and the other parameters by the
usual renormalization constants, and equip the fields and their shifts with
the usual wave-function renormalization constants (or field redefinitions).
The counterterms are uniquely specified, including the total derivatives, up
to topological terms. In the same way as the classical action is uniquely
specified by the classical variational problem (up to topological terms), so
is the renormalized action.

Often, nontrivial field redefinitions may be required, instead of
multiplicative wave-function renormalization constants, to absorb the
divergences proportional to the field equations. Actually, in the
coherent-state approach counterterms proportional to the field equations
appear more than often, because they are necessary to reduce the number of
time derivatives to one in the kinetic terms, and remove them completely
from the vertices, to match the structure (\ref{complac}) of the starting
action. In the presence of such types of counterterms, renormalization still
works as explained above.

\subsection{Power counting and locality of counterterms}

As far as power counting and the locality of counterterms are concerned, we
make some further remarks.

Power counting is not so transparent in the coherent-state variables $w$ and 
$\bar{w}$. Nevertheless, we can restore the usual power counting by
switching to the variables $P$ and $Q$ of (\ref{PQ}). Note that the endpoint
corrections $\mathcal{S}_{\text{e}}^{\prime }$ of (\ref{she}) are linear in $%
P$ and $Q$, so they can be ignored in this discussion.

In the case of quantum gravity with purely virtual particles, we must use
the two-derivative formulation of \cite{Absograv}, and include suitable
total derivatives, to make sure that no more than one derivative acts on
each field. The theory is renormalizable at $\tau =\infty $, $\Omega =%
\mathbb{R}^{3}$, but not manifestly: unwanted divergences may be generated
in the intermediate calculations. When we gather them together, we discover
that they \textquotedblleft miraculously\textquotedblright\ cancel out in
the physical quantities. This means they do not need any renormalization
(or, that they can be renormalized without introducing new physical
parameters). The cancelations survive the restrictions to $\tau <\infty $, $%
\Omega $ = compact manifold, because the divergent parts (and the field
equations, which are used to subtract certain divergences by means of field
redefinitions) do not depend on such restrictions.

The locality of counterterms can be proved by mimicking the standard
arguments, even without relating the diagrams to the usual ones. It is
sufficient to pretend that the external momenta are continuous variables,
and differentiate with respect to them a sufficient number times, and so
kill the overall divergences (in the variables $P$ and $Q$). We can take
care of the subdivergences by proceeding iteratively.

In a bounded box with Dirichlet boundary conditions, as well as in other
manifolds with boundary, the boundary reflections generate many copies of
similar diagrams. Consequently, there are many copies of similar
counterterms. Yet, the copies do not have to be added anew, since they are
just generated by the restrictions of the usual counterterms to a compact $%
\Omega $, due to the same boundary reflections.

\section{Unitarity}

\label{unitari}\setcounter{equation}{0}

Unitarity is the statement that the evolution operator $U(t_{\text{f}},t_{%
\text{i}})$ is unitary, i.e., 
\begin{equation}
U^{\dag }(t_{\text{f}},t_{\text{i}})U(t_{\text{f}},t_{\text{i}})=1,
\label{unitate}
\end{equation}%
for every $t_{\text{i}}$ and $t_{\text{f}}$. Equation (\ref{unitar}) is more
general, since it says that $U(t_{3},t_{2})U(t_{2},t_{1})$ is equal to $%
U(t_{3},t_{1})$ for arbitrary $t_{1}$, $t_{2}$ and $t_{3}$. Formula (\ref%
{unitate}) can be seen as a particular case of (\ref{unitar}) for $%
t_{3}=t_{1}=t_{\text{i}}$, $t_{2}=t_{\text{f}}$.

Equation (\ref{unitar}) holds under relatively mild assumptions. In the
functional integral approach, it just amounts to dividing the integral into
two portions, and integrating on all the configurations in between. A theory
with physical particles only (no ghosts) does satisfy (\ref{unitar}). Even a
theory with ghosts (particles with kinetic terms multiplied by the wrong
signs) satisfies it, but then (\ref{unitate}) is not interpreted as the
unitarity equation, due to the presence of negative-norm states, or a free
Hamiltonian not bounded from below\footnote{%
In that case, (\ref{unitate}) is called pseudounitarity equation.}. For the
time being, we assume that no ghosts are present. Later on (see section \ref%
{PV}) we explain how they can be included.

In this section we derive the diagrammatic version of the more general
equation (\ref{unitar}), and decompose it into thresholds and spectral
optical identities, by generalizing the results of \cite{diagrammarMio}. To
make the notation less heavy, we understand the subscripts $\mathbf{n}$, $%
\mathbf{n}^{\ast }$ everywhere, as well as the sums and products on $\mathbf{%
n}\in \mathcal{U}$. We denote the intermediate initial and final conditions
(i.e., those referring to the intermediate time $t_{2}$ of equation (\ref%
{unitar})), by means of variables $v$ and $\bar{v}$. Then, $\mathrm{d}\bar{v}%
\mathrm{d}v$ stands for $\prod_{\mathbf{n}\in \mathcal{U}}\mathrm{d}\bar{v}_{%
\mathbf{n}}\mathrm{d}v_{\mathbf{n}}$, $\omega \bar{v}v$ stands for $\sum_{%
\mathbf{n}\in \mathcal{U}}\bar{v}_{\mathbf{n}^{\ast }}\omega _{\mathbf{n}}v_{%
\mathbf{n}}$, etc. Similar notations are understood for the variables of the
functional integrals. The times $t_{1}$ and $t_{3}$ of (\ref{unitar}) are $%
t_{\text{i}}$ and $t_{\text{f}}$, respectively, while $t_{2}$ will be simply
denoted by $t$.

Although unitarity is obvious in the operatorial approach (if the
Hamiltonian is Hermitian, as we are assuming here), we take our time to
prove it directly in the functional-integral approach, assuming that the
Lagrangian is Hermitian, because the proof leads us straightforwardly to the
diagrammatic version of the unitarity equation itself.

Relabeling $t_{1}$, $t_{2}$ and $t_{3}$ as $t_{\text{a}}$, $t$ and $t_{\text{%
b}}$, respectively, we show that (\ref{unitar}) is equivalent to the identity%
\begin{equation}
\langle \bar{z}_{\text{b}},t_{\text{b}};z_{\text{a}},t_{\text{a}}\rangle
_{\zeta ,\bar{\zeta}}=\int \langle \bar{z}_{\text{b}},t_{\text{b}%
};v,t\rangle _{\zeta ,\bar{\zeta}}\mathrm{d}\mu _{\bar{v},v}\langle \bar{v}%
,t;z_{\text{a}},t_{\text{a}}\rangle _{\zeta ,\bar{\zeta}},\qquad \mathrm{d}%
\mu _{\bar{v},v}\equiv \frac{\omega \mathrm{d}\bar{v}\mathrm{d}v}{i\pi }%
\mathrm{e}^{-2\omega \bar{v}v},  \label{identa}
\end{equation}%
while (\ref{unitate})\ is equivalent to 
\begin{equation}
\int \langle \bar{z}_{\text{f}},t_{\text{f}};z_{\text{i}},t_{\text{i}%
}\rangle _{\zeta ,\bar{\zeta}}^{\ast }\mathrm{d}\mu _{\bar{z}_{\text{f}},z_{%
\text{f}}}\langle \bar{z}_{\text{f}},t_{\text{f}};z_{\text{i}},t_{\text{i}%
}\rangle _{\zeta ,\bar{\zeta}}=\mathrm{e}^{2\omega \bar{z}_{\text{i}}z_{%
\text{i}}}.  \label{unitatez}
\end{equation}

Formula (\ref{identa}) states that if we break the amplitude in two, and
integrate on all the intermediate possibilities as shown, we get the correct
result. What is nontrivial is the integration measure in between. The
initial condition $z(t^{+})=v$\ to the left and the final condition $\bar{z}%
(t^{-})=\bar{v}$ to the right show that we need to keep $z(t^{+})$ and $\bar{%
z}(t^{-})$ fixed. However, the extra integrals on $\bar{v}$\textrm{\ }and $v$
in between restore the missing integrals on $\bar{z}(t^{-})$\textrm{\ }and $%
z(t^{+})$. In the end, the integrals on the right-hand side of (\ref{identa}%
) exactly match the integrals on the left-hand side, and the trajectories
contributing to the functional integral on the right-hand side coincide with
those contributing to the normal integral of two functional integrals that
appears on the left-hand side.

Formula (\ref{unitatez}) is the functional-integral version of the
operatorial unitarity equation (\ref{unitate}), since $\mathrm{e}^{2\omega 
\bar{z}_{\text{i}}z_{\text{i}}}$ is the matrix element of the identity
matrix in the coherent-state approach. Because $\langle \bar{z}_{\text{b}%
},t_{\text{b}};v,t\rangle _{\zeta ,\bar{\zeta}}=\langle \bar{v},t;z_{\text{b}%
},t_{\text{b}}\rangle _{\zeta ,\bar{\zeta}}^{\ast }$ for $t>t_{\text{b}}$,
by (\ref{conj}), (\ref{unitatez}) can be seen as a particular case of (\ref%
{identa}) with $v=z_{\text{f}}$, $\bar{v}=\bar{z}_{\text{f}}$, $t=t_{\text{f}%
}$ and $z_{\text{b}}=z_{\text{a}}=z_{\text{i}}$, $\bar{z}_{\text{b}}=\bar{z}%
_{\text{a}}=\bar{z}_{\text{i}}$, $t_{\text{b}}=t_{\text{a}}=t_{\text{i}}$.
Thus, we can focus on the proof of (\ref{identa}).

\begin{figure}[t]
\begin{center}
\includegraphics[width=14truecm]{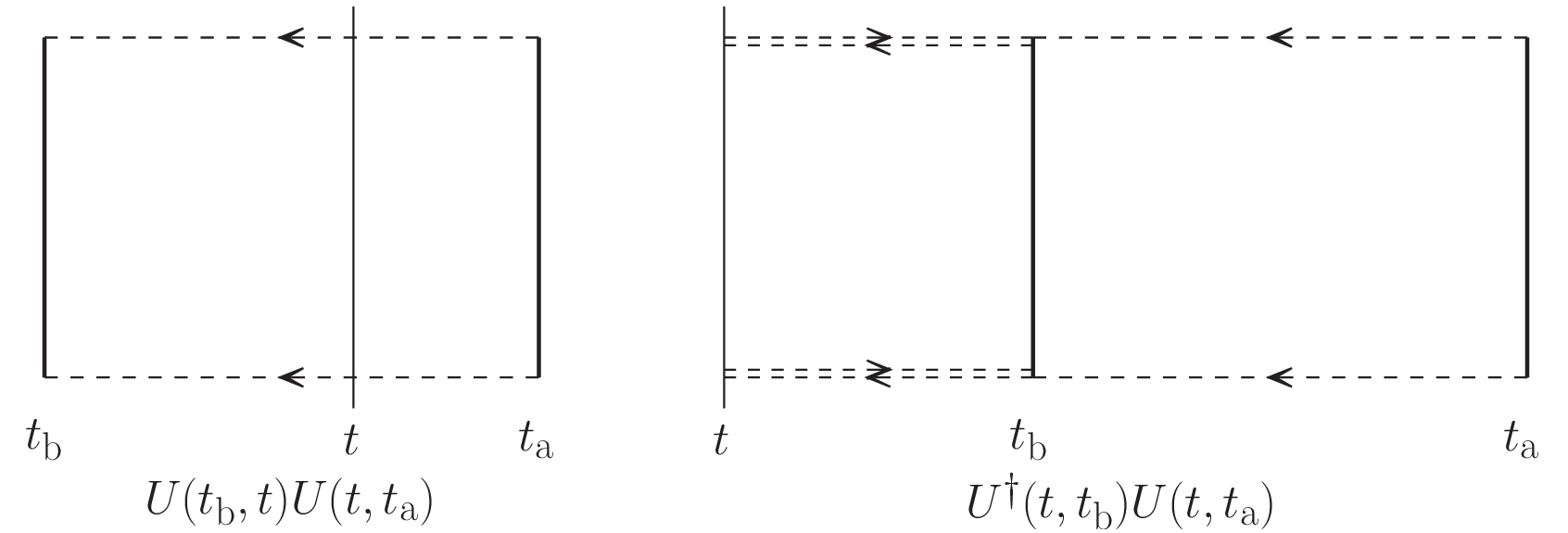}
\end{center}
\par
\vskip-.7truecm
\caption{The evolution operator $U(t_{\text{b}},t_{\text{a}})$ is equal to
the products shown here. The manifold $\Omega $ is depicted as a vertical
segment}
\label{unitario}
\end{figure}

We can distinguish three cases: $t_{\text{a}}<t<t_{\text{b}}$, $t>t_{\text{b}%
}$ and $t<t_{\text{a}}$. The third one is a mirror of the second one, so we
concentrate on the first two, illustrated in fig. \ref{unitario}.

\subsection{Proof of unitarity -- case I}

We start from the situation illustrated to the left in fig. \ref{unitario},
which is $t_{\text{a}}<t<t_{\text{b}}$. We first prove (\ref{identa}) in the
free limit $\mathcal{L}_{I}=0$ and later show that it can be extended to the
interacting case.

The identity%
\begin{equation}
\langle \bar{z}_{\text{b}},t_{\text{b}};z_{\text{a}},t_{\text{a}}\rangle
_{\zeta ,\bar{\zeta}}^{\text{free}}=\int \langle \bar{z}_{\text{b}},t_{\text{%
b}};v,t\rangle _{\zeta ,\bar{\zeta}}^{\text{free}}\mathrm{d}\mu _{\bar{v}%
,v}\langle \bar{v},t;z_{\text{a}},t_{\text{a}}\rangle _{\zeta ,\bar{\zeta}}^{%
\text{free}}  \label{identafree}
\end{equation}%
for $t_{\text{a}}<t<t_{\text{b}}$ easily follows from the formula%
\begin{equation}
\int \frac{\omega \mathrm{d}\bar{v}\mathrm{d}v}{i\pi }\mathrm{e}^{-2\omega (%
\bar{v}v-\bar{a}v-\bar{v}a)}=\mathrm{e}^{2\omega \bar{a}a},  \label{gau}
\end{equation}%
upon using the explicit expression (\ref{freeampl}). First, it is obvious
that 
\begin{equation}
\tilde{W}_{0}(t_{\text{b}},t_{\text{a}})=\tilde{W}_{0}(t_{\text{b}},t)+%
\tilde{W}_{0}(t,t_{\text{a}}),  \label{wot}
\end{equation}%
with self-evident notation. Second, at $\bar{\zeta}^{\prime }=\zeta ^{\prime
}=0$, we just have the identity%
\begin{equation}
\exp (2\bar{z}_{\text{b}}\omega \mathrm{e}^{-i\omega (t_{\text{b}}-t_{\text{a%
}})}z_{\text{a}})=\mathrm{e}^{2\omega \bar{z}_{0}(t)z_{0}(t)}=\int \mathrm{e}%
^{2\omega \bar{z}_{0}(t)v}\frac{\omega \mathrm{d}\bar{v}\mathrm{d}v}{i\pi }%
\mathrm{e}^{-2\omega \bar{v}v}\mathrm{e}^{2\omega \bar{v}z_{0}(t)},
\label{wott}
\end{equation}%
which is true by (\ref{gau}). Third, at nonvanishing sources $\bar{\zeta}%
^{\prime }$ and $\zeta ^{\prime }$, we have shifts of $a$ and $\bar{a}$ in (%
\ref{gau}), which complete the match. In particular, they provide the
correct two-point functions $\langle z(t_{1})\bar{z}(t_{2})\rangle $ for $t_{%
\text{a}}<t_{1}<t$, $t<t_{2}<t_{\text{b}}$ and $t<t_{1}<t_{\text{b}}$, $t_{%
\text{a}}<t_{2}<t$.

The interactions can be included by means of formula (\ref{amplint}), once
we observe that the right-hand side of (\ref{identa}) can be viewed as the
action of the operator%
\begin{equation}
\exp \left( i\int_{t}^{t_{\text{b}}}\!\!\mathrm{d}t^{\prime }\!\mathbf{%
\hspace{0.01in}}\mathcal{L}_{I}\left( \frac{\delta }{i\delta \bar{\zeta}%
(t^{\prime })},\frac{\delta }{i\delta \zeta (t^{\prime })}\right) \right)
\exp \left( i\int_{t_{\text{a}}}^{t}\!\!\mathrm{d}t\!\mathbf{\hspace{0.01in}}%
^{\prime }\mathcal{L}_{I}\left( \frac{\delta }{i\delta \bar{\zeta}(t^{\prime
})},\frac{\delta }{i\delta \zeta (t^{\prime })}\right) \right)  \label{proda}
\end{equation}%
on the right-hand side of (\ref{identafree}). The reason why we can move
these expressions outside the $\bar{v}$, $v$ integral is that they do not
depend on $\bar{v}$ and\textrm{\ }$v$, as shown by the definition $\mathcal{L%
}_{I}(z_{\mathbf{n}},\bar{z}_{\mathbf{n}})=L_{I}(\pi _{\mathbf{n}},\varphi _{%
\mathbf{n}},\phi _{0})$ given right below (\ref{complac}): the dependencies
on $\bar{v}$ and\textrm{\ }$v$, $\bar{z}_{0}$ and\textrm{\ }$z_{0}$ are
brought into $\mathcal{L}_{I}$ only after the shift (\ref{shift2}). Formula (%
\ref{proda}) is just the exponential of the integral between $t_{\text{a}}$
and $t_{\text{b}}$, which is the correct operator that gives the left-hand
side of (\ref{identa}) by acting on the left-hand side of (\ref{identafree}%
), as in (\ref{amplint}).

Incidentally, we remark that the correct measure $\mathrm{d}\mu _{\bar{v},v}$
can be derived by reversing the procedure just outlined. It is sufficient to
work in the free case (\ref{identafree}), starting from the most general
candidate for the measure.

\subsection{Useful identities for integrals on coherent states}

Before switching to the second part of the proof of unitarity, we derive
some useful identities for integrals with coherent states. First note the
formulas%
\begin{equation}
\int \mathrm{d}\mu _{\bar{v},v}v^{n}\bar{v}^{m}=\frac{n!\delta _{nm}}{%
(2\omega )^{n}},\qquad \int \frac{\omega \mathrm{d}\bar{\sigma}\mathrm{d}%
\sigma }{i\pi }\mathrm{e}^{2\omega (\bar{v}-\bar{\sigma})\sigma }g(\bar{%
\sigma})=g(\bar{v}),  \label{deltacoh}
\end{equation}%
where $g$ is an arbitrary function. The first identity is proved by
evaluating the integral in polar coordinates $v=\rho \mathrm{e}^{i\theta }$, 
$\bar{v}=\rho \mathrm{e}^{-i\theta }$, where $\mathrm{d}\mu _{\bar{\sigma}%
,\sigma }=2\omega \rho \mathrm{d}\rho \mathrm{d}\theta /\pi $. The second
identity is the delta function representation for coherent states, and
follows from the first one by expanding $\mathrm{e}^{2\omega \bar{v}\sigma
}g(\bar{\sigma})$ in powers of $\sigma $ and $\bar{\sigma}$.

Moreover, we have 
\begin{equation}
\int \frac{\omega \mathrm{d}\bar{v}\mathrm{d}v}{i\pi }\mathrm{e}^{-2\omega (%
\bar{v}v-\bar{a}v-\bar{v}a)+\varepsilon f(\bar{v},a)-\varepsilon f(\bar{a}%
,v)}=\mathrm{e}^{2\omega \bar{a}a+\mathcal{O}(\varepsilon ^{2})},
\label{gaua}
\end{equation}%
for every function $f$, where $\varepsilon $ is a small parameter. To prove
it, it is sufficient to expand the integrand in powers of $\varepsilon $ and
check that the first order vanishes by the second formula of (\ref{deltacoh}%
).

\subsection{Proof of unitarity -- case II}

Now we consider the situation illustrated to the right in fig. \ref{unitario}%
. For $t>t_{\text{b}}$, we have $\langle \bar{z}_{\text{b}},t_{\text{b}%
};v,t\rangle _{\zeta ,\bar{\zeta}}=\langle \bar{v},t;z_{\text{b}},t_{\text{b}%
}\rangle _{\zeta ,\bar{\zeta}}^{\ast }$, from (\ref{conj}), so the equation
we need to prove reads 
\begin{equation}
\langle \bar{z}_{\text{b}},t_{\text{b}};z_{\text{a}},t_{\text{a}}\rangle
_{\zeta ,\bar{\zeta}}=\int \langle \bar{v},t;z_{\text{b}},t_{\text{b}%
}\rangle _{\zeta ,\bar{\zeta}}^{\ast }\mathrm{d}\mu _{\bar{v},v}\langle \bar{%
v},t;z_{\text{a}},t_{\text{a}}\rangle _{\zeta ,\bar{\zeta}}.  \label{optid}
\end{equation}

Using (\ref{identa}), we may write the right-hand side as%
\begin{equation*}
\int \langle \bar{v},t;z_{\text{b}},t_{\text{b}}\rangle _{\zeta ,\bar{\zeta}%
}^{\ast }\mathrm{d}\mu _{\bar{v},v}\langle \bar{v},t;\sigma ,t_{\text{b}%
}\rangle _{\zeta ,\bar{\zeta}}\mathrm{d}\mu _{\bar{\sigma},\sigma }\langle 
\bar{\sigma},t_{\text{b}};z_{\text{a}},t_{\text{a}}\rangle _{\zeta ,\bar{%
\zeta}}.
\end{equation*}%
It is actually sufficient to prove the formula%
\begin{equation}
\int \langle \bar{v},t;z_{\text{b}},t_{\text{b}}\rangle _{\zeta ,\bar{\zeta}%
}^{\ast }\mathrm{d}\mu _{\bar{v},v}\langle \bar{v},t;\sigma ,t_{\text{b}%
}\rangle _{\zeta ,\bar{\zeta}}=\mathrm{e}^{2\omega \bar{z}_{\text{b}}\sigma
},  \label{suff}
\end{equation}%
because it turns the right-hand side of (\ref{optid}) into%
\begin{equation*}
\int \frac{\omega \mathrm{d}\bar{\sigma}\mathrm{d}\sigma }{i\pi }\mathrm{e}%
^{2\omega (\bar{z}_{\text{b}}-\bar{\sigma})\sigma }\langle \bar{\sigma},t_{%
\text{b}};z_{\text{a}},t_{\text{a}}\rangle _{\zeta ,\bar{\zeta}},
\end{equation*}%
which is equal to the left-hand side of (\ref{optid}) by the second identity
of (\ref{deltacoh}). Note that formula (\ref{suff}) is the unitarity
equation (\ref{unitatez}) after a suitable relabeling.

Having reduced the task to proving (\ref{suff}), we divide the interval $%
(t,t_{\text{b}})$ into $n+1$ intervals $(t_{k},t_{k+1})$, $k=0,1,\ldots n$,
where $t_{k}=t_{\text{b}}+k\varepsilon $, $\varepsilon =(t-t_{\text{b}%
})/(n+1)$, and apply (\ref{identa}) for every $k$. So doing, we obtain%
\begin{equation*}
\langle \bar{v},t;\sigma ,t_{\text{b}}\rangle _{\zeta ,\bar{\zeta}}=\int
\left( \prod_{k=1}^{n}\langle \bar{v}_{k+1},t_{k+1};v_{k},t_{k}\rangle
_{\zeta ,\bar{\zeta}}\mathrm{d}\mu _{\bar{v}_{k},v_{k}}\right) \langle \bar{v%
}_{1},t_{1};v_{0},t_{0}\rangle _{\zeta ,\bar{\zeta}},
\end{equation*}%
where $v_{0}=\sigma $, $\bar{v}_{n+1}=\bar{v}$. Doing the same for $\langle 
\bar{v},t;z_{\text{b}},t_{\text{b}}\rangle _{\zeta ,\bar{\zeta}}$ and
conjugating, formula (\ref{suff}) turns into 
\begin{equation}
\int E\left( \prod_{j=1}^{n}\mathrm{d}\mu _{\bar{\sigma}_{j},\sigma
_{j}}\langle \bar{\sigma}_{j+1},t_{j+1};\sigma _{j},t_{j}\rangle _{\zeta ,%
\bar{\zeta}}^{\ast }\right) \hspace{0.02in}\mathrm{d}\mu _{\bar{v},v}\left(
\prod_{k=1}^{n}\langle \bar{v}_{k+1},t_{k+1};v_{k},t_{k}\rangle _{\zeta ,%
\bar{\zeta}}\mathrm{d}\mu _{\bar{v}_{k},v_{k}}\right) F=\mathrm{e}^{2\omega 
\bar{z}_{\text{b}}\sigma },  \label{suff2}
\end{equation}%
where $\bar{\sigma}_{0}=\bar{z}_{\text{b}}$, $\sigma _{n+1}=v$, $E=\langle 
\bar{\sigma}_{1},t_{1};\sigma _{0},t_{0}\rangle _{\zeta ,\bar{\zeta}}^{\ast
} $, $F=\langle \bar{v}_{1},t_{1};v_{0},t_{0}\rangle _{\zeta ,\bar{\zeta}}$.
We can prove (\ref{suff2}) for the $n$ we want, since the left-hand side of (%
\ref{suff}) is equal to the left-hand side of (\ref{suff2}) for every $n$.
It is convenient to prove (\ref{suff2}) in the limit $n\rightarrow \infty $,
where $\varepsilon $ becomes infinitesimal. Then we can further reduce the
task to the one of proving 
\begin{equation}
\int \langle \bar{\sigma}_{j+1},t_{j+1};\sigma _{j},t_{j}\rangle _{\zeta ,%
\bar{\zeta}}^{\ast }\hspace{0.02in}\mathrm{d}\mu _{\bar{v}_{j+1},\sigma
_{j+1}}\langle \bar{v}_{j+1},t_{j+1};v_{j},t_{j}\rangle _{\zeta ,\bar{\zeta}%
}=\mathrm{e}^{2\omega \bar{\sigma}_{j}v_{j}+\mathcal{O}(\varepsilon ^{2})}.
\label{suff3}
\end{equation}%
Indeed, with the help of the relation 
\begin{equation*}
\int \mathrm{e}^{2\omega \bar{\sigma}_{n}v_{n}}\langle \bar{v}%
_{n},t_{n};v_{n-1},t_{n-1}\rangle _{\zeta ,\bar{\zeta}}\mathrm{d}\mu _{\bar{v%
}_{n},v_{n}}=\langle \bar{\sigma}_{n},t_{n};v_{n-1},t_{n-1}\rangle _{\zeta ,%
\bar{\zeta}},
\end{equation*}%
which follows from the second formula of (\ref{deltacoh}), the identity (\ref%
{suff3}) for $j=n$ allows us to turn (\ref{suff2}) into%
\begin{equation*}
\int \!\!E\!\left( \prod_{j=1}^{n-1}\mathrm{d}\mu _{\bar{\sigma}_{j},\sigma
_{j}}\langle \bar{\sigma}_{j+1},t_{j+1};\sigma _{j},t_{j}\rangle _{\zeta ,%
\bar{\zeta}}^{\ast }\!\right) \!\mathrm{d}\mu _{\bar{v},v}\!\!\left(
\prod_{k=1}^{n-1}\langle \bar{v}_{k+1},t_{k+1};v_{k},t_{k}\rangle _{\zeta ,%
\bar{\zeta}}\mathrm{d}\mu _{\bar{v}_{k},v_{k}}\!\right) \!F=\mathrm{e}%
^{2\omega \bar{z}_{\text{b}}\sigma +\mathcal{O}(\varepsilon ^{2})},
\end{equation*}%
with $\bar{v}_{n}=\bar{\sigma}_{n}=\bar{v}$, $\sigma _{n}=v$, which is the
same as (\ref{suff2}) with $n\rightarrow n-1$, $t\rightarrow t_{n}$, up to $%
\mathcal{O}(\varepsilon ^{2})$ in the exponent. Iterating in $n$, the last
step is (\ref{suff3}) with $j=0$. Taking $\varepsilon $ to zero, (\ref{suff2}%
) follows for $n\rightarrow \infty $, as desired.

It remains to prove (\ref{suff3}) for infinitesimal $\varepsilon $, which is
relatively easy. Using (\ref{amplint}) and (\ref{freeampl}), we have%
\begin{equation*}
\langle \bar{v}_{j+1},t_{j+1};v_{j},t_{j}\rangle _{\zeta ,\bar{\zeta}}=%
\mathrm{e}^{i\tilde{W}_{0j}+2\bar{v}_{j+1}\omega (1\!-i\omega \varepsilon
)v_{j}+i\varepsilon \mathbf{\hspace{0.01in}(}\bar{\zeta}_{j}^{\prime }v_{j}+%
\bar{v}_{j+1}\zeta _{j}^{\prime })+i\varepsilon \mathbf{\hspace{0.01in}}%
\mathcal{L}_{I}(v_{j},\bar{v}_{j+1})+\mathcal{O}(\varepsilon ^{2})},
\end{equation*}%
where $\tilde{W}_{0j}$, $\bar{\zeta}_{j}^{\prime }$ and $\zeta _{j}^{\prime
} $ are the restrictions of $\tilde{W}_{0}$, $\bar{\zeta}^{\prime }$ and $%
\zeta ^{\prime }$ to the $(j+1)$-th interval. We see that (\ref{suff3}) is
just a particular case of formula (\ref{gaua}), with $\bar{a}=\bar{\sigma}%
_{j}$, $a=v_{j}$, $\bar{v}=\bar{v}_{j+1}$, $v=\sigma _{j+1}$ (recalling that
the Lagrangian is assumed to the Hermitian). This concludes the proof.

Equation (\ref{optid}) turns into the unitarity equation $S^{\dag }S=1$
obeyed by the $S$ matrix when $t\rightarrow +\infty $, $t_{\text{b}%
}\rightarrow -\infty $, $t_{\text{a}}\rightarrow -\infty $, $\tau
\rightarrow 0$. Indeed, the left-hand side is equal to $\mathrm{e}^{2\omega 
\bar{z}_{\text{b}}z_{\text{a}}}$ for $\tau =0$, which is the matrix element
of the identity matrix in the coherent-state approach.

\section{Unitarity equations}

\label{unitareq}\setcounter{equation}{0}In this section we work out the
diagrammatic versions of the unitarity equation, which are also known as
Cutkosky-Veltman identities \cite{cutkoskyveltman}. To this purpose, it is
useful to define the cut correlation functions%
\begin{equation}
\langle \tilde{z}(t_{1})\cdots \tilde{z}(t_{k})\hspace{0.01in}|\hspace{0.01in%
}\tilde{z}(t_{k+1})\cdots \tilde{z}(t_{k+n})\rangle _{\zeta ,\bar{\zeta}%
}\equiv \!\!\int \!\!\frac{\delta ^{k}\langle \bar{v},t;z_{\text{b}},t_{%
\text{b}}\rangle _{\zeta ,\bar{\zeta}}^{\ast }}{i\delta \tilde{\zeta}%
(t_{1})\cdots i\delta \tilde{\zeta}(t_{k})}\mathrm{d}\mu _{\bar{v},v}\frac{%
\delta ^{n}\langle \bar{v},t;z_{\text{a}},t_{\text{a}}\rangle _{\zeta ,\bar{%
\zeta}}}{i\delta \tilde{\zeta}(t_{k+1})\cdots i\delta \tilde{\zeta}(t_{k+n})}%
\!\!\!,  \label{correcut}
\end{equation}%
where $\tilde{z}(t_{j})$ and $\tilde{\zeta}(t_{j})$ stand for $z(t_{j})$ and 
$\bar{\zeta}(t_{j})$, or $\bar{z}(t_{j})$ and $\zeta (t_{j})$, depending on
the case. We choose to write formula (\ref{correcut}) in the form that is
more convenient for $t>t_{\text{b}}$, since the simpler case $t<t_{\text{b}}$
can be easily reached by means of formula (\ref{conj}), i.e., by
understanding $\langle \bar{v},t;z_{\text{b}},t_{\text{b}}\rangle _{\zeta ,%
\bar{\zeta}}^{\ast }$ as $\langle \bar{z}_{\text{b}},t_{\text{b}};v,t\rangle
_{\zeta ,\bar{\zeta}}$. Strictly speaking, we should set $\zeta =\bar{\zeta}%
=0$ at the end, but it is not really necessary to do so for the validity of
the identities that we are going to study. It is understood that the
correlation functions we write vanish when an insertion $\tilde{z}(t_{j})$
lies outside the right time interval, which is identified by the functional
differentiation it originates from. In (\ref{correcut}), we have zero when $%
t_{j}\notin (t_{\text{b}},t)$ for $j\leqslant k$, or $t_{j}\notin (t_{\text{a%
}},t)$ for $j\geqslant k$.

\begin{figure}[t]
\begin{center}
\includegraphics[width=10truecm]{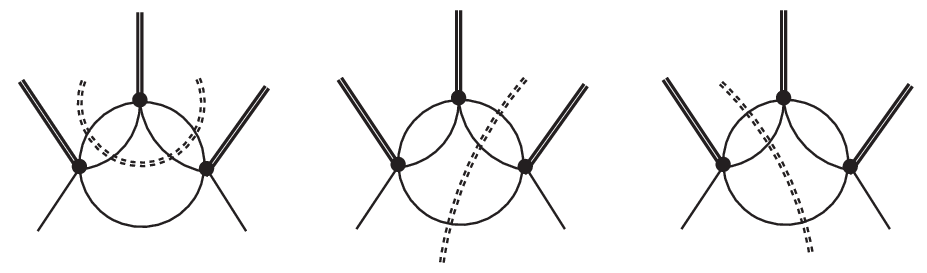}
\end{center}
\par
\vskip -.7truecm
\caption{Cut diagrams associated with the last diagram of fig. \protect\ref%
{BasicDiagra}. The cut is denoted by a dashed double line}
\label{CVDiagra}
\end{figure}

The correlation functions (\ref{correcut}) are made of two parts, identified
by two sets of insertions, which stand on the opposite sides of what we may
call a \textquotedblleft cut\textquotedblright , denoted by the vertical
bar. The diagrams contributing to (\ref{correcut}) are called
\textquotedblleft cut diagrams\textquotedblright . In fig. \ref{CVDiagra} we
list the cut diagrams associated with the last diagram of fig. \ref%
{BasicDiagra}.

Similarly, we can define cut correlation functions that contain insertions
of composite fields. Cut correlation functions with insertions of $w$ and $%
\bar{w}$ follow from the change of variables (\ref{split}), or (\ref{shift2}%
).

An important remark is that the correlation functions%
\begin{equation}
\langle \hspace{0.01in}|\hspace{0.01in}\tilde{z}(t_{1})\cdots \tilde{z}%
(t_{n})\rangle _{\zeta ,\bar{\zeta}},\qquad \langle \tilde{z}(t_{1})\cdots 
\tilde{z}(t_{n})\hspace{0.01in}|\hspace{0.01in}\rangle _{\zeta ,\bar{\zeta}%
},\qquad \langle \tilde{z}(t_{1})\cdots \tilde{z}(t_{n})\rangle _{\zeta ,%
\bar{\zeta}},  \label{tre}
\end{equation}%
do not coincide at finite $\tau $ on a compact $\Omega $. The first two are
cut correlation functions where all the external legs are located on the
same side with respect to the cut. The third one is the correlation function
defined in (\ref{corre}) (omitting $\bar{z}_{\text{b}},t_{\text{b}}$ and $%
\hspace{0.01in}z_{\text{a}},t_{\text{a}}$, for simplicity, and keeping
arbitrary sources $\zeta $, $\bar{\zeta}$), which has no cut. Recall that
every vertex is attached to an external source $K$, which takes care of the
restriction to finite $\tau $ and compact $\Omega $. We show below that
every propagator that crosses a cut (called \textquotedblleft cut
propagator\textquotedblright ) flows (positive) energy towards the same side
of the cut (as occurs at $\tau =\infty $, $\Omega =\mathbb{R}^{3}$). We
know, however, that there is no energy conservation at the vertices, because
a source $K$ can flow energy in or out. For example, a vertex can be cut out
of a diagram and still contribute: the left diagram of fig. \ref{CVDiagra}
is nontrivial, because energy may flow in and out through the upper source $%
K $. Diagrams like these make the difference between the third correlation
function of (\ref{tre}) and the other two. 

Now we show how to use the cut correlation functions to express unitarity
diagrammatically, by means of the cut diagrams. Differentiating (\ref{optid}%
), we obtain the identities%
\begin{equation}
\frac{\delta ^{n}\int \langle \bar{v},t;z_{\text{b}},t_{\text{b}}\rangle
_{\zeta ,\bar{\zeta}}^{\ast }\mathrm{d}\mu _{\bar{v},v}\langle \bar{v},t;z_{%
\text{a}},t_{\text{a}}\rangle _{\zeta ,\bar{\zeta}}}{i\delta \tilde{\zeta}%
(t_{1})\cdots i\delta \tilde{\zeta}(t_{n})}=\frac{\delta ^{n}\langle \bar{z}%
_{\text{b}},t_{\text{b}};z_{\text{a}},t_{\text{a}}\rangle _{\zeta ,\bar{\zeta%
}}}{i\delta \tilde{\zeta}(t_{1})\cdots i\delta \tilde{\zeta}(t_{n})}\equiv
\langle \tilde{z}(t_{1})\cdots \tilde{z}(t_{n})\rangle _{\zeta ,\bar{\zeta}%
},\qquad  \label{cut}
\end{equation}%
for every $n\geqslant 0$. Using the Leibniz rule on\ the left-hand side, we
obtain the Cutkosky-Veltman equation obeyed by the $n$-point function, which
reads%
\begin{equation}
\sum_{k=0}^{n}\sum_{\pi _{k}}(-1)^{k}\langle \tilde{z}(t_{\pi (1)})\cdots 
\tilde{z}(t_{\pi (k)})\hspace{0.01in}|\hspace{0.01in}\tilde{z}(t_{\pi
(k+1)})\cdots \tilde{z}(t_{\pi (k+n)})\rangle _{\zeta ,\bar{\zeta}}=\langle 
\tilde{z}(t_{1})\cdots \tilde{z}(t_{n})\rangle _{\zeta ,\bar{\zeta}},
\label{cuteq}
\end{equation}%
where $\pi _{k}$ denotes the set of $k$-combinations $(\pi (1),\ldots ,\pi
(k))$ of the set $(1,\ldots ,n)$.

The identities (\ref{cuteq}) are particularly interesting for $t>t_{\text{b}%
} $, which is the case of the unitarity equation (\ref{unitate}). When $%
t_{j}>t_{\text{b}}$ for some $j$ the right-hand sides of (\ref{cut}) and (%
\ref{cuteq}) vanish.

Now we explain how to build diagrams for the cut correlation functions (\ref%
{correcut}) and the identities (\ref{cuteq}). Each side of the cut is built
as explained in section \ref{volume}, so we can concentrate on the cut
itself, which is given by the integral on $v$ and $\bar{v}$. The left-hand
side of the cut depends on $v$, while the right-hand side depends on $\bar{v}
$. Formulas (\ref{freeampl}), (\ref{WW0}) and (\ref{amplint})\ show that, if
we treat the interaction Lagrangian $\mathcal{L}_{I}$ as in (\ref{amplint}),
the dependence on $v$ and $\bar{v}$ stems from the free generating
functional $W^{\text{free}}$ (which is at most linear in $v$, or $\bar{v}$,
on each side of the cut), and spreads around due to the functional
derivatives contained in $\mathcal{L}_{I}$.

We expand in powers of $v$ to the left of the cut, and in powers of $\bar{v}$
to the right of the cut. Then, the integration measure $\mathrm{d}\mu _{\bar{%
v},v}$ makes the $v$-$\bar{v}$ integrals convergent. Every $v$-$\bar{v}$
integral we obtain can be evaluated with the help of the first identity of (%
\ref{deltacoh}), which can be viewed as Wick's theorem for $v$ and $\bar{v}$%
. Indeed, $1/(2\omega )$ is the \textquotedblleft
elementary\textquotedblright\ $v$-$\bar{v}$ cut propagator, and the factor $%
n!$ takes care of all the possibilities of associating a $v$ to some $\bar{v}
$. In the end, the two sides of the cut are connected by the $v$-$\bar{v}$
integrals.

It is convenient to introduce different sources $\zeta ,\bar{\zeta}$ for $%
\langle \bar{v},t;z_{\text{b}},t_{\text{b}}\rangle _{\zeta ,\bar{\zeta}%
}^{\ast }$ and $\langle \bar{v},t;z_{\text{a}},t_{\text{a}}\rangle _{\zeta ,%
\bar{\zeta}}$ in (\ref{cut}). We denote them by means of subscripts $-$ and $%
+$, respectively. Then (\ref{cut}) can be written as%
\begin{equation}
\left. \mathcal{D}_{n}^{-+}\int \langle \bar{v},t;z_{\text{b}},t_{\text{b}%
}\rangle _{\zeta _{-},\bar{\zeta}_{-}}^{\ast }\mathrm{d}\mu _{\bar{v}%
,v}\langle \bar{v},t;z_{\text{a}},t_{\text{a}}\rangle _{\zeta _{+},\bar{\zeta%
}_{+}}\right\vert _{+=-}=\langle \tilde{z}(t_{1})\cdots \tilde{z}%
(t_{n})\rangle _{\zeta ,\bar{\zeta}},  \label{Dpm}
\end{equation}%
where%
\begin{equation*}
\mathcal{D}_{n}^{-+}\equiv \prod_{j=1}^{n}\left( \frac{\delta }{i\delta 
\tilde{\zeta}_{-}(t_{j})}+\frac{\delta }{i\delta \tilde{\zeta}_{+}(t_{j})}%
\right)
\end{equation*}%
and \textquotedblleft $+=-$\textquotedblright\ stands for $\zeta _{+}=\zeta
_{-}=\zeta $, $\bar{\zeta}_{+}=\bar{\zeta}_{-}=\bar{\zeta}$. Separating the
interactions $\mathcal{L}_{I}$ from the rest by means of (\ref{amplint}), we
also have 
\begin{equation}
\left. \mathcal{D}_{n}^{-+}\mathcal{D}_{I}^{-\ast }\mathcal{D}_{I}^{+}\int
\langle \bar{v},t;z_{\text{b}},t_{\text{b}}\rangle _{\zeta _{-},\bar{\zeta}%
_{-}}^{\text{free}\ast }\mathrm{d}\mu _{\bar{v},v}\langle \bar{v},t;z_{\text{%
a}},t_{\text{a}}\rangle _{\zeta _{+},\bar{\zeta}_{+}}^{\text{free}%
}\right\vert _{+=-}=\langle \tilde{z}(t_{1})\cdots \tilde{z}(t_{n})\rangle
_{\zeta ,\bar{\zeta}},  \label{DDD}
\end{equation}%
where%
\begin{equation*}
\mathcal{D}_{I}^{\pm }\equiv \exp \left( i\int_{t_{\pm }}^{t}\!\!\mathrm{d}%
t^{\prime }\!\mathbf{\hspace{0.01in}}\mathcal{L}_{I}\left( \frac{\delta }{%
i\delta \bar{\zeta}_{\pm }(t^{\prime })},\frac{\delta }{i\delta \zeta _{\pm
}(t^{\prime })}\right) \right) .
\end{equation*}%
$t_{+}$ standing for $t_{\text{a}}$, and $t_{-}$ standing for $t_{\text{b}}$%
. As explained after (\ref{proda}), we can move these expressions outside
the $\bar{v}$, $v$ integral, because $\mathcal{L}_{I}$ does not depend on $%
\bar{v}$ and\textrm{\ }$v$ before the shift (\ref{shift2}). Formula (\ref%
{DDD}) shows that it is sufficient to calculate 
\begin{equation}
\mathrm{e}^{iW_{-+}^{\text{free}}}\equiv \int \langle \bar{v},t;z_{\text{b}%
},t_{\text{b}}\rangle _{\zeta _{-},\bar{\zeta}_{-}}^{\text{free}\ast }%
\mathrm{d}\mu _{\bar{v},v}\langle \bar{v},t;z_{\text{a}},t_{\text{a}}\rangle
_{\zeta _{+},\bar{\zeta}_{+}}^{\text{free}},  \label{Wfree}
\end{equation}%
since everything else follows from it by means of repeated functional
differentiations. The calculation is straightforward. Basically, we have
already done it earlier to prove (\ref{identafree}). The result is%
\begin{eqnarray}
iW_{-+}^{\text{free}}=i &&\tilde{W}_{0}+2\omega \bar{z}_{0}(t)z_{0}(t)+%
\int_{t_{\text{b}}}^{t}\mathrm{d}t^{\prime }\int_{t_{\text{a}}}^{t}\mathrm{d}%
t^{\prime \prime }\bar{\zeta}_{-}^{\prime }(t^{\prime })\frac{\mathrm{e}%
^{-i\omega (t^{\prime }-t^{\prime \prime })}}{2\omega }\zeta _{+}^{\prime
}(t^{\prime \prime })  \notag \\
&&-i\int_{t_{\text{b}}}^{t}\mathrm{d}t^{\prime }\hspace{0.01in}\left. 
\mathbf{(}\bar{\zeta}_{-}^{\prime }z_{0}+\bar{z}_{0}\zeta _{-}^{\prime
})\right\vert _{t^{\prime }}-\int_{t_{\text{b}}}^{t}\mathrm{d}t^{\prime
}\int_{t_{\text{b}}}^{t}\mathrm{d}t^{\prime \prime }\bar{\zeta}_{-}^{\prime
}(t^{\prime })\theta (t^{\prime \prime }-t^{\prime })\frac{\mathrm{e}%
^{-i\omega (t^{\prime }-t^{\prime \prime })}}{2\omega }\zeta _{-}^{\prime
}(t^{\prime \prime })  \notag \\
&&+i\int_{t_{\text{a}}}^{t}\mathrm{d}t^{\prime }\hspace{0in}\hspace{0.01in}%
\left. \mathbf{(}\bar{\zeta}_{+}^{\prime }z_{0}+\bar{z}_{0}\zeta
_{+}^{\prime })\right\vert _{t^{\prime }}-\int_{t_{\text{a}}}^{t}\mathrm{d}%
t^{\prime }\int_{t_{\text{a}}}^{t}\mathrm{d}t^{\prime \prime }\bar{\zeta}%
_{+}^{\prime }(t^{\prime })\theta (t^{\prime }-t^{\prime \prime })\frac{%
\mathrm{e}^{-i\omega (t^{\prime }-t^{\prime \prime })}}{2\omega }\zeta
_{+}^{\prime }(t^{\prime \prime }),\qquad  \label{Wpm}
\end{eqnarray}%
where $\tilde{W}_{0}$ is the same as in (\ref{WW0}), the functions $\bar{z}%
_{0}(t)$ and $z_{0}(t)$ are the same as in (\ref{z0}), while $\zeta _{\pm
}^{\prime }=\zeta _{\pm }+A+i\omega B$, $\bar{\zeta}_{\pm }^{\prime }=\bar{%
\zeta}_{\pm }+A-iB\omega $, as in (\ref{zetap}).

Now we explain the meanings of the various terms that appear in (\ref{Wpm}).
The double integral in the last line encodes the usual propagator (\ref%
{propaw})-(\ref{prow}), which connects vertices placed to the right of the
cut. The double integral in the second line encodes the conjugate
propagator, which connects vertices placed to the left of the cut. The
double integral in the first line encodes the cut propagator, which connects
vertices located on opposite sides of the cut.

Differentiating with respect to $\bar{\zeta}_{-}(t_{1})$ and $\zeta
_{+}(t_{2})$, and concentrating on the connected part of the two-point
function (denoted by the subscript $c$), we find%
\begin{equation}
\langle z(t_{1})\hspace{0.01in}|\hspace{0.01in}\bar{z}(t_{2})\rangle _{c}^{%
\text{free}}=\frac{\mathrm{e}^{-i\omega (t_{1}-t_{2})}}{2\omega },\qquad
\langle z(e)\hspace{0.01in}|\hspace{0.01in}\bar{z}(-e)\rangle _{c}^{\text{%
free}}=(2\pi )\theta (e)\delta (e^{2}-\omega ^{2}),  \label{cutpro}
\end{equation}%
before and after the Fourier transform. Differentiating with respect to $%
\bar{\zeta}_{+}(t_{1})$ and $\zeta _{-}(t_{2})$, we find $\langle \bar{z}%
(t_{1})\hspace{0.01in}|\hspace{0.01in}z(t_{2})\rangle _{c}^{\text{free}}=0$,
instead. This proves that positive energies flow through the cut in a unique
direction (from the right to the left).

Note that $W_{-+}^{\text{free}}$ includes several contributions that are
linear in $\zeta ^{\pm }\ $and $\bar{\zeta}^{\pm }$. When the functional
derivatives act on those, the legs associated with them do not connect
vertices, but end into external sources (endpoints), built with $\phi _{0}$, 
$z_{0}$ and $\bar{z}_{0}$, which carry information about the restrictions to
finite $\tau $ and finite volume. 
\begin{figure}[t]
\begin{center}
\vskip-.7truecm\includegraphics[width=16truecm]{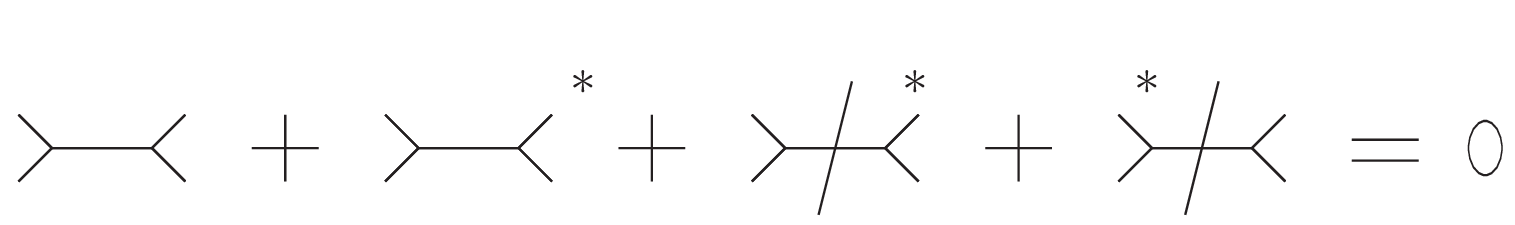}
\end{center}
\par
\vskip -.7truecm
\caption{Cutkosky-Veltman identity for the propagator}
\label{CVprop}
\end{figure}

The result confirms that the cut propagators (\ref{cutpro}) know nothing
about the restriction to finite times (and little enough about the
restriction to finite volume, which amounts to the discretization of the
frequencies $\omega $), like the (uncut) propagators (\ref{prow}). This
property ensures that the spectral optical identities expressing unitarity
coincide with the usual ones, internally (apart from the discretization of
the loop momenta), and differ only externally. Details on this are given in
the next section, where we use them to introduce purely virtual particles at
finite $\tau $ on a compact $\Omega $.

The simplest examples of equations (\ref{cuteq}) are those of the one- and
two-point functions. Formula (\ref{cut}) with $n=1$, $t>t_{1}>t_{\text{b}}$,
gives%
\begin{equation}
\langle z(t_{1})\hspace{0.01in}|\hspace{0.01in}\rangle =\langle \hspace{%
0.01in}|\hspace{0.01in}z(t_{1})\rangle ,\qquad \langle \bar{z}(t_{1})\hspace{%
0.01in}|\hspace{0.01in}\rangle =\langle \hspace{0.01in}|\hspace{0.01in}\bar{z%
}(t_{1})\rangle .  \label{cut1}
\end{equation}%
In the free limit $\mathcal{L}_{I}\rightarrow 0$, we find%
\begin{equation}
\mathrm{e}^{-iW_{-+}^{\text{free\hspace{0.01in}0}}}\langle z(t_{1})\hspace{%
0.01in}|\hspace{0.01in}\rangle ^{\text{free}}=\left. \frac{\mathrm{e}%
^{-iW_{-+}^{\text{free}}}\delta \mathrm{e}^{iW_{-+}^{\text{free}}}}{-i\delta 
\bar{\zeta}_{-}(t_{1})}\right\vert _{\zeta _{\pm }=\bar{\zeta}_{\pm
}=0}\!\!\!\!\!\!\!\!\!\!\!\!\!=z_{0}(t_{1})+i\!\int_{t_{\text{a}}}^{t}\!\!%
\mathrm{d}t^{\prime }\theta (t_{1}-t^{\prime })\frac{\mathrm{e}^{-i\omega
(t_{1}-t^{\prime })}}{2\omega }\left. (A+i\omega B)\right\vert _{t^{\prime
}},  \label{end}
\end{equation}%
where $W_{-+}^{\text{free\hspace{0.01in}0}}\equiv \left. W_{-+}^{\text{free}%
}\right\vert _{\zeta _{\pm }=\bar{\zeta}_{\pm }=0}$. The same result is
obtained for $\langle \hspace{0.01in}|\hspace{0.01in}z(t_{1})\rangle ^{\text{%
free}}$, which is calculated as $-i\delta \mathrm{e}^{iW_{-+}^{\text{free}%
}}/\delta \bar{\zeta}_{+}(t_{1})$. The second identity of (\ref{cut1}) is
verified similarly at $\mathcal{L}_{I}=0$. For $t_{1}<t_{\text{b}}$, we have 
$\langle \hspace{0.01in}|\hspace{0.01in}z(t_{1})\rangle =\langle
z(t_{1})\rangle $, $\langle \hspace{0.01in}|\hspace{0.01in}\bar{z}%
(t_{1})\rangle =\langle \bar{z}(t_{1})\rangle $, which are trivial in the
free limit.

Formula (\ref{cuteq}) for $n=2$, $\tilde{z}(t_{1})=z(t_{1})$, $\hspace{0.01in%
}\tilde{z}(t_{2})=\bar{z}(t_{2})$, gives 
\begin{equation*}
\langle \hspace{0.01in}|z(t_{1})\bar{z}(t_{2})\hspace{0.01in}\rangle
-\langle z(t_{1})\hspace{0.01in}|\hspace{0.01in}\bar{z}(t_{2})\rangle
-\langle \bar{z}(t_{2})\hspace{0.01in}|\hspace{0.01in}z(t_{1})\rangle
+\langle z(t_{1})\bar{z}(t_{2})\hspace{0.01in}|\hspace{0.01in}\rangle
=\langle z(t_{1})\bar{z}(t_{2})\rangle .
\end{equation*}%
In the free limit, the connected components give%
\begin{equation*}
\theta (t_{1}-t_{2})\frac{\mathrm{e}^{-i\omega (t_{1}-t_{2})}}{2\omega }%
\qquad -\frac{\mathrm{e}^{-i\omega (t_{1}-t_{2})}}{2\omega }\qquad -0\qquad
+\theta (t_{2}-t_{1})\frac{\mathrm{e}^{-i\omega (t_{1}-t_{2})}}{2\omega }%
\qquad =0,
\end{equation*}%
for $t>t_{1,2}>t_{\text{b}}$, after simplifying the common normalization
factor $\mathrm{e}^{iW_{-+}^{\text{free\hspace{0.01in}0}}}$. This identity
is illustrated in fig. \ref{CVprop}. For $t_{\text{b}}>t_{1,2}$, $t_{1}>t_{%
\text{b}}>t_{2}$ and $t_{2}>t_{\text{b}}>t_{1}$, we have $\langle \hspace{%
0.01in}|z(t_{1})\bar{z}(t_{2})\hspace{0.01in}\rangle =\langle z(t_{1})\bar{z}%
(t_{2})\rangle $, $\langle \hspace{0.01in}|z(t_{1})\bar{z}(t_{2})\hspace{%
0.01in}\rangle =\langle z(t_{1})\hspace{0.01in}|\hspace{0.01in}\bar{z}%
(t_{2})\rangle $ and $\langle \hspace{0.01in}|z(t_{1})\bar{z}(t_{2})\hspace{%
0.01in}\rangle =\langle \bar{z}(t_{2})\hspace{0.01in}|\hspace{0.01in}%
z(t_{1})\rangle $, respectively, which are also trivial in the free limit.

\begin{figure}[t]
\begin{center}
\includegraphics[width=10truecm]{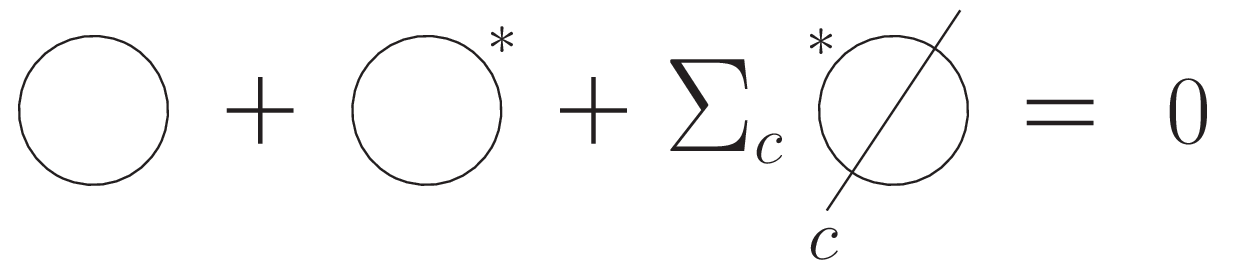}
\end{center}
\par
\vskip -.7truecm
\caption{Diagrammatic unitarity equations}
\label{Unitar}
\end{figure}

The unitarity equations (\ref{unitatez}) and (\ref{suff}) are studied by
assuming $t>t_{\text{b}}=t_{\text{a}}$ in (\ref{cut}). Then the right-hand
sides of (\ref{cut}) and (\ref{cuteq}) vanish for every $n>0$. Separating
the uncut diagram $G$ (and its conjugate diagram $G^{\ast }$) from every
other contributions to the left-hand side, equation (\ref{cuteq}) can be
written in the usual form, which is 
\begin{equation}
G+G^{\ast }+\sum_{c}G_{c}=0,  \label{CVid}
\end{equation}%
where the sum is on all the diagrams that contain nontrivial cuts, including
those coming from the first two correlation functions of (\ref{tre}).\ The
minus signs of (\ref{cuteq}) are included into the definitions of cut
diagrams. We illustrate the equation in fig. \ref{Unitar}. 

The diagrammatic rules for the identities (\ref{CVid}) are as follows.

--- Draw a vertical bar, which denotes the cut.

--- Distribute the external legs in all possible ways on the two sides, with
a minus sign for each leg to the left.

--- Do the same for the vertices and the endpoints (\textquotedblleft
one-leg vertices\textquotedblright ) that contribute to the order you are
interested in.

--- Draw the diagrams by connecting (internal and external) legs to vertices
and endpoints in all possible ways.

--- Two legs $z$, $\bar{z}$ are connected to each other by means of ordinary
propagators (when they both lie to the right of the cut), conjugate
propagators (when they both lie to the left), or cut propagators (when they
lie on opposite sides).

\bigskip

Ultimately, formulas (\ref{CVid}), and the rules just stated, are the same
as usual. The only differences are that: $i$) the loop momenta are
discretized; $ii$)\ each vertex has an external source $K$ attached to it; $%
iii$) there are endpoints, due to the restrictions to finite $\tau $ and
compact $\Omega $. Endpoints are actually common at $\tau =\infty $, $\Omega
=\mathbb{R}^{3}$ as well (for example, when the field is shifted by a
nontrivial background).

\begin{figure}[t]
\begin{center}
\includegraphics[width=16truecm]{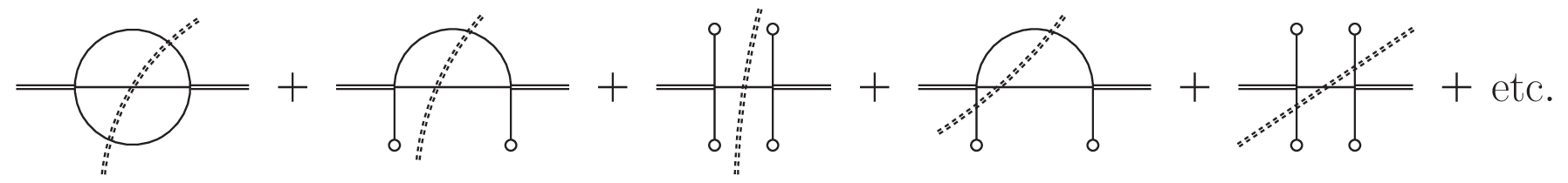}
\end{center}
\par
\vskip -.7truecm
\caption{Cut diagrams with two vertices}
\label{endpointdiagraCut}
\end{figure}

In fig. \ref{endpointdiagraCut} we show examples of cut diagrams with two
cubic vertices. Note that the cut may also cross legs attached to endpoints.
This is because the endpoints (\ref{end}) receive contributions from both
sides of the cut.

As said, the identities (\ref{CVid}) encode the unitarity equation (\ref%
{unitate}), which is also (\ref{unitatez}), (\ref{suff}), or the cases $t>t_{%
\text{b}}=t_{\text{a}}$ of (\ref{optid}) and (\ref{cut}). When we relax
these restrictions, we have other diagrammatic identities, which encode the
more general equation (\ref{unitar}). They look similar to (\ref{CVid}), but
for the following differences. First, the right-hand side needs not be
zero:\ it is replaced by the diagrammatic version of the right-hand side of (%
\ref{cuteq}). In addition, when $t_{\text{b}}>t$, no minus signs are
attached to the legs and the vertices located to the left of the cut, the
conjugate propagators are replaced by (unconjugate) propagators and the cut
propagators across the cut coincide with the same (uncut) propagators. 

\section{Purely virtual particles}

\label{PV}\setcounter{equation}{0}

In this section we introduce purely virtual particles at finite $\tau $ on a
compact space manifold $\Omega $, after briefly recalling what they are at $%
\tau =\infty $, $\Omega =\mathbb{R}^{3}$, and how they are introduced there,
following \cite{diagrammarMio}. For the time being, we assume that all the
particles have kinetic terms with the correct signs, and explain how to
render some physical particles purely virtual. Later, we explain how to
render ghosts purely virtual as well. We recall that tachyons cannot be
rendered purely virtual.

Consider an arbitrary Feynman diagram $G$ in momentum space, where, by
assumption, the propagators are defined by means of the usual $i\epsilon $
prescription. Label each internal leg by means of an index $a,b,\ldots $.
Let $m_{a}$ denote the mass of the $a$-th leg, and $k^{\mu }-p_{a}^{\mu }$
its four-momentum. Here $k^{\mu }=(k^{0},\mathbf{k})$ denotes a loop
four-momentum, or a combination of loop four-momenta, while $p_{a}^{\mu
}=(e_{a},\mathbf{p}_{a})$ is an external four-momentum. The frequency of the 
$a$-th leg is $\omega _{a}=\sqrt{m_{a}^{2}+(\mathbf{k}-\mathbf{p}_{a})^{2}}$%
. Note that each internal leg is equipped with its own $p_{a}^{\mu }$. This
redundant notation (various $p_{a}^{\mu }$ may depend on one another) makes
the formulas more symmetric and easier to handle.

We integrate on the loop energies $k^{0}$ with measure $\mathrm{d}%
k^{0}/(2\pi )$, by means of the residue theorem, and completely ignore the
integrals on the loop momenta $\mathbf{k}$. The reason is that the
identities we derive, which are crucial to switch to purely virtual
particles, hold for arbitrary values of the frequencies $\omega _{a}$.

After integrating on the loop energies, we rearrange the results in order to
remove the differences of frequencies from the denominators\footnote{%
We know that they must cancel out, thanks to the $i\epsilon $ prescription.
A quick proof is that differences of frequencies in denominators are not
well prescribed, while the diagram as a whole is well prescribed.}. We
remain with denominators of the form%
\begin{equation*}
\frac{1}{E-\sum_{i}\omega _{a_{i}}+i\epsilon },
\end{equation*}%
where $E$ is a linear combination of external energies. At this point, we
make the \textquotedblleft threshold decomposition\textquotedblright , to
separate the on-shell contributions from the off-shell ones, by carefully%
\footnote{%
Starting from the box diagram, certain caveats require further
rearrangements in order to make the decomposition properly. See \cite%
{diagrammarMio} for details. Nuisances like these can be avoided by
switching to the equivalent approach of ref. \cite{PVP20}.} applying the
identity%
\begin{equation}
\frac{i}{x+i\epsilon }=\mathcal{P}\frac{i}{x}+\pi \delta (x),  \label{P}
\end{equation}%
where $\mathcal{P}$ denotes the Cauchy principal value. The number of delta
functions is called \textquotedblleft level\textquotedblright\ of the
threshold decomposition.

What just said applies to the diagrams that contain ordinary, physical
particles, which we denote by $G_{\text{ph}}$. A certain surgical operation
on $G_{\text{ph}}$ allows us to define new diagrams $G_{\text{pv}}$, where
some internal legs are purely virtual. This is achieved by removing all the
on shell contributions that involve the particles we want to render purely
virtual\footnote{%
By \textquotedblleft on shell\textquotedblright\ we always mean
\textquotedblleft on the mass shell\textquotedblright\ here.}. Let $\omega _{%
\text{pv}}$ denote the frequencies of those particles in $G_{\text{ph}}$.
Given the threshold decomposition of $G_{\text{ph}}$, the threshold
decomposition of $G_{\text{pv}}$ is obtained by dropping every contribution
that involves delta functions with $\omega _{\text{pv}}$-dependent supports.
Finally, $G_{\text{pv}}$ itself is defined from its own threshold
decomposition.

At the tree level, the $w$-$\bar{w}$ propagator of formula (\ref{prow})
loses the delta function and becomes a principal value: 
\begin{equation}
\langle w(e)\bar{w}(-e)\rangle _{c}^{\text{free}}\rightarrow \mathcal{P}%
\frac{i}{2\omega (e-\omega )},\qquad \langle w(t)\hspace{0.01in}\bar{w}%
(t^{\prime })\rangle _{c}^{\text{free}}\rightarrow \text{sgn}(t-t^{\prime })%
\frac{\mathrm{e}^{-i\omega (t-t^{\prime })}}{4\omega },  \label{propv}
\end{equation}%
where sgn$(t)=\theta (t)-\theta (-t)$ is the sign function.

In a one-loop diagram, the identity (\ref{P}) separates quantities such as%
\begin{equation*}
\mathcal{P}^{ab}\equiv \mathcal{P}\frac{1}{e_{a}-e_{b}-\omega _{a}-\omega
_{b}},\qquad \Delta ^{ab}\equiv \pi \delta (e_{a}-e_{b}-\omega _{a}-\omega
_{b}).
\end{equation*}%
Once the threshold decomposition is completed, we remove all the
contributions that contain a $\Delta ^{ab}$ where $\omega _{a}$, or $\omega
_{b}$, or both, are the frequencies of particles that we want to render
purely virtual. In diagrams with more loops similar rules apply. Sums of
more than two frequencies may appear in the supports of the delta functions.

For example, the usual bubble diagram gives a result proportional to 
\begin{equation}
\frac{i}{e_{1}-e_{2}-\omega _{1}-\omega _{2}+i\epsilon }+\frac{i}{%
e_{2}-e_{1}-\omega _{1}-\omega _{2}+i\epsilon },  \label{bubble}
\end{equation}%
after integrating on the loop energy. If we want to render the particles
propagating in one or two internal legs purely virtual, we replace the
result by $i$ times 
\begin{equation}
\mathcal{P}^{12}+\mathcal{P}^{21},  \label{pvbubble}
\end{equation}%
by dropping the contributions $\Delta ^{12}$ and $\Delta ^{21}$. For details
of the triangle, the box, etc., and diagrams with more loops, see \cite%
{diagrammarMio,PVP20}.

The prescription just recalled takes care of the internal sectors of the
diagrams. For consistency, we must also restrict the external sectors, by
demanding that only physical particles lie on the external legs. So doing,
we project the set of states to the physical subspace. The physical
amplitudes are the amplitudes between incoming and outgoing physical
particles. Any other amplitude is dropped because unphysical.

The combination made by this projection and the prescription described above
defines a map $M_{\text{pv}}$ from a starting theory to a new theory. The
starting theory contains physical particles (and possibly ghosts). The final
theory contains physical particles and purely virtual particles\footnote{%
It may be useful to make a parallel with what we normally do to quantize
gauge theories. There, we drop all the scattering amplitudes with incoming
and outgoing Faddeev-Popov ghosts $C$, $\bar{C}$, and/or
temporal/longitudinal components $A_{0}$, $A_{L}$ of the gauge fields. The
internal sectors of the diagrams are automatically taken care of by the
gauge symmetry. In the case of purely virtual particles, where no symmetry
is helping us, we need to make the surgical operation described above, on
the internal sectors of the diagrams, which amounts to dropping the delta
functions with $\omega _{\text{pv}}$-dependent supports. Applied to gauge
theories, the prescription/projection operations return the same physical
results we obtain with the usual quantization method \cite{LambdaPaper}.}.

The map $M_{\text{pv}}$ is consistent with unitarity as follows: $i$) if the
starting theory is unitary (i.e., it has no ghosts), the final theory is
unitary; $ii$) if the starting theory has ghosts, the final theory is
unitary, provided all the ghosts are turned into purely virtual particles.
In other words, if we convert a subset of physical particles (and all the
ghosts, if present) into purely virtual particles, we preserve (or gain)
unitarity.

The reason why unitarity is preserved, or gained, relies on an important
fact: that the thresholds are independent from one another, so the unitarity
equations (\ref{CVid}) split into a large number of independent
\textquotedblleft spectral optical identities\textquotedblright\ \cite%
{diagrammarMio}, one for each threshold, which hold algebraically, before
integrating on the loop momenta. Suppressing certain types of thresholds
everywhere, the identities (\ref{CVid}) remain true. Moreover, the cut
diagrams where a cut crosses one or more legs of purely virtual particles
disappear entirely, because the cut propagators associated with those legs
are delta functions with $\omega _{\text{pv}}$-dependent supports. This is
consistent with projecting the purely virtual particles away externally.

In the case of theories with physical particles only (no ghosts), the
Cutkosky-Veltman identities (\ref{CVid}) encode the unitarity of the
starting theory. After implementing the prescription/projection to purely
virtual particles, they encode the unitarity of the final theory. In the
presence of ghosts, instead, the identities (\ref{CVid}) do not express
unitarity, since the starting theory is not unitary. Yet, they are still
valid (they are called pseudounitarity equations), and very useful. What is
important is that, after the prescription/projection, they encode the
unitarity of the final theory, provided \textit{all} the ghosts are rendered
purely virtual.

It is important to stress that the diagrammatics of purely virtual particles
is not governed by time ordering \cite{PVP20}, so the map $M_{\text{pv}}$
does not commute with the diagrammatic rules: it acts on the amplitudes and
the diagrams as such. More precisely, a loop diagram containing purely
virtual particles cannot be built by using the projected propagators of (\ref%
{Wpvfree}) inside an ordinary diagram: it should be built by projecting the
ordinary diagram as a whole. An explicit example can clarify this point. We
know that the tree propagator (\ref{prow}) is mapped into (\ref{propv}), and
the ordinary bubble diagram (\ref{bubble}) is mapped into (\ref{pvbubble}).
However, if we build a bubble diagram with two propagators (\ref{propv}), we
do not obtain (\ref{pvbubble}): we obtain something that is very different,
and not even consistent with unitarity (see \cite{Wheelerons}). For the same
reason, simple separations between the free and interaction parts, such as
those encoded in formulas (\ref{inta}), (\ref{amplint}) and (\ref{DDD}), do
not commute with the map $M_{\text{pv}}$ (beyond the tree level).

The extension to purely virtual particles is an interesting option that was
overlooked before. Nevertheless, it is allowed by quantum field theory, and
might be the solution to the problem of quantum gravity \cite{LWgrav,ABP}.

\subsection{Purely virtual particles in a finite interval of time and on a
compact manifold}

What we have just said holds at $\tau =\infty $, $\Omega =\mathbb{R}^{3}$.
We can generalize it to finite $\tau $ and compact $\Omega $ as follows.

A purely virtual particle is not associated with a dynamical degree of
freedom, since its on shell contributions are removed from the physical
quantities. In this respect, it is a sort of fake particle. Consequently, it
cannot have nontrivial initial or final conditions. This means that, at
finite $\tau $, on a compact $\Omega $, the projection to purely virtual
particles mentioned earlier is the set of conditions 
\begin{equation}
z(t_{\text{i}})=\bar{z}(t_{\text{f}})=0.  \label{pvbou}
\end{equation}%
The particles we want to render purely virtual are thereby removed from the
external sectors of the diagrams.


As far as the boundary conditions are concerned, we can keep them in the
general form (\ref{bou}), since they are not associated with degrees of
freedom. For example, the integral in between the right-hand side of (\ref%
{identa}) does not concern them. Besides, the function $f(t,\mathbf{x}%
_{\partial \Omega })$ might describe some property of our experimental
apparatus.

Next, we have to free the interior parts of the diagrams from the on shell
contributions due to the particles that we want to render purely virtual.
This goal can be achieved exactly as before, since the internal parts of the
diagrams are basically the same as at $\tau =\infty $, $\Omega =\mathbb{R}%
^{3}$.

Let us recapitulate what we know. We have shown that quantum field theory
(with physical particles and possibly ghosts) can be formulated
diagrammatically at finite $\tau $ and on a compact space manifold $\Omega $%
. The diagrams are the same as usual, internally, apart from a non invasive
change, which is the discretization of the loop momenta. Every other detail
about the restriction to finite $\tau $ and compact $\Omega $ is moved away
to the external sources $K$ (by which we also mean the endpoints). We have
seen that the Cutkovsky-Veltman identities (\ref{CVid}) are the same as
usual, apart from the external sources and the discretization of the loop
momenta. They encode the unitarity or pseudounitarity equation (\ref{unitate}%
) of the starting theory (depending on whether ghosts are absent or present).

We also know that at $\tau =\infty $, $\Omega =\mathbb{R}^{3}$ the
identities (\ref{CVid}) can be split into independent spectral optical
identities, one for every threshold, which hold for arbitrary frequencies $%
\omega $, before integrating on the loop momenta. In exactly the same way,
we can split the identities (\ref{CVid}) at finite $\tau $ and on a compact $%
\Omega $. The identities we obtain hold for arbitrary frequencies $\omega _{%
\mathbf{n}}$, before summing on $\mathbf{n}$. Similarly, the threshold
decomposition of a diagram can be performed at $\tau <\infty $, $\Omega $ =
compact manifold in the same way as it is performed at $\tau =\infty $, $%
\Omega =\mathbb{R}^{3}$.

Now that we have the threshold decomposition, we can apply the map $M_{\text{%
pv}}$ to it as before, by dropping all the delta functions that have $\omega
_{\text{pv}}$-dependent supports, where $\omega _{\text{pv}}$ denotes any
frequency of the particles that we want to render purely virtual. For this
operation to be meaningful, it does not matter whether the frequencies are
discretized or not. What we obtain is the threshold decomposition of the
diagrams containing physical and purely virtual particles. The identities (%
\ref{CVid}) remain true after the prescription/projection.

Ultimately, we define a new theory, which is unitary and contains physical,
as well as purely virtual particles. We gain or preserve unitarity, in the
form of equation (\ref{unitate}), depending on whether the starting theory
contains ghosts, or not.

\bigskip

We denote the evolution operator of the final theory by $U_{\text{ph}}(t_{%
\text{f}},t_{\text{i}})$ and its amplitudes by 
\begin{equation}
M_{\text{pv}}\left( \langle \bar{z}_{\text{f}},t_{\text{f}};z_{\text{i}},t_{%
\text{i}}\rangle _{\zeta ,\bar{\zeta}}\right) ,  \label{pvampl}
\end{equation}%
where it is understood that the initial and final conditions $z_{\text{i}}$, 
$\bar{z}_{\text{f}}$ only refer to the physical particles. The unitarity
equations (\ref{unitatez}), which express (\ref{unitate}), lose the
integrals $\mathrm{d}\mu _{\bar{z}_{\text{f}},z_{\text{f}}}$ on the
variables $\bar{z}_{\text{f}},z_{\text{f}}$ associated with the purely
virtual particles, and become 
\begin{equation}
\int M_{\text{pv}}^{\ast }\left( \langle \bar{z}_{\text{f}},t_{\text{f}};z_{%
\text{i}},t_{\text{i}}\rangle _{\zeta ,\bar{\zeta}}\right) \mathrm{d}\mu _{%
\bar{z}_{\text{f}},z_{\text{f}}}^{\text{ph}}M_{\text{pv}}\left( \langle \bar{%
z}_{\text{f}},t_{\text{f}};z_{\text{i}},t_{\text{i}}\rangle _{\zeta ,\bar{%
\zeta}}\right) =\mathrm{e}^{2\omega \bar{z}_{\text{i}}z_{\text{i}}},
\label{pvsuff}
\end{equation}%
where the measure $\mathrm{d}\mu _{\bar{z}_{\text{f}},z_{\text{f}}}^{\text{ph%
}}$ is restricted to the subspace of physical particles. A sum over the
physical particles in understood in the exponent of $\mathrm{e}^{2\omega 
\bar{z}_{\text{i}}z_{\text{i}}}$. Formula (\ref{pvsuff}) expresses the
unitarity equation $U_{\text{ph}}^{\dag }(t_{\text{f}},t_{\text{i}})U_{\text{%
ph}}(t_{\text{f}},t_{\text{i}})=1_{\text{ph}}$ of the final theory, where $%
1_{\text{ph}}$ denotes the identity matrix restricted to the subspace of
physical particles.

We can explain the disappearance of the $\bar{v},v$ integrals for purely
virtual particles as follows. For convenience, we relabel $z_{\text{f}}=v$, $%
\bar{z}_{\text{f}}=\bar{v}$, $t_{\text{f}}=t$, $z_{\text{i}}=z_{\text{b}}=z_{%
\text{a}}$, $\bar{z}_{\text{i}}=\bar{z}_{\text{b}}=\bar{z}_{\text{a}}$, $t_{%
\text{i}}=t_{\text{b}}=t_{\text{a}}$ in (\ref{pvsuff}), to match the
notation used in sections \ref{unitari} and \ref{unitareq} (equation (\ref%
{optid}) in particular). Using $t>t_{\text{b}}=t_{\text{a}}$ in formulas (%
\ref{Wfree}) and (\ref{Wpm}), we see that those integrals provide: $i$) the
cut propagators; and $ii$) contributions from the endpoint corrections (\ref%
{end}). The latter occur when the cut crosses a leg attached to an endpoint
(as in the last two drawings of fig. \ref{endpointdiagraCut}). Their
contributions can be of two types: $a$) the ones depending on the initial
and final conditions, through $z_{0}$ and $\bar{z}_{0}$; and $b$) the ones
depending on the boundary conditions, through $\phi _{0}$, $A$ and $B$. As
shown in (\ref{end}), the latter are attached to propagators, so they are
interested by the prescription/projection, while the former are not attached
to propagators\footnote{%
To be pedantic, the two types of contributions should be graphically
distinguished from each other. As long as we know what we are doing, it is
not really necessary to insist on this.}.

We know that the cut propagators of purely virtual particles vanish, because
of the prescription/projection. The mentioned corrections to the endpoints
also vanish: those of type $a$) vanish because of the conditions (\ref{pvbou}%
) (and their conjugates, for the conjugate amplitude); those of type $b$)
vanish because they are attached to cut propagators.

In the end, we can succinctly write $M_{\text{pv}}\left( \mathrm{d}\mu _{%
\bar{v},v}\right) =\mathrm{d}\mu _{\bar{v},v}^{\text{ph}}$ and $M_{\text{pv}%
}\left( 1\right) =1_{\text{ph}}$. In the coherent-state approach, $1_{\text{%
ph}}$ means $\mathrm{e}^{2\omega \bar{z}_{\text{i}}z_{\text{i}}}$, recalling
that $\bar{z}_{\text{i}}$ and $z_{\text{i}}$ are nontrivial only for
physical particles.

We first check our claims at the tree level in the case of a single purely
virtual particle, where the left-hand side of (\ref{pvsuff}) becomes just a
product. We can calculate it from (\ref{Wpm}). Setting $t_{\text{f}}=t>t_{%
\text{b}}=t_{\text{a}}=t_{\text{i}}$, using the conditions (\ref{pvbou})
(and their conjugates) and implementing the prescription with the help of (%
\ref{propv}), we obtain 
\begin{eqnarray}
iW_{-+\text{pv}}^{\text{free}} &=&-\int_{t_{\text{a}}}^{t}\mathrm{d}%
t^{\prime }\int_{t_{\text{a}}}^{t}\mathrm{d}t^{\prime \prime }\bar{\zeta}%
_{-}^{\prime }(t^{\prime })\text{sgn}(t^{\prime \prime }-t^{\prime })\frac{%
\mathrm{e}^{-i\omega (t^{\prime }-t^{\prime \prime })}}{2\omega }\zeta
_{-}^{\prime }(t^{\prime \prime })  \notag \\
&&-\int_{t_{\text{a}}}^{t}\mathrm{d}t^{\prime }\int_{t_{\text{a}}}^{t}%
\mathrm{d}t^{\prime \prime }\bar{\zeta}_{+}^{\prime }(t^{\prime })\text{sgn}%
(t^{\prime }-t^{\prime \prime })\frac{\mathrm{e}^{-i\omega (t^{\prime
}-t^{\prime \prime })}}{2\omega }\zeta _{+}^{\prime }(t^{\prime \prime }).
\label{Wpvfree}
\end{eqnarray}%
Clearly, $iW_{-+\text{pv}}^{\text{free}}=0$ for $\zeta _{+}=\zeta _{-}$, $%
\bar{\zeta}_{+}=\bar{\zeta}_{-}$. Moreover, (\ref{DDD}) is trivially
satisfied, if we restrict it to the tree diagrams.

In the case of the bubble diagram, we still obtain (\ref{pvbubble}) (with
discretized frequencies), when some internal leg belongs to purely virtual
particles. One proceeds similarly for the other loop diagrams.

In\ the operatorial language, the states on which we are summing in the
left-hand side of (\ref{pvsuff}) (which are the states of the physical
subspace) are built by means of creation operators of physical particles
only, acting on the vacuum state $|0\rangle $:\ there are no creation
operators for purely virtual particles.

There is an important (to some extent unexpected) turn of events, though.
The starting theory also satisfies the more general identity (\ref{unitar}),
i.e., $U(t_{\text{f}},t)U(t,t_{\text{i}})=U(t_{\text{f}},t_{\text{i}})$ for
arbitrary $t_{\text{f}}$, $t$ and $t_{\text{i}}$. What is the fate of that
identity under the map $M_{\text{pv}}$? The answer is that it is lost. We
can check this claim already in the free-field limit, with a single purely
virtual particle. Applying the map $M_{\text{pv}}$ to (\ref{freeampl}) and (%
\ref{WW0}), we get $U(t_{\text{f}},t_{\text{i}})=\exp \left( iW^{\text{free}%
}(t_{\text{f}},t_{\text{i}})\right) $, where%
\begin{equation}
iW^{\text{free}}(t_{\text{f}},t_{\text{i}})=\tilde{W}_{0}(t_{\text{f}},t_{%
\text{i}})-\int_{t_{\text{i}}}^{t_{\text{f}}}\mathrm{d}t^{\prime }\int_{t_{%
\text{i}}}^{t_{\text{f}}}\mathrm{d}t^{\prime \prime }\bar{\zeta}^{\prime
}(t^{\prime })\text{sgn}(t^{\prime }-t^{\prime \prime })\frac{\mathrm{e}%
^{-i\omega (t^{\prime }-t^{\prime \prime })}}{2\omega }\zeta ^{\prime
}(t^{\prime \prime }).  \label{w1}
\end{equation}%
However, applying the map $M_{\text{pv}}$ to (\ref{identa}) with $t_{\text{a}%
}=t_{\text{i}}<t<t_{\text{f}}=t_{\text{b}}$, we get $U(t_{\text{f}},t)U(t,t_{%
\text{i}})=$ $\mathrm{\exp }\left( iW^{\text{free}}(t_{\text{f}},t)+iW^{%
\text{free}}(t,t_{\text{i}})\right) $. We see that the missing contribution,
which is equal to $iW^{\text{free}}(t_{\text{f}},t_{\text{i}})-iW^{\text{free%
}}(t_{\text{f}},t)-iW^{\text{free}}(t,t_{\text{i}})$, is precisely the one
associated with the missing integral in between.

The reason why $U(t_{\text{f}},t)U(t,t_{\text{i}})=U(t_{\text{f}},t_{\text{i}%
})$ cannot be preserved for $t_{\text{f}}>t>t_{\text{i}}$ is actually
intuitive: the identity (\ref{identa}) means that we can break an amplitude
into a sum on all the intermediate states. However, those intermediate
states must be built with physical particles (or at most ghosts: see below),
which are associated with arbitrary initial and final conditions,\ on which
we must integrate in the middle. They cannot be built with purely virtual
particles, because the only physical state that is acceptable for a purely
virtual particle is the vacuum state. When we break the amplitude, we just
get the vacuum in the middle of (\ref{identa}) (for purely virtual
particles), which makes us unable to recover the right result. The case $t_{%
\text{f}}\rightarrow t_{\text{i}}$, $t\rightarrow t_{\text{f}}$, is
different, in this respect, because the right-hand side of (\ref{unitate})
is trivial.

\subsection{Hamiltonian for purely virtual particles?}

We have shown that the theories of physical and purely virtual particles
admit a unitary evolution operator $U_{\text{ph}}(t_{\text{f}},t_{\text{i}})$%
, built from the evolution operator $U(t_{\text{f}},t_{\text{i}})$ of
ordinary theories by means of a certain map $M_{\text{pv}}$. A natural
question, at this point, is: can we define a Hamiltonian for $U_{\text{ph}%
}(t_{\text{f}},t_{\text{i}})$? This is not an easy task. We could, for
example, differentiate $U_{\text{ph}}(t,t^{\prime })$ with respect to $t$,
or $t^{\prime }$, but the result, 
\begin{equation}
H_{\text{ph}}(t,t^{\prime })\equiv i\frac{\partial U_{\text{ph}}(t,t^{\prime
})}{\partial t}U_{\text{ph}}^{\dag }(t,t^{\prime }),  \label{h1}
\end{equation}%
depends on both $t$ and $t^{\prime }$. Then we would not know how to
reconstruct $U_{\text{ph}}(t,t^{\prime })$ from it. The time-ordered
exponential cannot be the right answer, since time ordering does not apply
to the diagrammatics of purely virtual particles \cite{PVP20}.

The question might have no answer, or multiple answers: each candidate
Hamiltonian, such as (\ref{h1}), must be equipped with a procedure to
reconstruct the evolution operator $U_{\text{ph}}(t_{\text{f}},t_{\text{i}})$
from it. What is the correct definition of energy, then? And what is the
fate of energy conservation? What we can say at present is that energy is
conserved at $\tau =\infty $, and is approximately conserved any time $\tau $
is longer enough than the duration $\Delta t$ of the interactions, as well
as when $\tau $ is longer than $1/m_{\text{pv}}$, where $m_{\text{pv}}$ is
the mass of the lightest purely virtual particle. When $\tau $ violates
these restrictions, micro violations are not excluded: the energy might be
conserved only upon averaging on time.

Be that as it may, the answer to this and any other question is encoded into
the unitary evolution operator $U_{\text{ph}}(t_{\text{f}},t_{\text{i}})$.
From a strictly physical point of view, $U_{\text{ph}}(t_{\text{f}},t_{\text{%
i}})$ is everything we need: it allows us to make (and hopefully test)
physical predictions for processes between arbitrary initial and final
states, with arbitrary initial and final times $t_{\text{i}}$ and $t_{\text{f%
}}$, and arbitrary boundary conditions (\ref{bou}).

\subsection{From ghosts to purely virtual particles}

We have mentioned ghosts, but we did not give enough details about them, and
their diagrammatic rules. Ghosts are particles $\phi $ with negative kinetic
terms. The correct way to treat them, by means of the functional integral,
is as follows. We split the set of couplings $\lambda $ into $\lambda _{%
\text{odd}}$ and $\lambda _{\text{even}}$, where the couplings $\lambda _{%
\text{odd}}$ multiply the vertices that contain an odd number of ghosts,
while the couplings $\lambda _{\text{even}}$ multiply the vertices that
contain an even number of ghosts. Denoting the action by $S(\phi ,\lambda _{%
\text{odd}},\lambda _{\text{even}})$, we perform the non-Hermitian change of
variables $\phi =i\tilde{\phi}$, and the non-Hermitian redefinition $\lambda
_{\text{odd}}=i\tilde{\lambda}_{\text{odd}}$, and switch to the action 
\begin{equation*}
\tilde{S}(i\tilde{\phi},i\tilde{\lambda}_{\text{odd}},\lambda _{\text{even}%
})\equiv S(\phi ,\lambda _{\text{odd}},\lambda _{\text{even}}),
\end{equation*}%
which has no ghosts and is itself Hermitian. Next, we treat $\tilde{\phi}$
as an ordinary physical particle, including its initial, final and boundary
conditions, and derive its diagrammatics, the unitarity equation (\ref%
{unitate}), as well as equation (\ref{unitar}), as before. The
coherent-state approach and every other tool we have used for ordinary
physical particles extend straightforwardly to $\tilde{\phi}$.\ 

We switch $\tilde{\phi},\tilde{\lambda}_{\text{odd}},\lambda _{\text{even}}$
back to $-i\phi ,-i\lambda _{\text{odd}},\lambda _{\text{even}}$ in the
final results. At that point, of course, the converted version of equation (%
\ref{unitate}) can no longer be interpreted as the unitarity equation. Note
that the ghost propagator turns out to be 
\begin{equation*}
-\frac{i}{p^{2}-m^{2}+i\epsilon }
\end{equation*}%
in the variables $\phi $. The $i\epsilon $ prescription shown here is
implied by the convergence of the functional integral in the variables $%
\tilde{\phi}$. Also note that the functional integral is not convergent is
the original ghost variables $\phi $ (which is the reason why we need to
switch to $\tilde{\phi}$).

The map $M_{\text{pv}}$ must be applied while working in the parametrization 
$\tilde{\phi},i\tilde{\lambda}_{\text{odd}},\lambda _{\text{even}}$, where
it is the same as before. This gives the identity (\ref{pvsuff}) in those
variables. Then, the conversion $\tilde{\phi},\tilde{\lambda}_{\text{odd}%
},\lambda _{\text{even}}\rightarrow -i\phi ,-i\lambda _{\text{odd}},\lambda
_{\text{even}}$ gives the right unitarity equation (\ref{pvsuff}) obeyed by
the evolution operator $U_{\text{ph}}(t_{\text{f}},t_{\text{i}})$. As far as
the initial and final conditions of the ghosts are concerned, they are
trivialized by the map $M_{\text{pv}}$. As far as the integral in between (%
\ref{pvsuff}) is concerned, it is trivialized as well, so we do not even
need to worry about its convergence before the switch $\phi \rightarrow i%
\tilde{\phi}$.

\section{Conclusions}

\label{conclusions}\setcounter{equation}{0}

Perturbative quantum field theory can be formulated in a finite interval of
time $\tau $ and on a compact space manifold $\Omega $ by expressing the
transition amplitudes between arbitrary initial and final states, with
arbitrary boundary conditions on $\partial \Omega $, in terms of diagrams,
which coincide internally with the ones we commonly use for the $S$ matrix
amplitudes at $\tau =\infty $, $\Omega =\mathbb{R}^{3}$ (apart from the
discretization of the loop momenta), and differ externally by the presence
of sources attached to every vertex (including endpoints). The sources take
care of the other details about the restriction to finite $\tau $ and
compact $\Omega $. The usual diagrammatic properties apply, or can be
generalized with little effort, provided we use the approach based on
coherent states. Every other approach exhibits remarkable complications,
which can be avoided if we reach it from the coherent-state approach through
a change of basis.

We have extended the dimensional and analytic regularization techniques to
finite $\tau $ and compact $\Omega $, by attaching an evanescent
(noncompact) manifold $\mathbb{R}^{-\varepsilon }$ to $\Omega $. We have
proved, under general assumptions, that renormalizability holds whenever it
holds at $\tau =\infty $, $\Omega =\mathbb{R}^{3}$, and that the divergences
are removed by the same counterterms.

Unitarity can be studied by means of the diagrammatic version of the
unitarity equation $U^{\dag }(t_{\text{f}},t_{\text{i}})U(t_{\text{f}},t_{%
\text{i}})=1$, obeyed by the evolution operator $U(t_{\text{f}},t_{\text{i}%
}) $, and its threshold decomposition into spectral optical identities. The
more general identity $U(t_{3},t_{2})U(t_{2},t_{1})=U(t_{3},t_{1})$ is also
studied diagrammatically.

Purely virtual particles are introduced by rendering some physical particles 
$\chi $ and ghosts $\chi _{\text{gh}}$ purely virtual. This is done as
follows: $i$) the $\chi $, $\chi _{\text{gh}}$ initial and final conditions
are trivialized, while their boundary conditions can stay nontrivial; and, $%
ii$) the on-shell contributions involving $\chi $ and $\chi _{\text{gh}}$
are removed from the diagrams.

If all the ghosts are rendered purely virtual, we obtain a theory of
physical and purely virtual particles, and its evolution operator $U_{\text{%
ph}}(t_{\text{f}},t_{\text{i}})$ is unitary. However, $U_{\text{ph}}(t_{%
\text{f}},t_{\text{i}})$ does not satisfy the more general identity $U_{%
\text{ph}}(t_{3},t_{2})U_{\text{ph}}(t_{2},t_{1})=U_{\text{ph}}(t_{3},t_{1})$%
.

The breakdown of this property is not totally upsetting, because, on a
second thought, it is inherent to the very concept of purely virtual
particle. Yet, it is a remarkable fact, because it implies that $U_{\text{ph}%
}(t_{\text{f}},t_{\text{i}})$ cannot be derived from a Hamiltonian in a
standard way. In this context, it is interesting to explore the fate of
energy conservation at microscales. Microviolations might make the pair with
the violations of microcausality, which are typical of the theories with
purely virtual particles.

\vskip.4 truecm \noindent {\large \textbf{Acknowledgments}} \nopagebreak %
\vskip.3 truecm \nopagebreak
We thank U. Aglietti, M. Bochicchio and D. Comelli for helpful discussions.

\vskip.6truecm

\noindent {\textbf{\huge Appendix}} \renewcommand{\thesection}{%
\Alph{section}} \renewcommand{\theequation}{\thesection.\arabic{equation}} %
\setcounter{section}{0}

\vskip.3truecm

\section[]{Calculation of $W_{0}(0)$}

\label{diagrammar20}\setcounter{equation}{0}

In this appendix, we calculate the quantity $W_{0}(0)$ of (\ref{proje}) and
the quantity $W_{0}(0,0)$ of (\ref{Wcohe}) in quantum mechanics. In the
approach based on position eigenstates, we have 
\begin{equation*}
Z_{0}(0)=\mathrm{e}^{iW_{0}(0)}=\!\!\!\!\!\!\!\!\!\underset{_{q(t_{\text{i}%
})=q(t_{\text{f}})=0}}{\int }\!\!\!\!\!\!\!\!\![\mathrm{d}q]\exp \left(
i\int_{t_{\text{i}}}^{t_{\text{f}}}\mathrm{d}t\mathbf{\hspace{0.01in}}%
L_{0}(q(t))\right) =\langle 0_{q}\hspace{0.01in}|\mathrm{e}^{-iH_{0}\tau }|%
\hspace{0.01in}0_{q}\rangle ,
\end{equation*}%
where $H_{0}$ is the free Hamiltonian of the harmonic oscillator (of unit
mass) and $|\hspace{0.01in}0_{q}\rangle $ is the position eigenstate with
eigenvalue zero. Inserting complete sets of $H_{0}$ eigenstates $\hspace{%
0.01in}|\hspace{0.01in}n\rangle $, $|\hspace{0.01in}m\rangle $ we can also
write%
\begin{equation*}
\langle 0_{q}\hspace{0.01in}|\mathrm{e}^{-iH_{0}\tau }|\hspace{0.01in}%
0_{q}\rangle =\sum_{n,m=0}^{\infty }\langle 0_{q}\hspace{0.01in}|\hspace{%
0.01in}n\rangle \langle n\hspace{0.01in}|\mathrm{e}^{-iH_{0}\tau }|\hspace{%
0.01in}m\rangle \langle m\hspace{0.01in}|\hspace{0.01in}0_{q}\rangle
=\sum_{n=0}^{\infty }|\hspace{0.01in}\psi _{n}(0)|^{2}\mathrm{e}%
^{-iE_{0n}\tau },
\end{equation*}%
where 
\begin{equation*}
\psi _{n}(q)=\frac{\omega ^{1/4}}{\sqrt{2^{n}n!}\pi ^{1/4}}\mathrm{e}%
^{-\omega q^{2}/2}\mathcal{H}_{n}(\omega ^{1/2}q)
\end{equation*}%
is the normalized $H_{0}$ eigenfunction with eigenvalue $E_{0n}=(2n+1)\omega
/2$, $\mathcal{H}_{n}$ denoting the $n$th Hermite polynomial. We have%
\begin{equation*}
\mathcal{H}_{n}(0)=\left\{ 
\begin{tabular}{l}
$(-2)^{n/2}(n-1)!!$ for $n$ even, \\ 
$0$ for $n$ odd.%
\end{tabular}%
\right.
\end{equation*}%
Hence,%
\begin{equation*}
\!\!\!\!\!\!\!\!\!Z_{0}(0)=\left( \frac{\omega }{\pi }\right) ^{1/2}\mathrm{e%
}^{-i\omega \tau /2}\sum_{n=0}^{\infty }\frac{((2n-1)!!)^{2}}{(2n)!}\mathrm{e%
}^{-2in\omega \tau }.
\end{equation*}%
Normally, the result is written as $(\omega /\pi )^{1/2}/\sqrt{2i\sin
(\omega \tau )}$ for $|\sin (\omega \tau )|<1$. To be more general, we keep
the expansion explicit. Actually, it is more convenient to write it as $%
\mathrm{e}^{iW_{0}(0)}$, where the expansion simplifies. We easily find%
\begin{equation*}
\!\!\!\!\!\!\!\!\!W_{0}(0)=-\frac{\omega \tau }{2}-\frac{i}{2}\ln \frac{%
\omega }{\pi }-i\sum_{n=1}^{\infty }\frac{\mathrm{e}^{-2in\omega \tau }}{2n}.
\end{equation*}

In the case of coherent states, we have%
\begin{equation*}
\exp \left( iW_{0}(0,0)\right) =\!\!\!\!\!\!\!\!\!\!\!\!\underset{w(t_{\text{%
i}})=\bar{w}(t_{\text{f}})=0}{\int }\!\!\!\!\!\!\!\!\!\![\mathrm{d}w\mathrm{d%
}\bar{w}]\mathrm{\exp }\left( i\int_{t_{\text{i}}}^{t_{\text{f}}}\mathrm{d}t%
\hspace{0.01in}\mathcal{L}_{0}(w,\bar{w})\right) =\frac{C}{\sqrt{\det Q}},
\end{equation*}%
where $Q$ is given in (\ref{Q}) and $C$ is a numerical factor, which we
choose so that $W_{0}(0,0)=0$ at $\tau =0$. It is convenient to first work
out the $Q$ eigenstates%
\begin{equation*}
i\dot{w}_{n}-\omega w_{n}=\lambda _{n}\bar{w}_{n},\qquad -i\dot{\bar{w}}%
_{n}-\omega \bar{w}_{n}=\lambda _{n}w_{n},\qquad w_{n}(t_{\text{i}%
})=0,\qquad \bar{w}_{n}(t_{\text{f}})=0,
\end{equation*}%
and the $Q$ eigenvalues $\lambda _{n}$. We find that $\sigma _{n}(\omega
)\equiv \sqrt{\lambda _{n}^{2}-\omega ^{2}}$ is the solution of the equation%
\begin{equation*}
2\tau \sigma _{n}-\ln (\omega +i\sigma _{n})+\ln (\omega -i\sigma _{n})=0.
\end{equation*}%
At $\omega =0$ we have $\sigma _{n}(0)=(2n+1)i\pi /(2\tau )$, $n\in \mathbb{Z%
}$. At $\omega \neq 0$, we work out $\sigma _{n}(\omega )$ as a series
expansion in powers of $\omega $, around $\sigma _{n}(0)$. We find%
\begin{equation*}
\ln \sqrt{\frac{\det Q(0)}{\det Q(\omega )}}=-\frac{1}{2}\sum_{n\in \mathbb{Z%
}}\ln \frac{\lambda _{n}(\omega )}{\lambda _{n}(0)}=-\frac{i\omega \tau }{2}.
\end{equation*}%
Fixing $C$ as said, we conclude that $W_{0}(0,0)=-\omega \tau /2$.

\end{document}